\documentclass[12pt, journal, onecolumn]{IEEEtran}
\usepackage{amsmath,amssymb,amsfonts}
\usepackage{graphicx}
\usepackage{textcomp}
\usepackage{xcolor}
\usepackage{cases}
\usepackage{algorithm}
\usepackage[noend]{algpseudocode}
\usepackage{tikz} 
\usepackage{cite}
\usepackage{comment}
\usepackage{balance}
\usepackage{subfigure}
\usepackage{color}

\usepackage[thmmarks, amsmath, thref]{ntheorem}

\theoremstyle{plain}
\theoremheaderfont{\upshape\bfseries}
\theoremseparator{.}
\theorembodyfont{\upshape}
\newtheorem{theorem}{Theorem}
\newtheorem{lemma}{Lemma}
\newtheorem{corollary}{Corollary}
\newtheorem{assumption}{Assumption}
\newtheorem{definition}{Definition}
\newtheorem{remark}{Remark}
\newtheorem{proposition}{Proposition}

\theoremstyle{nonumberplain}
\theoremseparator{:}
\theorembodyfont{\upshape}
\theoremsymbol{\ensuremath{\color{black}\blacksquare}}
\newtheorem{proof}{Proof}

\newtheorem{example}{Example}

\usepackage{caption}
\usepackage{yhmath}
\usepackage{enumitem}

\newcommand{\bs}{\boldsymbol}

\newcommand\independent{\protect\mathpalette{\protect\independenT}{\perp}}
\def\independenT#1#2{\mathrel{\rlap{$#1#2$}\mkern2mu{#1#2}}}

\renewcommand{\thepage}{\arabic{page}}
\renewcommand{\thesection}{\arabic{section}}
\renewcommand{\theequation}{\arabic{equation}}
\renewcommand{\thefigure}{\arabic{figure}}
\renewcommand{\thetable}{\arabic{table}}

\allowdisplaybreaks

\usepackage{xspace}
\usepackage{bbm}
\usepackage{mathrsfs}

\def\Var{\mathop{\rm Var}\nolimits}%
\def\Cov{\mathop{\rm Cov}\nolimits}%
\newcommand{\Ec}{\mathcal{E}}

\newcommand{\Pcal}{\mathcal{P}}

 \def\E{\mathsf{E}}

\newcommand{\U}{\mathrm{Unif}}

\def\textiid{i.i.d.\@\xspace}
\newcommand\iid{\ifmmode\text{ i.i.d. } \else \textiid \fi}

\newcommand{\Real}{\mathbb{R}}

\makeatletter
\def\thickhline{%
  \noalign{\ifnum0=`}\fi\hrule \@height \thickarrayrulewidth \futurelet
   \reserved@a\@xthickhline}
\def\@xthickhline{\ifx\reserved@a\thickhline
               \vskip\doublerulesep
               \vskip-\thickarrayrulewidth
             \fi
      \ifnum0=`{\fi}}
\makeatother

\newlength{\thickarrayrulewidth}
\usepackage{hyperref}

\begin{document}
%
\title{Learning Invariant Representations under General Interventions on the Response}
%
%
%

\author{%
 Kang~Du and Yu~Xiang
 
\thanks{K. Du and Y. Xiang are with the Department of Electrical and Computer Engineering, University of Utah, Salt Lake City,
UT, 84112, USA. This paper has supplementary downloadable material available at http://ieeexplore.ieee.org., provided by the authors. The material includes all the proofs. Contact \{kang.du, yu.xiang\}@utah.edu for further questions about this work.
}}

\maketitle

\begin{abstract}
It has become increasingly common nowadays to collect observations of feature and response pairs from different environments. As a consequence, one has to apply learned predictors to data with a different distribution due to distribution shifts. One principled approach is to adopt the structural causal models to describe training and test models, following the invariance principle which says that the conditional distribution of the response given its predictors remains the same across environments. However, this principle might be violated in practical settings when the response is intervened. A natural question is whether it is still possible to identify other forms of invariance to facilitate prediction in unseen environments. To shed light on this challenging scenario, we focus on linear structural causal models (SCMs) and introduce \emph{invariant matching property (IMP)}, an explicit relation to capture interventions through an additional feature, leading to an alternative form of invariance that enables a unified treatment of general interventions on the response as well as the predictors. We analyze the asymptotic generalization errors of our method under both the discrete and continuous environment settings, where the continuous case is handled by relating it to the semiparametric varying coefficient models. We present algorithms that show competitive performance compared to existing methods over various experimental settings including a COVID dataset. 

\end{abstract}


%
\IEEEpeerreviewmaketitle

\section{Introduction}
\label{sec:introduction}
%
%
%
%
\IEEEPARstart{H}{ow} to make reliable predictions in unseen environments that are different from training environments is a challenging problem, which is fundamentally different from the classical machine learning settings~\cite{quinonero2008dataset,weiss2016survey,csurka2017domain}. Modeling these distribution shifts in a principled way is of great importance in many fields including robotics, medical imaging, and environmental science. Apparently, this problem is ill-posed without any constraints on the relationship between training and test distributions, as the test distribution may be arbitrary. Consider the problem of predicting the response $Y$ given its predictors $X = (X_1,...,X_d)^\top$ in unseen environments. 
To model distribution changes across different environments (or training and test distributions), we follow the approach of using \emph{structural causal models} (SCMs)~\cite{pearl2009causality,peters2017elements} to model different data-generating mechanisms. The common assumption is that the assignment for $Y$ does not change across environments (or $Y$ is not intervened), which allows for natural formulations of the invariant conditional distribution of $Y$ given a subset of $X$~\cite{scholkopf2012causal,zhang2013domain,peters2016causal,peters2017elements,rojas2018invariant,heinze2017conditional,buhlmann2020invariance,ghassami2018multi,ghassami2017learning}. The underlying principle is known as invariance, autonomy or modularity~\cite{haavelmo1944probability,aldrich1989autonomy,pearl2009causality,imbens2015causal}. 

Following this principle, the invariance-based causal prediction initiated in~\cite{peters2016causal} (also see~\cite{meinshausen2018causality} and~\cite{buhlmann2020invariance} and references therein) assumes that the conditional distribution of $Y$ given a set of predictors $X_S\subseteq\{X_1,...,X_d\}$ is invariant in all environments, i.e., $\Pcal_e(Y|X_S) =  \Pcal_h(Y|X_S)$ for environments $e$ and $h$, where $(X, Y)$ is generated according to the joint distribution $\Pcal_e:=\Pcal_e^{X,Y}$. Focusing on linear SCMs, it assumes the existence of a linear model that is invariant across environments, with an unknown noise distribution and arbitrary dependence among predictors (see extensions to nonlinear~\cite{heinze2018invariant} and time series~\cite{pfister2019invariant} settings). Following this framework, theoretical guarantees for domain adaptation have been developed in~\cite{rojas2018invariant,magliacane2018domain}. More recently, a multi-environment regression method for domain adaptation called the \emph{stabilized regression}~\cite{pfister2021stabilizing} explicitly enforces stability (based on a weaker version of invariance $\E_{\Pcal_e}[Y|X_S=x_s]=\E_{\Pcal_h}[Y|X_S=x_s]$) by introducing the \emph{stable blanket}, which is a refined version of the \emph{Markov blanket} to promote generalization. The tradeoff between predictive performance on training and test data has been studied via regularization under shift interventions~\cite{rothenhausler2021anchor}. Motivated by~\cite{peters2016causal}, the invariant risk minimization (IRM)~\cite{arjovsky2019invariant} only uses data from the training environments (i.e., the out-of-distribution generalization setting), and imposes $\Pcal_e(Y|\phi(X)) =  \Pcal_h(Y|\phi(X))$, where $\phi$ is invariant across environments, leading to a bi-leveled optimization problem that is not practical. Several relaxed versions of IRM have been proposed in~\cite{arjovsky2019invariant}, but they behave very differently from the original IRM~(see, e.g.,~\cite{rosenfeld2021risks,kamath2021does}). For a framework of the out-of-distribution setting from a causal perspective with a focus on minimizing the worst-case risk, see~\cite{christiansen2021causal} and references therein. \emph{In this line of invariance-based work, the fundamental assumption is that interventions on the target variable $Y$ are not allowed}. 

In practical settings, however, the structural assignment of $Y$ might change across environments, namely, $Y$ might be intervened. How to relax this assumption in a principled way is one of the main motivations in our work. We propose to explore alternative forms of invariance and make an attempt in this direction by focusing on linear SCMs. Concretely, the assignment for $Y$ allows general interventions  
\begin{align*}
 Y^{e} = (\beta^{e})^{\top}X^{e} +  \varepsilon_{Y}^{e},
\end{align*}
where $Y$ can be intervened through coefficient $\beta^e$ and/or the noise $\varepsilon_Y^e$, to capture the dependence of structural assignment across different environments (preliminary results have been reported by the same authors in~\cite{du2022invariant}). We consider a multi-environment regression setting for domain adaption: There are multiple training data $(X^e, Y^e)$ for $e\in\mathcal{E}^{\text{train}}$ that are generated from a training model and one test data (indexed by  $e^{\text{test}}$) from a test model; we assume the training model and test model follow SCMs with the same \emph{but unknown} graph structure, but we allow $\beta^{e}$ and the mean and variance of $\varepsilon_Y^e$ to be arbitrarily different under the two models. To avoid the setting to be ill-posed, a key necessary condition is that $Y^e$ needs to have at least one child in the SCMs, as prediction is not possible otherwise given that $Y^e$ may change arbitrarily over environments. The main challenge lies in whether it is still possible to identify other forms of invariance to facilitate prediction in the test environment. We propose an alternative form of invariance $\Pcal_e(Y|\phi_e(X)) =  \Pcal_h(Y|\phi_h(X))$ that is enabled by a family of conditional invariant transforms $\Phi\ni\phi_e,\phi_{h}$. Under general interventions on $Y$, we provide explicit constructions of such transforms by developing \emph{invariant matching property (IMP)}, a deterministic relation between an estimator of $Y$ and $X$ along with an additional predictor constructed from $X$. To enable a systematic way of constructing the IMP, we provide a natural decomposition of it and demonstrate this when only $Y$ is intervened or both $X$ and $Y$ are intervened. The IMP comes with several appealing features: (1) it does not vary over environments, making it applicable in unseen environments, and (2) the identification of the IMP follows directly from the fact that the training data contains multiple environments. We study the asymptotic generalization error for both the discrete environment setting and continuous counterpart, which is the more challenging setting. Interestingly, we reveal a connection between the continuous environment setting with the semiparametric varying models, which makes the asymptotic generalization analysis possible. We believe that our results open up new possibilities for multi-environment regression methods for domain adaptation under the structure causal models.


\subsection{Further Related Work} \label{sec:related_work}
In~\cite{chen2021domain}, the authors have provided a systematic treatment of domain adaptation using the SCMs to enable analysis and comparisons of domain adaption methods, which leads to the conditional invariant residual matching (CIRM) method. The CIRM and its variants combine the domain invariant projection (DIP)-type methods~(see \cite{baktashmotlagh2013unsupervised,ganin2016domain} and the generalized label shift to handle target label perturbation~\cite{li2019target,tachet2020domain}) with the idea of conditional invariance penalty (appeared in~\cite{gong2016domain,heinze2021conditional} under slightly different settings) that assumes the existence of conditional invariant components (CICs) in the anticausal setting where $Y$ causes $X$. Theoretical guarantees have been provided for the prediction performance under shift interventions on $Y$, while numerical studies are provided for interventions on the noise variance of $Y$~\cite{chen2021domain}. \emph{It has also been pointed out that the general mixed-causal-anticausal domain adaptation problem remains open}. We aim to shed light on this challenging setting by constructing explicit conditional invariant transforms. 


The role of causality in facilitating domain adaptation problems is first articulated in~\cite{scholkopf2012causal}, focusing on causal and anticausal predictions. Reweighting methods have been extensively studied for covariate shift~\cite{quinonero2008dataset,storkey2009training,sugiyama2012machine}, which assumes that only the feature distribution changes over environments while the conditionals remain the same. The label shift, which aligned with the anticausal setting, has attracted much attention recently~\cite{lipton2018detecting,azizzadenesheli2018regularized,garg2020unified}. Many other interesting domain adaption methods have been developed but they are less related to this work. The performance bounds using Vapnik-Chervonenkis (VC) theory have been initiated in~\cite{ben2006analysis}. There are fundamental works from the robust statistics perspective including distributional robust learning~\cite{bagnell2005robust,hu2013kullback,sinha2018certifying,gao2017distributional,lee2018minimax,duchi2021learning} and adversarial machine learning~\cite{goodfellow2018making,raghunathan2018semidefinite}. 

\subsection{Contribution and Structure}
There are four main contributions in this work. (I) We formulate a general invariance property and the corresponding conditional invariant transforms for analyzing general interventions on linear SCMs (Section~\ref{sec:formulation}). (II) We tackle this problem by introducing the invariant matching property (IMP) and providing a systematic approach for establishing explicit characterization of it (Section~\ref{sec:invariance} and Section~\ref{sec:chara_mpm}). (III) To handle the continuous environment setting, we bridge our framework with the profile likelihood estimators developed in the semiparametric literature, leading to asymptotic performance guarantees under this challenging setting (Section~\ref{sec:profile}). (IV) Motivated by our theoretical results, we develop efficient algorithms that show competitive empirical performance over a variety of simulation settings including a COVID dataset (Section~\ref{sec:alg} and Section~\ref{sec:exp}). All the technical details are deferred to Appendix.

\section{Background and Problem Formulation} 
\label{sec:formulation}
 Consider a response $Y \in \mathbb{R}$ and a vector of predictors  $X=(X_{1},\ldots,X_{d})^{\top} \in \mathcal{X}\subseteq \mathbb{R}^{d}$ following an acyclic linear SCM, 
\begin{numcases}{\mathcal{M}:}
    X = \gamma Y + BX+\varepsilon_{X} \nonumber\\
    Y = \beta^{\top}X +  \varepsilon_{Y}, \nonumber
\end{numcases}
where $\beta,\gamma \in \mathbb{R}^{d}$, $B \in \mathbb{R}^{d \times d}$, $\varepsilon_{X}=(\varepsilon_{X_1},\ldots, \varepsilon_{X_d})^\top$, and the noise variables $\{\varepsilon_{X_1},\ldots, \varepsilon_{X_d}\}$ and $\varepsilon_{Y}$ are jointly independent. We use $\mathcal{G}(\mathcal{M})$ to denote the directed acyclic graph induced by $\mathcal{M}$, with edges determined by the non-zero coefficients in $\mathcal{M}$. We denote the parents, children, descendants, and Markov blanket of a variable $Z \in \{X_{1},\ldots,X_{d},Y\}$ as $PA(Z)$, $CH(Z)$, $DE(Z)$, and $MB(Z)$, respectively. When $(X,Y)$ is observed in different environments (e.g., different experiment settings for data collection), the parameters $\{\beta, \gamma, B\}$ and the distributions of $\{\varepsilon_{X},\varepsilon_{Y}\}$ may change. In the following, we use interventions on the SCM $\mathcal{M}$ to model such changes.

Let $\mathcal{E}^{\text{all}}$ denote the set of all possible environments~\footnote{We use training (or test) environments and observable (or unseen) environments interchangeably.}, which are partitioned into \emph{multiple} training environments $\mathcal{E}^{\text{train}}$ and one test environment $\{e^{\text{test}}\}$ such that $\mathcal{E}^{\text{all}} =  \mathcal{E}^{\text{train}}\cup\{e^{\text{test}}\}$. In each environment $e\in \mathcal{E}^{\text{all}}$, $(X, Y)$ is generated according to the joint distribution $\Pcal_e:=\Pcal_e^{X,Y}$, and to simplify notation, we write $\Pcal_{\text{test}}:=\Pcal_{e^\text{test}}^{X,Y}$. A variable from $\{X_{1},\ldots,X_{d}, Y\}$ is intervened if the parameters or noise distribution in its assignment changes over different $e \in \mathcal{E}^{\text{all}}$. For instance, the changes of $\beta$ and/or the distribution of $\varepsilon_{Y}$ correspond to the intervention on $Y$ (see different intervention cases below). Importantly, we allow both $Y$ and any subset of $\{X_{1},\ldots,X_{d}\}$ to be intervened in $\mathcal{E}^{\text{all}}$.

Now, we introduce the linear SCM $\mathcal{M}$ with parameters that change with environments. For each $e \in \mathcal{E}^{\text{all}}$, the linear SCM $\mathcal{M}$ is modified to be
\begin{numcases}
{\mathcal{M}^{e}:}
    X^{e} = \gamma^{e} Y^{e} + B^{e} X^{e}+ \varepsilon_{X}^{e} \\
    Y^{e} = (\beta^{e})^{\top}X^{e} +  \varepsilon_{Y}^{e}. 
    \label{eq:linear_scm_u}
\end{numcases}
This formulation is fairly general. From the structural perspective, this consists of causal, anticausal, and mixed-causal-anticausal settings~\cite{scholkopf2012causal}. It should be noted that we only adopt the linear SCM rather than the fully specified SCMs as in~\cite{pearl2022external}, since learning the functional forms can be more complicated than the prediction problem we aim to solve. Regarding the intervention types, we discuss several special cases to put them into perspective.

\begin{enumerate}[leftmargin=*]
\item \emph{Shift interventions on $X$ or $Y$:}
A variable $X_{j}$ is intervened through a shift if the mean of the noise variable $\varepsilon_{X,j}^{e}$ changes with $e\in \mathcal{E}^{\text{all}}$. For the shift intervention on $Y$, the mean of $\varepsilon_{Y}^{e}$ changes. 

\item \emph{Interventions on the coefficients of $X$ or $Y$:}
A variable $X_{j}$ is intervened through coefficients if the coefficients $\{\gamma_{j}^{e},B_{j\cdot}^{e}\}$ change with $e\in \mathcal{E}^{\text{all}}$. For $Y$, the change is on the coefficient vector $\beta^{e}$. 

\item \emph{Interventions on the noise variance of $X$ or $Y$:} Similar to shift interventions, a variable $X_{j}$ or $Y$ is intervened if its noise variance changes.
\end{enumerate} 

We observe $n^e$ \iid samples $\{(x_1, y_1), ..., (x_{n^e}, y_{n^e})\}$ from each training environment distribution $\Pcal_{e}$ for $e\in\mathcal{E}^{\text{train}}$, but in the test environment $e^\text{test}$ we only observe $m$ \iid samples $\{x_1, ..., x_{m}\}$ from $\Pcal_{\text{test}}^X$. The goal is to learn a function $f: \mathcal{X}\to\hat{\mathcal{Y}}$ that works well on $e^\text{test}$ in the sense that it minimizes the \emph{test population loss}
\begin{equation}
\mathcal{L}_{\text{test}}(f):=\E_{(X, Y)\sim\Pcal_{\text{test}}}[l(Y,f(X))].   \label{mini_every_u}
\end{equation}
where $l$ is the square loss function $l(y,\hat{y})= (y-\hat{y})^{2}$. The optimal function is $f(x) = \E_{\Pcal_{\text{test}}}[Y|X=x]$, which cannot be learned from the observed data in general when $X$ and/or $Y$ is intervened. Without any constraints on the relationship between $\Pcal_{\text{test}}$ and $\Pcal_{e}$ for $e\in\mathcal{E}^{\text{train}}$, the test population loss can be arbitrarily large. To make this problem tractable, we assume that $(X,Y)$ under $\Pcal_{\text{test}}$ and $\Pcal_{e}$ are generated according to the SCMs described in~\eqref{eq:linear_scm_u} but we \emph{do not assume that the causal graph is known} and we \emph{allow for general types of interventions}. 

It is well-known that if $Y$ is not intervened, a general form of invariance principle applies, assuming the existence of some subset $S\subseteq\{1,...,d\}$ such that
\begin{equation}
    \Pcal_e(Y|X_S) =  \Pcal_h(Y|X_S)
    \label{eq:ICP}
\end{equation}
holds for any $e,h \in \mathcal{E}^{\text{all}}$. Under this assumption, the \emph{causal function} $\E_{\Pcal_{e}}[Y|X_{PA(Y)}]$ is invariant across different environments and minimax under the class of all possible interventions on $X$~\cite{rojas2018invariant}. If not every predictor is intervened arbitrarily, previous works that are motivated by~\eqref{eq:ICP} (as mentioned in Section~\ref{sec:introduction}) aim to improve upon the causal function. The main challenge in our setting comes from the general interventions on $Y$, making the traditional invariance principle not applicable. Importantly, the causal function can change arbitrarily with environments. In this work, we propose to exploit an alternative form of invariance to tackle this problem.
\smallskip

\begin{definition}
A function $\phi : (\mathcal{E}^{\text{all}},\mathbb{R}^{d}) \to \mathbb{R}^{q}$ is called a \emph{conditional invariant transform} if the following \emph{invariance property} holds for any $e,h \in \mathcal{E}^{\text{all}}$
\begin{equation}
    \Pcal_e(Y|\phi_{e}(X)) =  \Pcal_h(Y|\phi_{h}(X)). \label{inva_trans}
\end{equation}
\end{definition}
\smallskip
Under general intervention settings, we denote this class of conditional invariant transforms as $\Phi$, and we provide \emph{explicit} characterizations of it via the invariant matching property (IMP) (see Definition~\ref{def:imp}). For each $\phi \in \Phi$, the invariance property~\eqref{inva_trans} enables us to compute
\begin{equation}
    f_{\phi_e}(x) = g\circ \phi_e(x)=\E_{\Pcal_{\text{e}}}[Y|\phi_{e}(x)], \label{def_f_phi}
\end{equation}
for any $e\in \mathcal{E}^{\text{all}}$, where the function $g: \Real^q\to \hat{\mathcal{Y}}$ is invariant across environments and is nonlinear in general. 
Equivalently, this solves a relaxed version of~\eqref{mini_every_u} by minimizing $\mathcal{L}_{\text{test}}(f_{\phi})$ over $\{\phi \in \Phi \}$. This formulation allows us to treat the general mixed-causal-anticausal problem under general interventions on $Y$ in a unified manner.

\section{Invariant Matching Property and Theoretical Guarantees} 
\label{sec:invariance}
Among the intervention settings below~\eqref{eq:linear_scm_u}, the interventions on only $Y$ are important but rarely studied, which concerns the changes of $\beta^{e}$ and the distribution of $\varepsilon^{e}_{Y}$. Our method can be motivated in this setting by the following observation: If $X$ includes any descendants of $Y$, then $\beta$ and $\varepsilon_{Y}$ (that may change arbitrarily in the unseen environments) will be passed on to the descendants. Thus, the changes might be revealed by the changes of certain statistical properties of $X$, leading to our proposed invariant matching property detailed in this section. We start with a few toy examples. 
\subsection{Motivating Examples}
\label{subsec:example}

\smallskip
\begin{example}
\label{example1}
Consider $(Y^{e},X^{e}),e \in \mathcal{E}^{\text{toy}}=\{1,2\}$, with $X^{e} := (X_{1}^{e},X_{2}^{e},X_{3}^{e})^{\top}$ satisfying the following linear acyclic SCM (illustrated in Fig.~\ref{fig_toy}), \\
\begin{figure}[h]
\centering
 \vspace{-2em}
\includegraphics[width=0.2\linewidth]{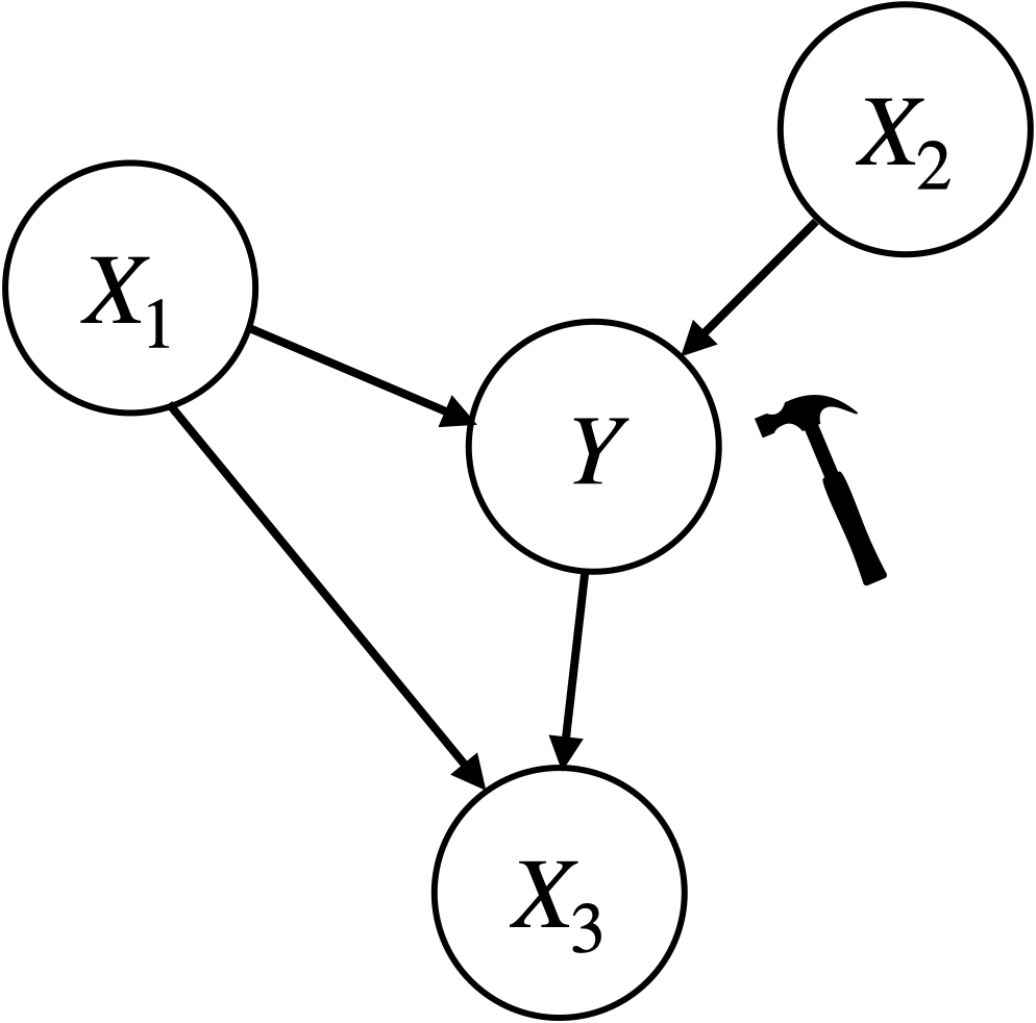}
\caption{Directed acyclic graph $\mathcal{G}(\mathcal{M}_{\text{toy}}^{e})$. \label{fig_toy}
}
 \vspace{-1em}
\end{figure}
\begin{align}
\mathcal{M}_{\text{toy}}^{e}:\label{scm_toy}
 \begin{cases}
    Y^{e} = a^{e} X_1^{e} + X_2^{e} + N_{Y}^{e}\\
    X_3^{e} = Y^{e} + X_1^{e} + N_3^{e},
\end{cases}
\end{align}
where $X_1^{e},X_2^{e},N_{3}^{e}$, and $N_{Y}^{e}$ are independent and $\mathcal{N}(0,1)$-distributed for every $e \in \mathcal{E}^{\text{toy}}$. Since $(Y^{e},X^{e})$ is multivariate Gaussian, the MMSE estimator of $Y^{e}$ given $X^{e}$ is
\begin{align}
\E_{\Pcal_{e}}[Y|X]  =& X^{\top}\left(\E_{\Pcal_{e}}[XX^{\top}]\right)^{-1}\E_{\Pcal_{e}}[XY]\nonumber\\
=& \frac{1}{2} (a^{e}-1)X_{1}^{e} + \frac{1}{2}X_{2}^{e} + \frac{1}{2}X_{3}^{e}, \nonumber 
\end{align}
\begin{figure*}[h]\centering
\begin{subfigure}{}\includegraphics[scale=0.21]{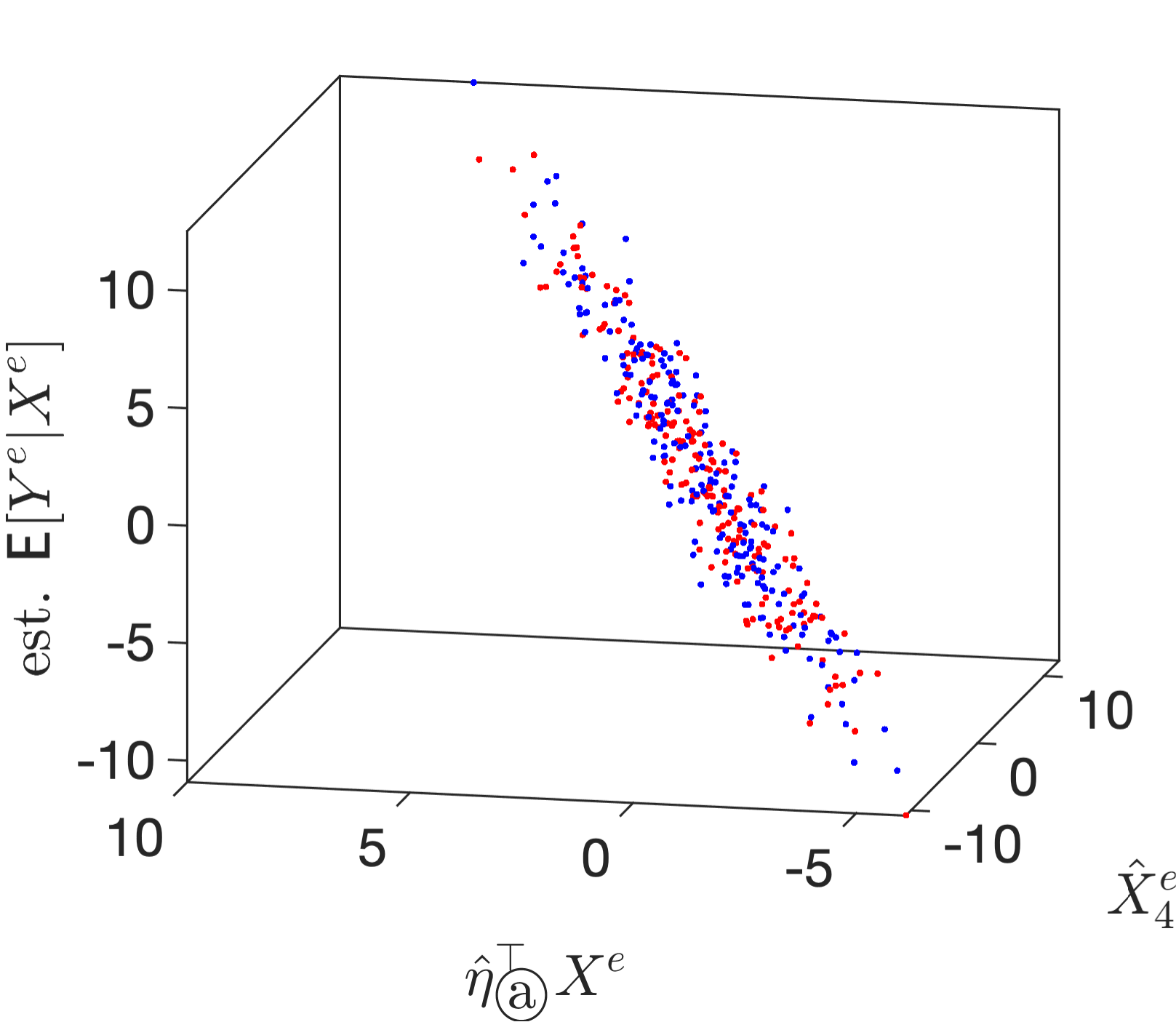}\end{subfigure}
\begin{subfigure}{}\includegraphics[scale=0.21]{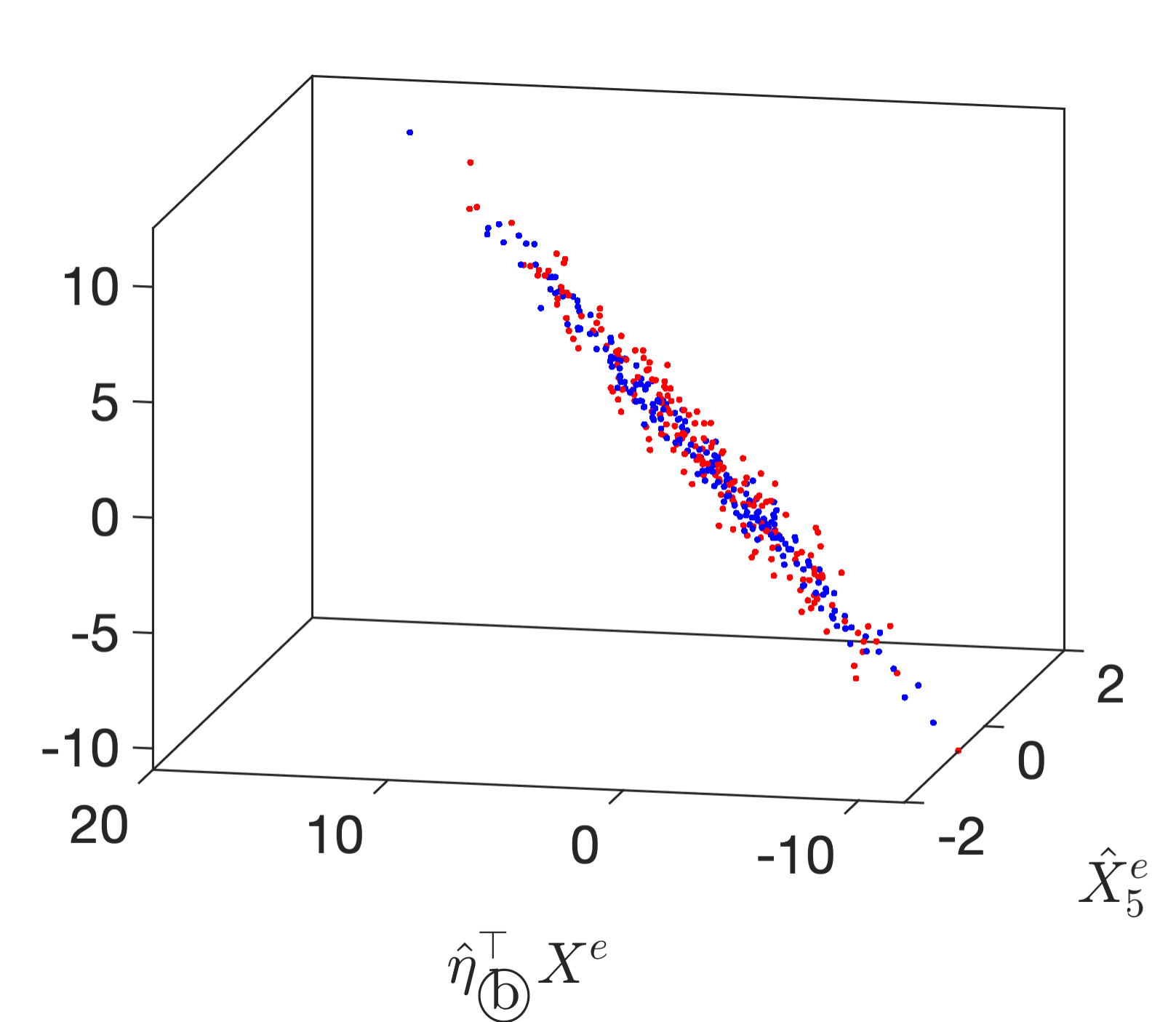}\end{subfigure}
\begin{subfigure}{}\includegraphics[scale=0.21]{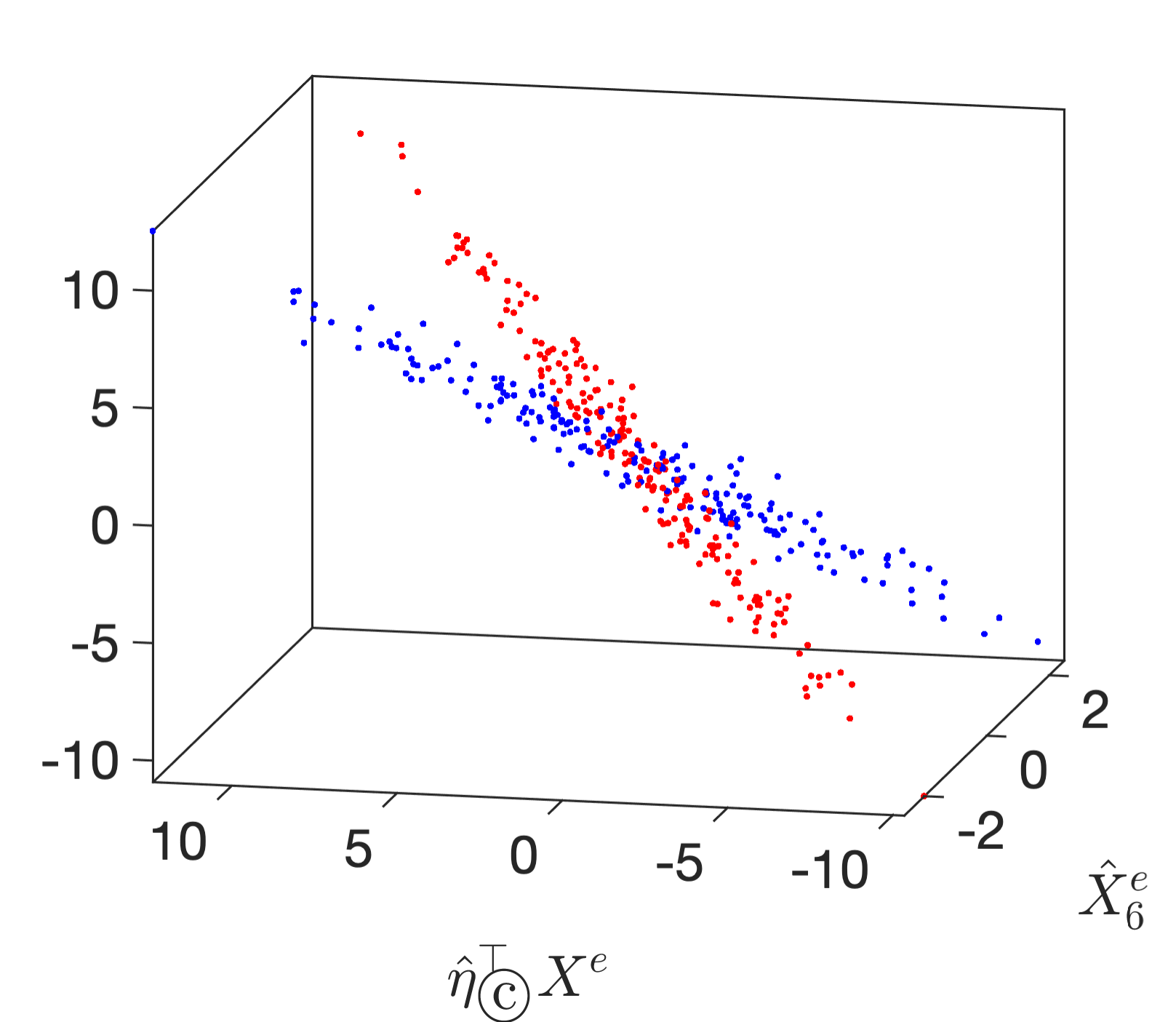}\end{subfigure}
\caption{\label{toy_plot}Illustrations of invariant relations in Example~\ref{example1}. In (a), we illustrate the linear invariant relation~\eqref{toy_determin} by visualizing estimates\protect\footnotemark~of the tuple $(X_{4}^{e}$, $\eta^{\top}X^{e}, \E_{\Pcal_e}[Y|X])$ (corresponding to the $\{x,y,z\}$ axes).
 Similarly, for (b) and (c), we verify such a relation for $X_{5}^{e}$ and $X_{6}^{e}$. The overlaps between the red dots ($e=1$) and the blue dots ($e=2$) in (a) and (b) indicate the invariant linear relations. However, the red and blue dots in (c) are not aligned, implying that no invariant relations as in~\eqref{toy_determin} and~\eqref{inv_rel_x2} hold for $(\E_{\Pcal_e}[Y|X], X_{6}^{e})$.
} 
\end{figure*}which is not directly applicable for predicting $Y^{e^{\text{test}}}$ as $a^{e}$ can change arbitrarily. Similarly, one can compute $\E_{\Pcal_{e}}[X_{3}|X_{1},X_{2}] = (1+a^{e})X_{1}^{e} + X_{2}^{e}$. It is noteworthy that $\E_{\Pcal_{e}}[Y|X]$ and $\E_{\Pcal_{e}}[X_{3}|X_{1},X_{2}]$ can each be generated by a linear combination of $\E_{\Pcal_{e}}[Y|X_{PA(Y)}] = a^{e}X_{1}^{e}+X_{2}^{e}$ and $\{X_{1}^{e},X_{2}^{e},X_{3}^{e}\}$, and we will highlight this observation by introducing two invariant relations (see Def.~\ref{def:first} and Def.~\ref{def:second}). As a result, there exists a deterministic linear relation, which we refer to as \emph{matching},
\begin{align}
\E_{\Pcal_{e}}[Y|X]   &= \lambda\E_{\Pcal_{e}}[X_{3}|X_{1},X_{2}] +\eta^{\top} X^{e},
\label{toy_determin}
\end{align}
with coefficients $\lambda = 1/2$ and $\eta = (-1,0,1/2)^{\top}$ that are \emph{invariant} with respect to the environment (illustrated in Fig.~\ref{toy_plot}.(a)-(b)). For simplicity of notation, we denote $X_{4}^{e} := \E_{\Pcal_{e}}[X_{3}|X_{1},X_{2}]$ in Fig.~\ref{toy_plot}. Moreover, one can verify that $\Pcal_{e}(Y|X,X_{4})$ is invariant since $\E_{\Pcal_{e}}[Y|X,X_{4}]$ and $\Var_{\Pcal_{e}}(Y|X,X_{4})$ are invariant. Thus $\phi_{e}(X)=(X^{\top},X_{4})^{\top}$ satisfies the invariance property~\eqref{inva_trans}. A prediction model in~\eqref{toy_determin} with invariant coefficients is often not unique when it exists. One can show that
\begin{align}
    \E_{\Pcal_{e}}[Y|X]= - X_{1}^{e} + \frac{1}{2}X_{2}^{e}+X_{3}^{e} -\frac{3}{2}X_{5}^{e}, \label{inv_rel_x2}
\end{align}
with $X_{5}^{e} := \E_{\Pcal_{e}}[X_{2}|X_{1},X_{3}]$. 
However, invariant relations in~\eqref{toy_determin} and~\eqref{inv_rel_x2} do not hold for $X_{6}^{e} := \E_{\Pcal_{e}}[X_{1}|X_{2},X_{3}]$, since
\begin{align}
   X_{6}^{e} = -\frac{a^{e}+1}{(a^{e}+1)^2+2}X_2^{e} + 
    \frac{a^{e}+1}{(a^{e}+1)^2+2} X_{3}^{e}  \nonumber
\end{align}
depends on $e$ in a more complicated form so that there is no linear invariant relation between $(X_{6}^{e},X^{e})$ and   
$\E_{\Pcal_{e}}[Y|X]$, as illustrated in Fig.~\ref{toy_plot}.(c).
\end{example}
\smallskip

\begin{figure}[h]
\centering
\includegraphics[width=0.6\linewidth]{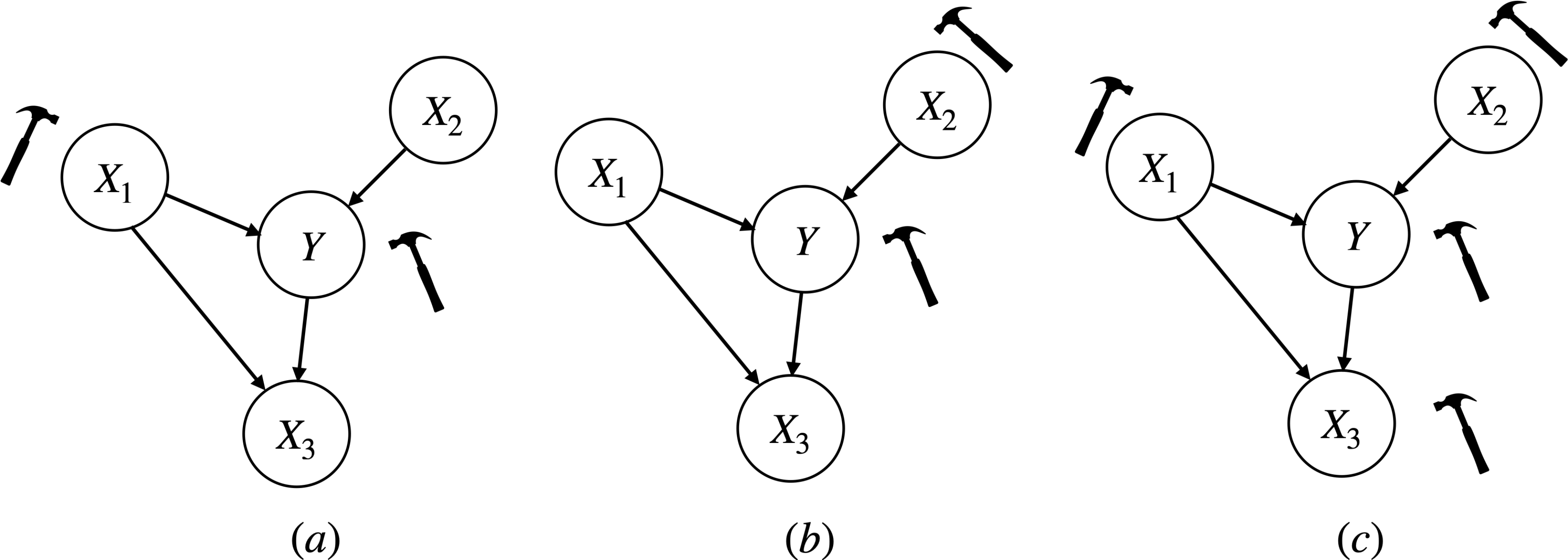}
\caption{$\mathcal{G}(\mathcal{M}_{\text{toy}}^{e})$ with interventions on $(X_{1},Y)$ in (a), $(X_{2},Y)$ in (b), and $(X_{1},X_{2},X_{3},Y)$ in (c), respectively.\label{fig_toy2}
 }
\end{figure}

In the next example, we extend Example~\ref{example1} to allow for interventions on $X_{1}$, $X_{2}$, and $Y$ through the means and/or variances of the noise variables (see Fig.~\ref{fig_toy2}).  
\smallskip

\smallskip
\begin{example}[additional settings]
\label{exam2}
\footnotetext{For $e \in\{1,2\}$, $X_{4}^{e}$ and $\E_{\Pcal_e}[Y|X]$ are estimated using OLS. Then, $\eta$ (denoted as $\eta_{\text{\textcircled{a}}}$ in Figure 2.(a)) is estimated using OLS by regressing $\E_{\Pcal_e}[Y|X]$ on $(X_{4}^{e},X^{e})$ using the pooled data.}
Consider model $\mathcal{M}_{\text{toy}}^{e}$ in Example~\ref{example1} with additional shift interventions and/or interventions on the noise variances. The results are summarized in the following.
\begin{enumerate}[leftmargin=*]
    \item Under shift interventions on $Y$ and $X_{1}$ and an intervention on the variance of $X_{1}$ as illustrated in Fig.~\ref{fig_toy2}(a), the two invariant relations~\eqref{toy_determin} and~\eqref{inv_rel_x2} hold.
    
    \item When $Y$ is intervened through the variance of $N_{Y}$ as illustrated in Fig.~\ref{fig_toy}, the relations~\eqref{toy_determin} and~\eqref{inv_rel_x2} will not hold. In this case, new relations can be established if $\E_{\Pcal_{e}}[Y|X]$ in~\eqref{toy_determin} and~\eqref{inv_rel_x2} is replaced by $\E_{\Pcal_{e}}[Y|X_{PA(Y)}] = \E_{\Pcal_{e}}[Y|X_{1},X_{2}]$.
    
    \item When $X_{2}$ is intervened through either the mean or variance as illustrated in Fig.~\ref{fig_toy2}(b), a relation as in~\eqref{inv_rel_x2} that is based on $X_{5}^{e}$ will not hold. 
    \item Combining the interventions above along with an intervention on $X_{3}$ through the noise variance as illustrated in Fig.~\ref{fig_toy2} (c), there will be one invariant relation left,
\begin{equation}
    \E_{\Pcal_{e}}[Y|X_{1},X_{2}] = X_{4}^{e}-X_{1}^{e}+b, \label{invar_rela_toy2}
\end{equation}
for some intercept $b \in \mathbb{R}$. 
It is noteworthy that~\eqref{invar_rela_toy2} will fail to hold if $X_{3}$ is intervened through the coefficients or noise mean. However, due to the intervention on the noise variance of $Y$,  $\Pcal_{e}(Y|X_{S},X_{4})$ is not invariant for any $S \subseteq \{1,2,3\}$, since $\Var_{\Pcal_{e}}(Y|X_{S},X_{4})$ changes with $e$.
\end{enumerate}

\begin{remark}
In~\cite{du2020causal}, the authors have shown that a form of varying filter connecting the features and response (as a special case of the varying coefficients captured by $\beta^e$ in~\eqref{eq:linear_scm_u}) is effective for causal inference tasks, by adopting estimators from~\cite{xiang2019estimation}.
\end{remark}
\end{example}

\begin{figure}[h]
\centering
\includegraphics[width=0.5\linewidth]{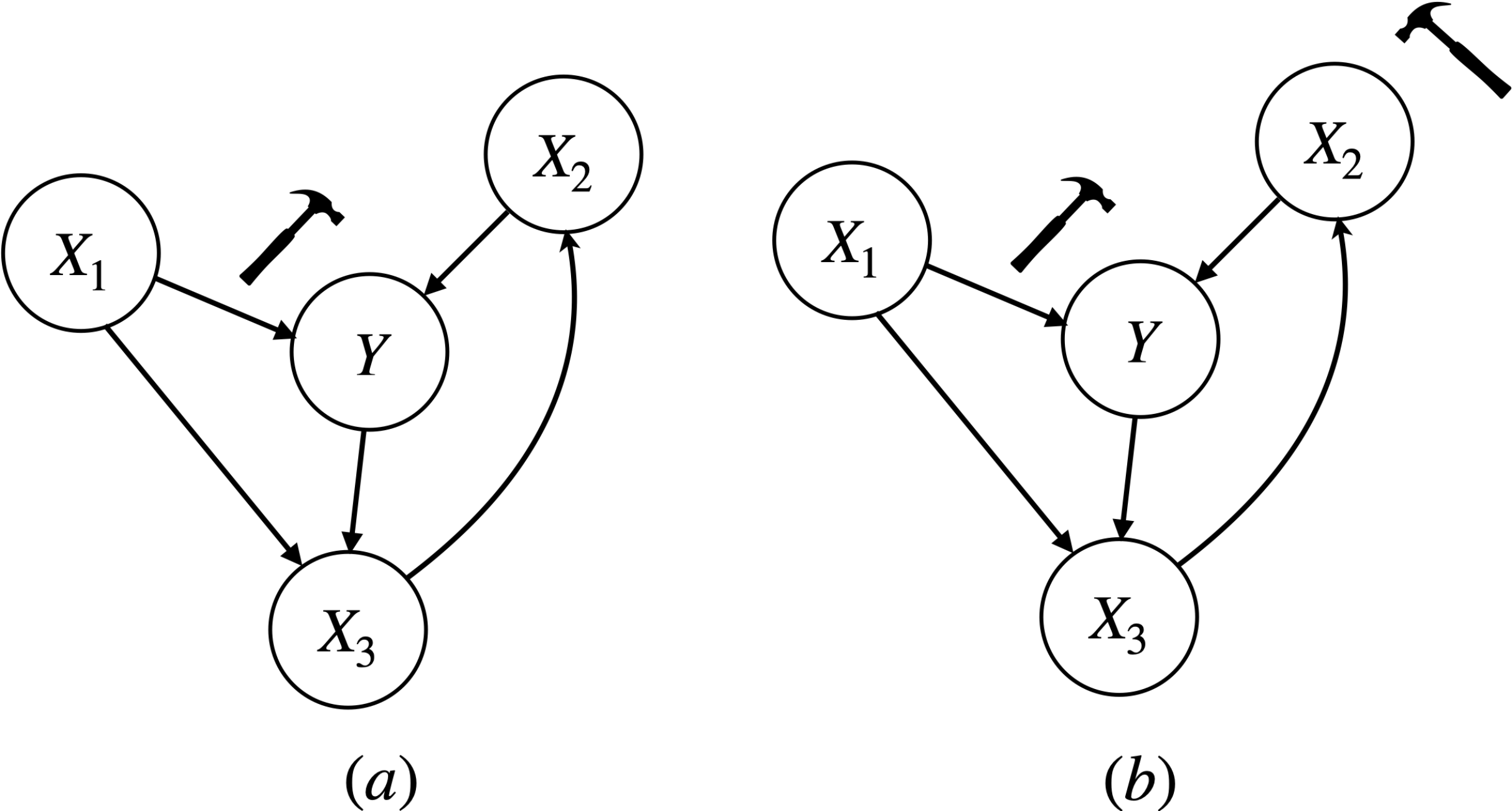}
\caption{A cyclic model with interventions on $Y$ in (a) and $(X_{2},Y)$ in (b), respectively.\label{fig_toy4}
 }
\end{figure}

\begin{example}[cyclic graph setting]
Consider model $\mathcal{M}_{\text{toy}}^{e}$ in Example~\ref{example1} with $X_{2}^{e}$ generated by $X^{e}_{2} = aX^{e}_{3}+N^{e}_{2}$ such that $a\neq 1$, which creates a cyclic model as illustrated in Fig.~\ref{fig_toy4}(a). One can compute that the relation in~\eqref{toy_determin} still holds. Moreover, the same relation holds even when $X_{2}$ is intervened through the coefficient $a$ (i.e., when it changes over $e$, denoted by $a^e$) and/or the noise $N^{e}_{2}$, illustrated in Fig.~\ref{fig_toy4}(b); this example can be further generalized to more complicated cyclic models. This cyclic setting highlights the point that the matching described in~\eqref{toy_determin} can hold without assuming \emph{acyclic} causal models, which will be left for future work. Focusing on acyclic causal models, in the subsequent sections, we formalize the matching conditions and provide a systematic analysis of sufficient conditions through a natural decomposition. 
\end{example}

\begin{remark}
The condition on $a$ avoids the cyclic model to be ill-defined, since combining the structural equations under $a=1$ yields a contradiction where the independent noise variables satisfy a linear constraint.
\end{remark}

\subsection{Invariant Matching Property}
In this section, we generalize the invariant relations observed in Example~\ref{example1} and Example~\ref{exam2} to a class of such relations for $\mathcal{M}^{e}$, $e \in \mathcal{E}^{\text{all}}$ without assuming joint Gaussian distributions, and connect this with the invariance property in~\eqref{inva_trans}. For the identification of such relations, we show that even \emph{two} training environments suffice (see Proposition~\ref{prop:unique_theta}). 

To handle non-Gaussian cases (beyond Example~\ref{example1}), we choose to adopt the \emph{linear MMSE} (or LMMSE) estimators for constructing \emph{linear} invariant relations. For a target variable $Y \in \mathbb{R}$ given a vector of predictors $X\in \mathbb{R}^{p}$, the LMMSE estimator is defined as 
\begin{equation*}
    \E_{l}[Y|X] := (\theta^{\text{ols}})^{\top}(X-\E[X])+\E[Y],
\end{equation*} 
where $\theta^{\text{ols}}:= \Cov(X,X)^{-1}\Cov(X,Y)$ is called the population ordinary least squares (OLS) estimator. With a slight abuse of notation, we write $\E_{l, \Pcal_e}[Y\,|\,X]$ to denote the LMMSE of $Y$ given $X$ with respect to $(X,Y)\sim\Pcal_e$. To simplify presentation, we focus on $(X^{e},Y^{e})$ with zero means for each $e \in \mathcal{E}^{\text{all}}$ (or equivalently all the noise variables have zero means), while the non-zero mean settings can be handled by introducing the constant one as an additional predictor.

\begin{definition}
\label{def:imp}
For $k \in \{1,\ldots,d\}$, $R \subseteq \{1,\ldots,d\}\setminus k$, and $S \subseteq \{1,\ldots,d\}$, we say that the tuple $(k,R,S)$ satisfies the \emph{invariant matching property} (IMP) if, for every $e \in \mathcal{E}^{\text{all}}$,
 \begin{align}
    \E_{l, \Pcal_e}[Y\,|\,X_{S}] 
    = \lambda \E_{l, \Pcal_e}[X_{k}\,|\,X_{R}] + \eta^{\top}X^{e}  \label{invar_coeff} ,
 \end{align}
for some $\lambda \in \mathbb{R} $ and $\eta \in \mathbb{R}^{d}$ that do not depend on $e$. We denote $\mathcal{I}_{\mathcal{M}}:= \{(k,R,S):~\eqref{invar_coeff} \text{ holds}\}$ for model $\mathcal{M}$, and we call $(\eta^\top, \lambda)^\top$ the \emph{matching parameters.}
\end{definition}

For a tuple $(K,R,S)$ such that the IMP does not hold, we simply call it a non-IMP. Observe that $\E_{l, \Pcal_e}[Y|X_{S}]$ is not directly applicable to the test environment due to its components depending on $e$, but those components are fully captured by $\E_{l, \Pcal_e}[X_{k}|X_{R}]$. If the matching parameters are identified from the training environments, the IMP is applicable to the test environment since $\E_{l, \Pcal_\text{test}}[X_{k}|X_{R}]$ is completely determined by the distribution of $X^{e^\text{test}}$ without the need of the target $Y^{e^\text{test}}$.  Since computing the additional feature $\E_{l, \Pcal_e}[X_{k}|X_{R}]$ is simply the prediction of $X_{k}$ (as if $X_{k}$ is not observed), the IMP indicates that the prediction of $Y$ can benefit from the predictions of certain predictors. We formally define this class of additional features as follows. 
   

\smallskip

\begin{definition}
\label{def:module}
For any $k \in \{1,\ldots,d\}$ and $R \subseteq \{1,\ldots,d\}\setminus k$,  we call $\E_{l, \Pcal_e}[X_{k}\,|\,X_{R}]$ a \emph{prediction module}. If a prediction module satisfies an IMP for some $S \subseteq \{1,\ldots,d\}$, we call it a \emph{matched prediction module} for $S$.
\end{definition}
\smallskip

Now we discuss the relationship between the IMP and the invariance property $\Pcal_e(Y|\phi_{e}(X)) =  \Pcal_h(Y|\phi_{h}(X))$ in~\eqref{inva_trans}, we rewrite~\eqref{invar_coeff} in a compact form as 
\begin{equation}
      \E_{l, \Pcal_e}[Y|X_{S}] =\theta^{\top}  \widetilde{X}^{e}, \label{imp_theta}
\end{equation}
where
\begin{equation}
   \widetilde{X}^{e} := (X_{1}^{e},\ldots,X_{d}^{e},\E_{l,\Pcal_e}[X_{k}|X_{R}])^{\top}, \label{IMP_trans}
\end{equation}
and $\theta = (\eta^{\top},\lambda)^{\top}$ denotes the matching parameter. Define
\begin{equation}
    \phi_{e}^{(k,R,S)}(X^{e}) := (X_{S'}^{e},\E_{l,\Pcal_e}[X_{k}|X_{R}])^{\top},  
\end{equation}
where $X_{S'}^{e}$ is a row vector for some $S' \subseteq \{1,\ldots,d\}$. For notational convenience, we introduce the shorthand $\phi_{e}(X^{e}) :=\phi_{e}^{(k,R,S)}(X^{e})$. Note that $\phi_{e}(X^{e})$ is a \emph{linear} transform of $X^{e}$. 

In general, however, the invariance of the matching parameter $\theta$ does not imply the invariance property~\eqref{inva_trans}. In Section~\ref{sec:chara_mpm}, we will characterize a class of IMPs that each satisfies~\eqref{inva_trans}.  When the invariance property holds, one can apply the general conditional expectation $f_{\phi_e}(x)=\E_{\Pcal_{\text{e}}}[Y|\phi_{e}(x)]$ as in~\eqref{def_f_phi}, since the linear estimator from the IMP is in general sub-optimal for the non-Gaussian cases. We will focus on linear estimators in this work as the extension can be handled via nonlinear regression methods in a straightforward manner. 

It is noteworthy that since $\E_{l, \Pcal_e}[X_{k}|X_{R}]$ is a linear function of $X_{R}^{e}$, the matching parameter $\theta$ is not unique given a single environment $e \in \mathcal{E}^{\text{train}}$. This causes issues when one aims to identify the possible IMPs given the distribution of $(X^{e},Y^{e})$. However, we show that two environments in the training data are sufficient to identify $\theta$ under a mild assumption on the matched prediction module. 
\smallskip

\begin{proposition}\label{prop:unique_theta}
For a tuple $(k,R,S)$ that satisfies an IMP, the matching parameter $\theta$ can be uniquely identified in $\mathcal{E}^{\text{train}}$ if $|\mathcal{E}^{\text{train}}|\geq 2$ and
\begin{equation}
    \E_{l, \Pcal_e}[X_{k}|X_{R}=x] \neq \E_{l, \Pcal_h}[X_{k}|X_{R}=x] 
    \label{eq_assup_unique}
\end{equation}
for some $e,h \in \mathcal{E}^{\text{train}}$ and $x \in \mathcal{X}_{R}$.  
\end{proposition}
\smallskip

This proposition shows how the heterogeneity of the data generating process can be helpful for identifying important invariant relations.

\subsection{A Decomposition of the IMP}
In our toy examples, recall that the IMPs are derived by first computing $\E_{\Pcal_{e}}[Y|X_{S}]$ and $\E_{\Pcal_{e}}[X_{k}|X_{R}]$ separately and then fitting a linear relation from $(\E_{\Pcal_{e}}[X_{k}|X_{R}],X_{S})$ to $\E_{\Pcal_{e}}[Y|X_{S}]$. These two steps reveal a natural decomposition of the IMP, which we term as the first and second matching properties below. 

\begin{definition}
\label{def:first}
We say that $S\subseteq \{1,\ldots,d\}$ satisfies the \emph{first matching property} if, for every $e \in \mathcal{E}^{\text{all}}$,
\begin{equation}
    \E_{l, \Pcal_e}[Y|X_{S}] = \lambda_{Y}\E_{\Pcal_e}[Y|X_{PA(Y)}] + \eta_{Y}^{\top}X^{e}, \label{eq:fmp}
\end{equation}
for some $\lambda_{Y}\in \mathbb{R} $ and $\eta_{Y} \in \mathbb{R}^{d}$ that do not depend on $e$. 
\end{definition}
\smallskip

First, observe that the first matching property holds for $S= PA(Y)$ since 
\begin{align*}
    \E_{l, \Pcal_e}[Y|X_{PA(Y)}] = \E_{\Pcal_e}[Y|X_{PA(Y)}] =(\beta^{e})^{\top}X^{e}.
\end{align*}
The first matching property concerns the set $S$ such that the components in $\E_{l, \Pcal_e}[Y|X_{S}]$ that depend on $e$ are fully captured by the causal function $\E_{\Pcal_e}[Y|X_{PA(Y)}]$. However, this invariant relation is not directly useful for the prediction of $Y^{e^{\text{test}}}$, since the causal function can change arbitrarily with $e$. To this end, we identify another invariant relation from $\mathcal{M}^{e}$ which is called the second matching property.  

\smallskip
\begin{definition}
\label{def:second}
For $k \in \{1,\ldots,d\}$ and $R \subseteq \{1,\ldots,d\}\setminus k$, we say that a tuple $(k,R)$ satisfies the \emph{second matching property} if, for every $e \in \mathcal{E}^{\text{all}}$, 
\begin{equation}
    \E_{l, \Pcal_e}[X_{k}|X_{R}] = \lambda_{X}\E_{\Pcal_e}[Y|X_{PA(Y)}] + \eta_{X}^{\top}X^{e},\label{eq:smp}
\end{equation}
for some $\lambda_{X}\in \mathbb{R}$ and $\eta_{X} \in \mathbb{R}^{d}$ that do not depend on $e$.  
\end{definition}
\smallskip

It is straightforward to see that, if $\lambda_{X}\neq 0$ in the second matching property, the first and second matching properties imply the IMP as follows,
\begin{align}
     \E_{l, \Pcal_e}[Y|X_{S}] &= \frac{\lambda_{Y}}{\lambda_{X}} \E_{l, \Pcal_e}[X_{k}|X_{R}]+\left(\eta_{Y}-\frac{\lambda_{Y}}{\lambda_{X}}\eta_{X}\right)^{\top}X^{e}\nonumber\\
        &:= \lambda \E_{l, \Pcal_e}[X_{k}|X_{R}] + \eta^{\top}X^{e}  \nonumber.
\end{align}

\begin{figure}[h]
\centering
\includegraphics[width=0.4\linewidth]{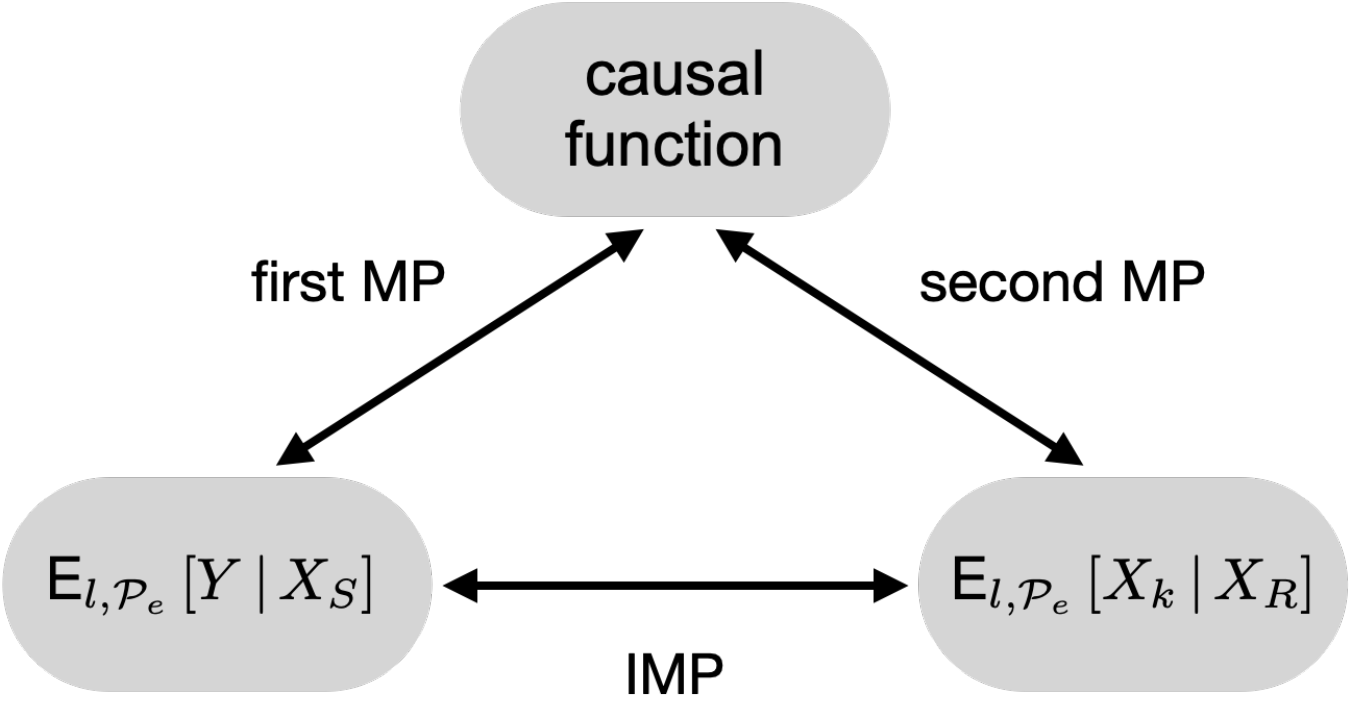}
\caption{A triangular relation consists of the first, second, and invariant matching properties.\label{fig_triangle}
 }
\end{figure}

For prediction tasks under SCMs, the causal function often plays a central role.  Our first and second matching properties show how the LMMSE estimator $\E_{l, \Pcal_e}[Y|X_{S}]$ and the matched prediction module $\E_{l, \Pcal_e}[X_{k}|X_{R}]$ are connected with the causal function, respectively. Together, the two individual connections make up the IMP (illustrated in Fig.~\ref{fig_triangle}). 

\smallskip

\section{Characterization of Invariant Matching Properties}\label{sec:chara_mpm}

\subsection{Interventions on the Response}

\label{sec:characterize}
First, we consider model $\mathcal{M}^{e}$ with interventions only on $Y$ through the coefficients, i.e.,
\begin{numcases}{\mathcal{M}^{e,1}:}
    X^{e} = \gamma Y^{e} + B X^{e}+ \varepsilon_{X}^{e} \label{eq:train_m_X} \\
    Y^{e} = (\alpha^{e}+\beta)^{\top}X^{e} +  \varepsilon_{Y}^{e}. \label{eq:train_m_Y}
\end{numcases}
To distinguish the parents of $Y$ with varying and invariant coefficients, we decompose $\beta^{e}$ in $\mathcal{M}^{e}$ into two parts $\alpha^{e}$ and $\beta$. Without loss of generality, we assume that $\alpha_{j}^{e}\neq 0$ if and only if $\alpha_{j}^{e}$ is a non-constant function of $e$, and we define the following subset of parents of $Y$, 
\begin{equation}
    PE =\{j\in \{1,\ldots,d\}:\alpha_{j}^{e}\neq 0\}. \nonumber
\end{equation}

\begin{remark}
Note that, in the non-zero mean settings, this model covers the shift intervention on $Y$ through the varying coefficient of the predictor which is a constant one. 
\end{remark}
Recall that prediction modules do not rely on the response $Y$ but on the relations between the predictors for each environment. When $Y$ is unobserved (or equivalently, substituting $Y$ in~\eqref{eq:train_m_Y} into~\eqref{eq:train_m_X}), the relations between the predictors are as follows,
\begin{equation}
    X^{e} =\left( \gamma\left(\alpha^{e}+\beta\right)^{\top} + B \right)X^{e}+ \gamma \varepsilon_{Y}+\varepsilon_{X}, \label{eq_only_X}
\end{equation}
where $\gamma \varepsilon_{Y}+\varepsilon_{X}$ a vector of dependent random variables when $\gamma$ is not a zero vector.  If $\alpha^{e}$ vanishes from~\eqref{eq_only_X}, the distribution of $X^{e}$ becomes invariant with respect to environments.
As a consequence, the condition~\eqref{eq_assup_unique} in Proposition~\ref{prop:unique_theta} will not be satisfied. Observe that $\alpha^{e}$ is non-vanishing in~\eqref{eq_only_X} only if $\gamma$ is not a zero vector, which brings up the following key assumption. 

\smallskip
\begin{assumption} \label{assump_one_child}
When $Y$ is intervened, we assume that $Y$ has at least one child.
\label{assump1}
\end{assumption}
\smallskip

Note that if $Y$ has no children, by the Markov property of SCMs~\cite{pearl2009causality}, the test data sampled from $\mathcal{P}_{\text{test}}^{X}$ provides no information about $\mathcal{P}_{\text{test}}^{Y|X}$ and thus the observed $X^{e^{\text{test}}}$ may correspond to two arbitrarily different  $Y^{e^{\text{test}}}$'s, which makes the problem ill-posed. 



The first and second matching properties enable us to characterize the tuples $(k,R,S)$'s that satisfy IMPs through the characterizations of $S$ (for the first matching property) and $(k,R)$ (for the second matching property) separately.  In the following theorem, we show that a class of IMPs implied by the first and second invariant matching properties satisfy the invariance property~\eqref{inva_trans}.

\smallskip
\begin{theorem} For model $\mathcal{M}^{e,1}$, the first and second matching properties hold in the following cases.
\label{thm1}\begin{enumerate}
    \item \emph{On the first MP:} For each $S\subseteq \{1,\ldots,d\}$ such that $PE \subseteq S$, the first matching property holds.
    
\item  \emph{On the second MP:} For each $k \in \{1,\ldots,d\} \setminus PE$ and $R \subseteq \{1,\ldots,d\} \setminus k$ such that $PE \subseteq R$, the second matching property holds.
\end{enumerate}

For any tuple $(k,R,S)$ above such that $R \subseteq S$, if $\lambda_{X}\neq 0$ in the second matching property, then  $\phi_{e}(X^{e})= (X_{S}^{e},\, \E_{l,\Pcal_e}[X_{k}|X_{R}])^{\top}$ satisfies~\eqref{inva_trans}. Furthermore, $\mathcal{L}_{\text{test}}(f_{\phi})$ is minimized by any $\phi$ with $S=\{1,...,d\}$. 
\end{theorem}

\smallskip


As a concrete example, recall~\eqref{toy_determin} from our motivating example, where $R \subseteq S$ and $\lambda_{X}\neq 0$ are satisfied, $\E_{\Pcal_{e}}[Y|X]$ can be represented using invariant coefficients. It is noteworthy that Assumption~\ref{assump_one_child} is a necessary condition for $\lambda_X\ne 0$, and we provide a sufficient condition for $\lambda_X\ne 0$ in a concrete setting with $S = \{1,\ldots,d\}$ below. 



\smallskip

\begin{corollary}\label{coro1} For model $\mathcal{M}^{e,1}$, the first and second matching properties hold in the following cases.
\begin{enumerate}
    \item \emph{On the first MP:} The first matching property holds for $S= \{1,\ldots,d\}$.
\item  \emph{On the second MP:} For each  $k \in \{j \in MB(Y): \alpha_{j}^{e} = 0\}$ and $R=-k:= \{1,\ldots,d\}\setminus k$, the second matching property holds.
\end{enumerate}
\end{corollary}

\begin{proposition}
\label{prop_l}
Under Assumption~\ref{assump_one_child}, for each $(k,R)$ in Corollary~\ref{coro1}, we have
$\lambda_{X}\neq 0$ in the second matching property if $B_{-k,k}$ is not in the following hyperplane,
\begin{equation}
   w^{\top}x + b= 0,   \label{eq:hyper}
\end{equation}
where $w \in \mathbb{R}^{d-1}$ and $b \in \mathbb{R}$ are determined by the parameters in $\mathcal{M}^{e,1}$ other than $B_{-k,k}$. 
\end{proposition}

\smallskip
The explicit expressions of $w$ and $b$ using the parameters in $\mathcal{M}^{e,1}$ are provided in the proof of Proposition~\ref{prop_l}. For generic choices of the parameters, the second matching property holds with $\lambda_{X} \neq 0$ since $B_{k,-k}$ is not necessarily on the hyperplane described in~\eqref{eq:hyper}. 

\smallskip 

 When $Y$ is additionally intervened through the noise variance, the proof of Theorem~\ref{thm1} will break down in general (see Remark~\ref{rem_int_noise_var} in Appendix~\ref{app:thm1_proof}). However, recall that the first matching property holds for $S= PA(Y)$ by definition. In this case, we provide an example for the second matching property in the following corollary.
 \smallskip

\begin{corollary}\label{coro:int_noise_var} Under Assumption~\ref{assump1}, if $Y$ is intervened through the noise variance in model $\mathcal{M}^{e,1}$, the second matching property holds for $k \in CH(Y)$ such that $k \not \in DE(i)$ for any $i \in CH(Y)\setminus k$, and $R = \{1,\ldots,d\}\setminus DE(Y)$.
\end{corollary}
\smallskip

The resulting IMPs no longer satisfy the invariance property~\eqref{inva_trans}, but we can use the IMP directly for the prediction of $Y^{e^{\text{test}}}$.
\begin{remark} \label{rmk:imp_inva}
To sum up, the class of IMPs constructed under interventions on $Y$ only through the coefficients and shifts will in general imply the invariance property~\eqref{inva_trans}, but it is not the case under interventions on $Y$ through the noise variance. For the characterizations of IMP, we focus on sufficient conditions that are relatively simple to evaluate, while they are not necessary conditions in general. One way to find necessary and sufficient conditions for the first/second IMP is through explicit (but tedious) calculations as in the proof of Proposition~\ref{prop_l}. However, such conditions are hard to evaluate and provide limited insights into our methodology. We thus do not pursue them in this work.  When $Y$ is intervened through the coefficients of all its parents, i.e., $PE=PA(Y)$, a recent work~\cite{du2023identifying} shows that $PA(Y) \subseteq S$ and $PA(Y) \subseteq R$ are also necessary for the IMP under mild assumptions, leading to a novel algorithm for identifying direct causes of $Y$ in the multi-environment setting. 

\end{remark}

\subsection{Interventions on both Predictors and Response} \label{ext_interven}

To generalize the setting when only $Y$ is intervened to the general setting when $X$ and $Y$ are both intervened, an idea is to merge the setting when only $Y$ is intervened with the one when only $X$ is intervened. The latter setting has been studied in the stabilized regression framework~\cite{pfister2021stabilizing}. The following set of predictors is identified (see Definition~3.4 therein), 
\begin{align}
    &X^{\text{int}}(Y) = CH^{I}(Y)\cup\bigl\{j \in \{1,\ldots,d\}\,|\, \exists i \in CH^{I}(Y)\text{ such that } j\in DE(X_{i}) \bigr\}, \nonumber
\end{align}
which contains the intervened children of $Y$ (denoted by $CH^{I}(Y)$) and the descendants of such children. This useful notion can be defined for each $X_{j} \in \{X_{1},\ldots,X_{d}\} $, denoted by  $X^{\text{int}}(X_{j})$ for each $X_j$. When only $X$ is intervened, the invariance principle~\eqref{eq:ICP} holds for $S^{*}= \{1,\ldots,d\}\setminus X^{\text{int}}(Y)$; the Markov blanket of $Y$ defined with respect to $X_{S^{*}}$ is called the \emph{stable blanket} of $Y$ in~\cite{pfister2021stabilizing}. In other words, by excluding the predictors in $X^{\text{int}}(Y)$, the target $Y$ is blocked from the interventions on $X$ when conditioning on $X_{S^{*}}$. This holds when $Y$ is additionally intervened, as if only $Y$ is intervened given $X_{S^{*}}$. In order for $S^{*}$ to include as least one child of $Y$ as in Assumption~\ref{assump_one_child}, we need the following assumption. 

\smallskip
\begin{assumption} \label{assump_second_child}
When $Y$ is intervened, we assume that $Y$ has at least one child that is not intervened and that child is not a descendant of some intervened child of $Y$. 
\label{assump2}
\end{assumption}
\smallskip

Based on the observation above, we identify an important class of IMPs for the general setting in the following Theorem.     


\begin{theorem} \label{thm:int_xy} For the training model $\mathcal{M}^{e}$ without the intervention on the noise variance of $Y$, the first and second matching properties hold in the following cases. 
\begin{enumerate}\item \emph{On the first MP:}  For $S \subseteq \{1,\ldots,d\}\setminus X^{\text{int}}(Y)$ such that $PA(Y) \subseteq S$,  the first matching property holds. \item \emph{On the second MP:}  For each $k \in \{1,\ldots,d\}\setminus \{PE \cup X^{\text{int}}(Y) \}$, 
 and $R \subseteq \{1,\ldots,d\} \setminus\{k, X^{\text{int}}(X_{k})\cup X^{\text{int}}(Y) \}$ such that $PA(X_{k})\setminus Y \subseteq R$, the second matching property holds.
\end{enumerate}
 Furthermore, if $\lambda_{X}\neq 0$ in the second matching property, then $ \phi_{e}(X^{e})= (X_{S}^{e},\, \E_{l,\Pcal_e}[X_{k}|X_{R}])^{\top}$ satisfies~\eqref{inva_trans}. 

\end{theorem}

\begin{remark}\label{rmk:thm_xy}

   In general, there exist subsets of $S$ and $R$ from Theorem~\ref{thm:int_xy} such that the first and second matching properties hold, respectively. In particular, $PA(Y) \subseteq S$ and $PA(X_{k}) \subseteq R$ are not necessarily satisfied. Due to the Markov property of SCMs, $Y$ is independent of its ancestors when conditioning on $PA(Y)$, thus we focus on the IMPs that are more predictive by including $PA(Y)$ in $S$. 
   

\end{remark}

When $Y$ is intervened through its noise variance, recall again that the first matching property always holds for $S=PA(Y)$ by definition. The following proposition follows from Remark~\ref{rmk:int_var_coro2} in Appendix~\ref{app:coro2}.

\begin{proposition} 
   When $Y$ is additionally intervened through the noise variance in Theorem~\ref{thm:int_xy}, the second matching property in Theorem~\ref{thm:int_xy} still holds.
\end{proposition}

 Similar to the argument in the proof of Theorem~\ref{thm1}, the class of $\phi$'s from Theorem~\ref{thm:int_xy} will lead to the same test population loss, as they depend on the same $S$ that is fixed in this setting. Assumption~\ref{assump_second_child} is necessary for $\lambda_{X} \neq 0$, while sufficient conditions for $\lambda_{X} \neq 0$ can be found similarly as in Proposition~\ref{prop_l}.  

\section{Algorithms}
\label{sec:alg}
For each $e \in \mathcal{E}^{\text{train}}$, we are given the \iid training data $\bs{X}_{e}\in  \mathbb{R}^{n_{e} \times d}, \bs{Y}_{e} \in \mathbb{R}^{n_{e}}$, and we observe the \iid test data $\bs{X}^{\tau} \in \mathbb{R}^{m \times d}$ and aim to predict $\bs{Y}^{\tau} \in \mathbb{R}^{m}$. Let $\bs{X} \in \mathbb{R}^{n \times d}$ with $n := \sum_{i=1}^{|\mathcal{E}^{\text{train}}|}n_{e}$ denote the pooled data of  $\bs{X}_{e}$'s. In this section, we present the implementation of our method starting with the case when $e$ is sampled from a discrete distribution with a finite support. In this setting, we expect to have $n_{e} \gg 1$ for every $e \in \mathcal{E}^{\text{train}}$ in general, thus it is possible to do estimation based on the data from each single environment. The challenging setting of continuous environments will be handled afterward. 

\subsection{Discrete Environments}

 To implement our method, the main task is to identify the set of IMPs $\mathcal{I}_{\mathcal{M}}$ from the training data. For each tuple $(k,R,S)$ in Algorithm~\ref{alg1}, we test the following null hypothesis
\begin{equation}
    \mathcal{H}_0 \text{ : There exists  }\theta \in \mathbb{R}^{d+1} \text{ such that}~\eqref{imp_theta} \text{ holds}. \nonumber 
\end{equation}
We propose two test procedures. 
\begin{enumerate}[leftmargin=*]
\item \emph{Test of the Deterministic Relation:} Since the IMPs are linear and deterministic (i.e., noiseless), we test whether the residual vector $\bs{R} \in \mathbb{R}^{n}$ of fitting an IMP on $(k,R,S)$ is a zero vector or not using the test statistics,
\begin{equation}
    T =\frac{1}{n}\boldsymbol{R}^{\top}\boldsymbol{R}, \nonumber
\end{equation}
where $\boldsymbol{R}$ is a pooled data vector of $\boldsymbol{R}_{e}$'s defined below. To fit an IMP, we first estimate the two LMMSE estimators in~\eqref{invar_coeff} using OLS for each environment,
\begin{align*}
    \boldsymbol{\hat{L}}_{e,1}&:=(\boldsymbol{X}_{e,S}^{\top}\boldsymbol{X}_{e,S})^{-1}\boldsymbol{X}_{e,S}^{\top}\boldsymbol{Y}_{e}\,,\\
    \boldsymbol{\hat{L}}_{e,2}&:=(\boldsymbol{X}_{e,R}^{\top}\boldsymbol{X}_{e,R})^{-1}\boldsymbol{X}_{e,R}^{\top}\boldsymbol{X}_{e,k}\,.
\end{align*}
 Let $\boldsymbol{\hat{L}}_{1} \in \mathbb{R}^{n}$ and $\boldsymbol{\hat{L}}_{2}  \in \mathbb{R}^{n}$ denote the pooled data of $\boldsymbol{\hat{L}}_{e,1}$'s and $\boldsymbol{\hat{L}}_{e,2}$'s, respectively. \emph{It is noteworthy that $\bs{\hat{L}}_{2}$ only depends on $\bs{X}$, thus $\bs{\hat{L}}_{2}^{\tau}$ for the test data can be computed similarly using $\bs{X}^{\tau}$}. Next, we estimate the matching parameter using OLS on the pooled data (recall that the matching parameter cannot be identified using the data from a single environment). The OLS estimator of the matching parameter is 
    \begin{equation}
    \hat{\theta} :=    (\hat{\eta}^{\top},\hat{\lambda})^{\top} = ([\bs{X}_{S},\bs{\hat{L}}_{2}]^{\top}[\bs{X}_{S},\bs{\hat{L}}_{2}])^{-1}[\bs{X}_{S},\bs{\hat{L}}_{2}]^{\top}\bs{\hat{L}}_{1}\,. \nonumber
    \end{equation}
For each $e \in \mathcal{E}^{\text{train}}$, we obtain the residual vector of fitting an IMP 
  \begin{align}
        \boldsymbol{R}_{e} = \boldsymbol{\hat{L}}_{e,1} - \hat{\lambda}\boldsymbol{\hat{L}}_{e,2}-  \boldsymbol{X}_{e,S}\hat{\eta}\,. \nonumber
    \end{align}
    
\item \emph{Approximate Test of Invariant Residual Distributions}:
According to the invariance property~\eqref{def_f_phi}, we test whether the residual when regressing $\bs{Y}$ on $[\bs{X}_{S},\bs{L}_{2}]$ has constant mean and variance. Specifically, we use the t-test and F-test with corrections for multiple hypothesis testing from~\cite{peters2016causal} (see Section 2.1 Method II). The test yields a p-value.

\end{enumerate}

\smallskip

The test statistic from the first procedure and the p-value from the second procedure quantify how likely an IMP holds (i.e., the smaller the more likely), and thus we will refer to either one of them as an IMP score denoted by $s_{\text{IMP}}$. Let $\hat{\mathcal{I}} = \{(k,R,S): s_{\text{IMP}}(k,R,S) < c_{\text{IMP}} \}$ denote the set of IMPs identified from the training data, where $c_{\text{IMP}}$ is some cutoff parameter. Then, since IMPs are not equally predictive in general, we focus on the most predictive ones by introducing the mean squared prediction error as a prediction score $s_{\text{pred}}$, and we select the set of IMPs that are more predictive $\widehat{\mathcal{I}_{\text{pred}}} = \{(k,R,S) \in \hat{\mathcal{I}} :  s_{\text{pred}}(k,R,S) < c_{\text{pred}}\}$ with some cutoff parameter $c_{\text{pred}}$. For the second IMP score that is a p-value, the cutoff parameter $c_{\text{IMP}}$ is simply a significance level that is fixed to $0.05$ in this work. For choosing the rest of the cutoff parameters, we follow a  bootstrap procedure from~\cite{pfister2019invariant} with one subtle difference: We sample the same amount of bootstrap samples from each environment rather than sampling over the pool data as in~\cite{pfister2019invariant} since our procedure involves estimations using the data from each environment.

\begin{algorithm}[H]
\caption{\label{alg1} Invariant  Prediction using the IMP (discrete)
}
\begin{algorithmic}
\Procedure{Identify IMPs from the Training data}{}	
    	\For{$k \in \{1,\ldots,d\}$, $S \subseteq \{1,\ldots,d\}$, $R \subseteq S \setminus k $} 
        	    \State {Compute the IMP score $s_{\text{imp}}$ and the prediction }
        	    \State{score $s_{\text{pred}}$ for  $i=(k,R,S)$ }
        	    \State{Regress $\bs{Y}$ on $[\bs{X}_{S},\bs{\hat{L}}_{2}]$ to obtain $f_{i}$}
        	    \EndFor
        	    \State{Identify $\hat{\mathcal{I}}$ and $\widehat{\mathcal{I}_{\text{pred}}}$}
\EndProcedure	
\Procedure{Prediction on the Testing data}{}	
\State{$\hat{\bs{Y}}^{\tau} =\frac{1}{|\widehat{\mathcal{I}_{\text{pred}}}|} \sum_{i \in \widehat{\mathcal{I}_{\text{pred}}}} f_{i}(\bs{X}_{S}^{\tau},\bs{\hat{L}}_{2}^{\tau})$}
\EndProcedure	
\end{algorithmic}
\end{algorithm}

 In practice, there can be spurious IMPs that have extremely small IMP scores but have large prediction scores, e.g., when $Y$ is independent of $X_{S}$ and $X_{k}$ is independent of $X_{R}$. To this end, we will pre-select $(k,R,S)$'s with prediction scores smaller than the median of all the computed prediction scores before identifying $\hat{\mathcal{I}}$.

If the regression function $f_{i}$ in Algorithm~\ref{alg1} is chosen to be linear, one can use the IMP directly, i.e.,
\begin{equation}
    \bs{\hat{Y}}_{\text{IMP}}(k,R,S) =  [\bs{X}_{S},\bs{\hat{L}}_{2}]\hat{\theta}, \label{imp_esti}
    \end{equation}
which we call the \emph{discrete IMP estimator} denoted by $\text{IMP}_{\text{d}}$. To make use of all the IMPs selected in $\widehat{\mathcal{I}_{\text{pred}}}$, we use an averaging step for the prediction of $\bs{Y}^{\tau}$ in Algorithm~\ref{alg1}.  

 Now we discuss the computation complexity of our method. For a given graph of size $d \geq 3$, the number of $(k,R,S)$'s with nonempty $(R,S)$'s is given by 
\begin{equation}
 \sum_{j=1}^{d} \binom{d}{j}\biggl( j\cdot(2^{j-1}-1)+(d-j)\cdot(2^{j}-1)\biggr) = d\cdot (2\cdot3^{d-1}-2^{d}),\nonumber
\end{equation}
which follows by first choosing a set $S$ with $j$ elements, and then considering two settings $k \in S$ and $k \not \in S$ since $R$ has to satisfy $R \subseteq S \setminus k$. This calculation implies that the exhaustive search step in our method is not applicable for relatively large graphs (e.g., $d\geq 15$) due to the exponential time complexity. It is noteworthy that in the ICP framework~\cite{peters2016causal}, a similar issue occurs due to an exhaustive search over $S \subseteq \{1,\ldots,d\}$. We discuss several options to alleviate this issue. (1) One can adopt a preprocessing step for feature selection to reduce the dimension $d$, using Lasso~\cite{tibshirani1996regression} or Boosting~\cite{bartlett1998boosting}, as proposed in ICP~\cite{peters2016causal}. (2) When prior information about the graph structure (e.g., the maximum number of parents of the nodes) is available or the graph structure can be estimated, the search space can be reduced. For instance, one can first adopt existing causal discovery methods (e.g.,~\cite{shimizu2006linear}) for graph structure estimation and then apply our methods according to the sufficient conditions in Theorem~\ref{thm1} and~\ref{thm:int_xy}. (If no IMP is found, one can still perform an exhaustive search over other $(k,R,S)$'s). 
(3) One can also exploit some intrinsic sparsity regarding IMPs, observe that there is only one matched prediction module (recall Definition~\ref{def:module}) in the IMP, while the number of prediction modules grows as $d\cdot 3^{d}$. This sparsity has been studied when only $Y$ is intervened~\cite{du2023generalized}, leveraging a variant of Lasso.

\subsection{Continuous Environments}
To model continuous environments, we introduce an environmental variable $U$ that is a \emph{continuous} random variable with support $\mathcal{U}$. Apparently, this is a much more challenging setting compared with the discrete environment case, as we only have one training data sample for each $u \in \mathcal{U}$, making the OLS a poor estimate of $\E_{l,\Pcal_{u}}[Y|X_{S}]$. Fortunately, it turns out that we can leverage the \emph{semi-parametric varying coefficient} (SVC) models~\cite{fan2005profile} (see Appendix~\ref{PLS_estimate}) to remedy this issue. In particular, we estimate $\E_{l,\Pcal_{u}}[Y|X_{S}]$ by fitting,
\begin{equation}
        Y = M + \beta^{\top}Z + N \;\text{ with }\; M =\alpha^{\top}(U) W, \label{svc_y}
\end{equation}
where $N$ is independent of $U$ and the two vectors of predictors $W \in \mathbb{R}^{p}$ (for the varying coefficient) and $Z \in \mathbb{R}^{q}$ (for the invariant coefficients) with $p+q = |S|$. Since we assume  $N \independent U$, we focus on the settings when $Y$ is not intervened through the noise variance.

\begin{remark}
Our estimation procedure for the discrete environments can also be formulated under the SVC model with a discrete random variable $U$, where we treat all the coefficients as varying coefficients (i.e., $\beta=\bs{0}$), and $\text{IMP}_{\text{d}}$ becomes an estimate of $M$. 
\end{remark}
An SVC model over $(Y^{\tau},W^{\tau}, Z^{\tau},N^{\tau})$ for the test data can be defined similarly, where $\sigma^2 = \E[(N^{\tau})^{2}]$ is the population generalization error of the IMP estimator. Observe that the linear SCM $\mathcal{M}^{u} $ in~\eqref{eq:linear_scm_u} can be viewed as a collection of SVC models parameterized by $U=u$. Thus the estimation tasks for the linear SCMs from continuous environments can greatly benefit from the existing theories developed for SVC models. More precisely, we employ the following estimate
\begin{equation}
   \widehat{ \E_{l,\Pcal_{u}}[Y|X_{S}]} = \hat{M}|_{U=u} + \hat{\beta}^{\top}Z, \nonumber
\end{equation}
where the profile least-squares estimation of $\beta$ and $M$ proposed in~\cite{fan2005profile} can be found in Appendix~\ref{PLS_estimate}. Similarly, $\E_{l,\Pcal_{u}}[X_{k}|X_{R}]$ can be estimated by fitting another semi-parametric varying coefficient model 
\begin{align}
      V= M_{V} + \beta_{V}^{\top}Z_{V} + N_{V} \;\text{ with }\; M_{V} =\alpha_{V}^{\top}(U)W, \label{svc_v}
\end{align}
where $V$ denotes any $X_{k}$, and $X_{R}$ is divided into $Z_{V}\in \mathbb{R}^{r}$ and $W$. 

It is noteworthy that the two SVC models share the same set of predictors with varying coefficients, which we explain below. A challenge for fitting such models is that the vector of predictors with varying coefficients, namely $W$, needs to be known. For continuous environments, we focus on discovering IMPs that can be decomposed into the first and second matching properties. Thus, since the causal function captures the predictors with varying coefficients, the first and second matching properties imply that the vector $W$ is simply $X_{PE}$, i.e., the parents of $Y$ with varying coefficients in $\mathcal{M}^{u}$, for both models. 

Based on this observation and Theorem~\ref{thm1}, we replace the exhaustive search over $(k,R,S)$ in Algorithm~\ref{alg1} by a search over $(P,k,R,S)$ according to the conditions in Theorem~\ref{thm1} with $PE=P$. That is, we choose $(P,k,R,S)$ from
\begin{align}
  &  P\subseteq \{1,\ldots,d\}, \quad k \in \{1,\ldots,d\}\setminus P,  \nonumber\\ 
 & P \subseteq  S \subseteq  \{1,\ldots d\}, \quad  P \subseteq R \subseteq S \setminus k \nonumber, 
\end{align}
such that $W= X_{P}$, $Z = X_{S \setminus P}$, $V= X_{k}$, and $Z_{V}= X_{R \setminus P}$.

Unlike $\text{IMP}_{\text{d}}$ in~\eqref{imp_esti}, we make use of the fact that $\beta$ is invariant and propose the \emph{continuous IMP estimator} denoted by $\text{IMP}_{\text{c}}$ as follows
\begin{equation}
\bs{\hat{Y}}_{\text{IMP}}(P,k,R,S) = [\bs{W},\bs{\hat{M}}_{V}]\hat{w} + \bs{Z}\hat{\beta},\label{imp_c}
\end{equation}
with the matching parameter $w \in \mathbb{R}^{p+1}$ estimated by 
\begin{equation}
    \hat{w} = ( [\bs{W},\bs{\hat{M}}_{V}]^{\top} [\bs{W},\bs{\hat{M}}_{V}])^{-1} [\bs{W},\bs{\hat{M}}_{V}]^{\top}\bs{\hat{M}}. \label{estimate_w}
\end{equation}
The data matrices for the two models (e.g., $\bs{W} \in \mathbb{R}^{ n \times p}$) can be defined accordingly and we provide the details in Appendix~\ref{pf_age}. In this case, the residual vector for the first IMP score is given by
\begin{equation}
    \bs{R} = \bs{M} - [\bs{W},\bs{\hat{M}}_{V}]\hat{w}. \nonumber
\end{equation} 
Note the second IMP score is not applicable for continuous environments due to the small sample size in each environment, thus we focus on the first IMP score. 
\begin{remark}
The $\text{IMP}_{\text{d}}$ and $\text{IMP}_{\text{c}}$ are similar in spirit, the $\text{IMP}_{\text{c}}$ relies on the first and second matching properties for identifying $W$ (so we can reuse $Z$ in~\eqref{imp_c}), whereas  $\text{IMP}_{\text{d}}$ directly tests the IMP since the estimation process treats all the coefficients as varying coefficients (also see the proof of Corollary~\ref{discrete_IMP}). 
\end{remark}

\section{Asymptotic Generalization Error} 
\label{sec:profile}

In this section, we provide the asymptotic generalization errors (as $n,m \to \infty$) of the $\text{IMP}_{\text{c}}$ and $\text{IMP}_{\text{d}}$ estimators for $(k,R,S)$'s that satisfy IMPs, i.e., $(k,R,S) \in \mathcal{I}_{\mathcal{M}}$. Recall that $\sigma^2 = \E[(N^{\tau})^{2}]$ is the population generalization error of the IMP estimators. Due to the estimations on both training and test data, the asymptotic generalization error can be decomposed to the error terms depending on the training data size $n$ and the test data size $m$ as follows. Let $c_{n} = \left\{\frac{\log(1/h)}{nh}\right\}^{1/2}+h^{2}$, where $h$ is the kernel bandwidth (see Appendix~\ref{PLS_estimate} for details). 

\smallskip

\begin{theorem}\label{thm_general}
For any $(k,R,S) \in \mathcal{I}_{\mathcal{M}}$, 
under the technical assumptions in Appendix~\ref{app:tech_lem}, the asymptotic generalization error of the $\text{IMP}_{\text{c}}$ estimator is given by
\begin{equation}
    \frac{1}{m} \sum_{i=1}^{m} (\hat{Y}^{\tau}_{i}- Y_{i}^{\tau})^2 =  \sigma^{2} +  O_{p}(c_{n} \vee n^{-1/2}) + O_{p}(c_{m} \vee m^{-1/2}) \nonumber.
\end{equation}
\end{theorem}
\smallskip

The following corollary considers the setting when the amount of unlabeled training and test data grows in a higher order than that of labels in the training data. The generalization error due to the estimation on the test data disappears. 

\smallskip

\begin{corollary}\label{coro_thm_gen}
Given \iid training data of size $n$ with $0<l_n<n$ labels and test data of size $m$, if $\max(\frac{l_n}{n}, \frac{l_n}{m}) \to 0$ as $\min(m,n) \to \infty$, under the technical assumptions in Appendix~\ref{app:tech_lem}, the asymptotic generalization error of the $\text{IMP}_{\text{c}}$ estimator is given by
\begin{equation}
      \frac{1}{m} \sum_{i=1}^{m} (\hat{Y}^{\tau}_{i} - Y_{i}^{\tau})^2 =\sigma^{2} 
      + O_{p}(c_{l_n} \vee l_n^{-1/2}). \nonumber
\end{equation}
\end{corollary}

\smallskip
The setting of discrete environments can be viewed as a special case of continuous environments, where the error term $c_{n} \vee n^{-1/2}$ due to the kernel estimation procedure is replaced by an error term from multiple OLS estimations. 

\begin{corollary} \label{discrete_IMP}
For any $(k,R,S) \in \mathcal{I}_{\mathcal{M}}$, 
under the technical assumptions in Appendix~\ref{app:tech_lem}, the asymptotic generalization error of the $\text{IMP}_{\text{d}}$ estimator is given by
\begin{equation}
    \frac{1}{m} \sum_{i=1}^{m} (\hat{Y}^{\tau}_{i}-Y_{i}^{\tau})^2 =  \sigma^{2} +  O_{p}(a_{n}) + O_{p}(a_{m}), \nonumber
\end{equation}
where $a_{n} = (\min_{e\in\Ec^{\text{train}}}n_{e})^{-1/2}$ and $a_{m} = m^{-1/2}$.
\end{corollary}

This asymptotic generalization error heavily depends on the environment with the smallest sample size, which also supports the fact that $\text{IMP}_{\text{d}}$ should not be employed for continuous environment settings.

\section{Experiments}
\label{sec:exp}
The prediction performance is measured by the mean residual sum of squares (RSS) on the test environments. We compare our method with several baseline methods: Ordinary Least Squares (OLS), stabilized regression (SR)~\cite{pfister2021stabilizing}, anchor regression (AR)~\cite{rothenhausler2021anchor}. We have also compared with domain invariant projection (DIP)~\cite{baktashmotlagh2013unsupervised},  conditional invariance penalty (CIP)~\cite{heinze2021conditional}, conditional invariant residual matching (CIRM)~\cite{chen2021domain}, and invariant risk minimization (IRM)~\cite{arjovsky2019invariant}; it turns out that the empirical performance of these methods is not as competitive as the other baselines in our experimental settings, thus we do not report them below.

The two IMP scores lead to two versions of our algorithm, and we refer to the first one as IMP and the second one as $\text{IMP}_{\text{inv}}$ (since it tests the invariance of the noise mean and variance). We focus on linear functions $f_{i}$'s in Algorithm~\ref{alg1}, namely, we use the IMP estimators. For the profile likelihood estimation, we adopt the Epanechnikov kernel $k(u) = 0.75\max(1-u^2,0)$ with the bandwidth fixed to be $0.1$. We test DIP, CIP, CIRM, and their variants provided in~\cite{chen2021domain} with the default parameters. For the anchor regression, we use a $5$-fold cross-validation procedure to select the hyper-parameter $\gamma $ from $\{0,0.05,0.1,\ldots,0.5\}$. The significance levels are fixed to be $0.05$ for all methods. We randomly simulate $500$ data sets for each experiment, if not mentioned otherwise. 

\subsection{Discrete environments}

First, we generate linear SCMs $\mathcal{M}^{e}$'s without interventions. 
For each $e \in \mathcal{E}^{\text{train}} = \{1,\ldots,5\}$ or $e \in \mathcal{E}^{\text{test}} = \{6,\ldots,10\}$, we randomly generate a linear SCM with 9 variables as follows. The graph $\mathcal{G}(\mathcal{M}^{e})$ is specified by a lower triangular matrix of \iid $\mathrm{Bernoulli}(1/2)$ random variables. The response $Y$ is randomly selected from the $9$ variables and we require that $Y$ has a least one parent and one child in $\mathcal{G}(\mathcal{M}^{e})$.  When $X$ and $Y$ are both intervened, we randomly choose a child of $Y$ to not be intervened. For each linear SCM, the non-zero coefficients are sampled from $\U[-1.5,-0.5] \cup[0.5,1.5]$ and the noise variables are standard normal. For each training or test environment, we simulate \iid data of sample size $300$.
\subsubsection{Interventions on $X$}\label{sec:int_X}
\begin{figure}[h]
\centering
\includegraphics[width=0.55\linewidth]{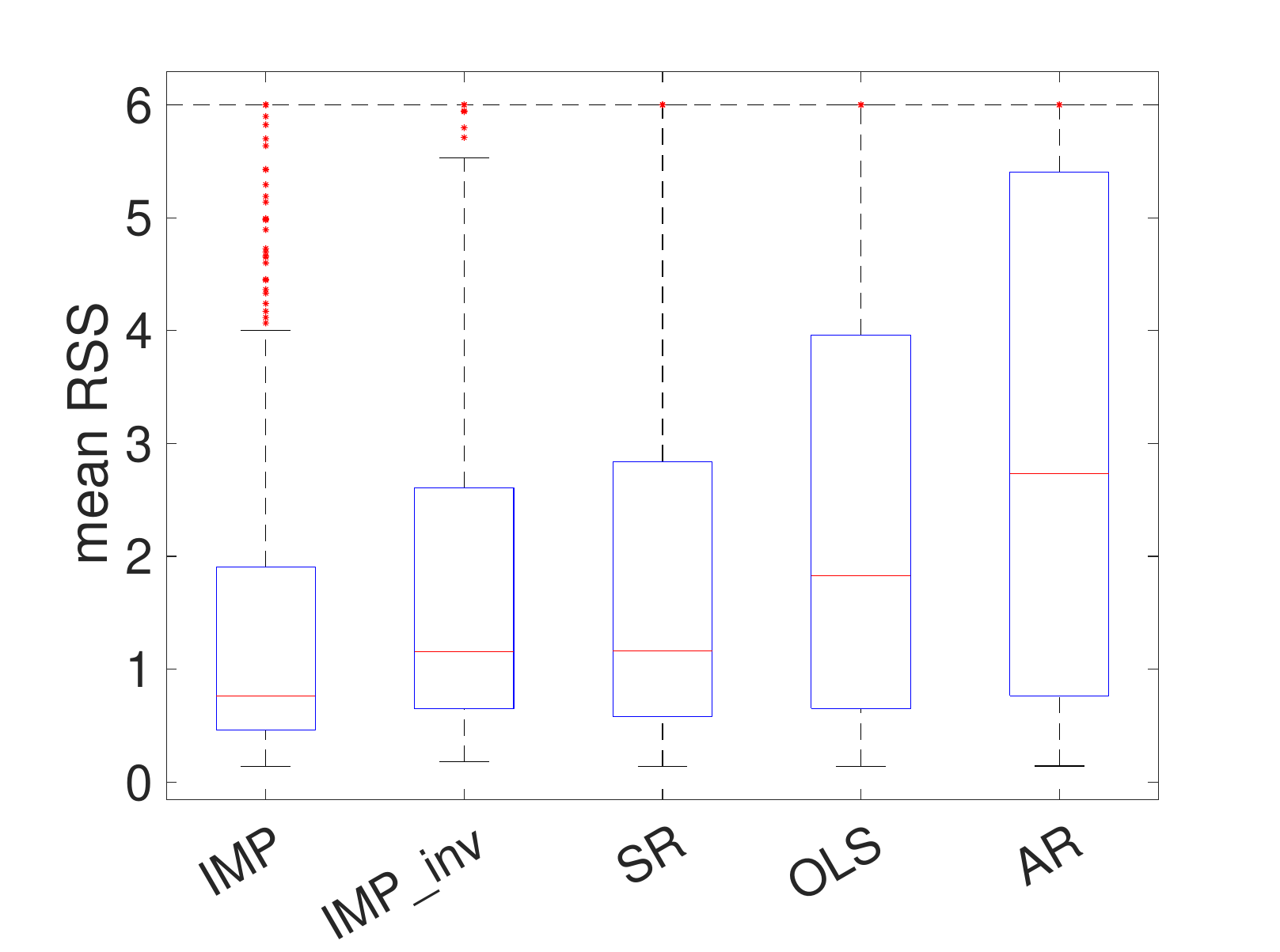}
\caption{Experiment 7.A.1 with discrete environments: Only a subset of $X$ is intervened. \label{fig1_1}
 } 
\end{figure}

Since the baseline methods have been examined extensively under shift interventions, we focus on shift interventions on $X$ for comparison. The general interventions on $X$ will be considered in Section~\ref{exp:robust}. Specifically, for each training environment, we randomly selected $4$ predictors to be intervened through shifts sampled from $\U[-2,2]$. For each test environment, the shifts are sampled from $\U[-10,10]$.
In Fig.~\ref{fig1_1}, $\text{IMP}_{\text{inv}}$ performs similarly to SR since they share a similar idea when only $X$ is intervened. Due to the averaging steps of IMP, $\text{IMP}_{\text{inv}}$, and SR, they have smaller variances compared with OLS and AR (a similar result has been reported in~\cite{pfister2019invariant}).

\subsubsection{Interventions on $Y$}

\begin{figure}[h]
\centering
\includegraphics[width=0.55\linewidth]{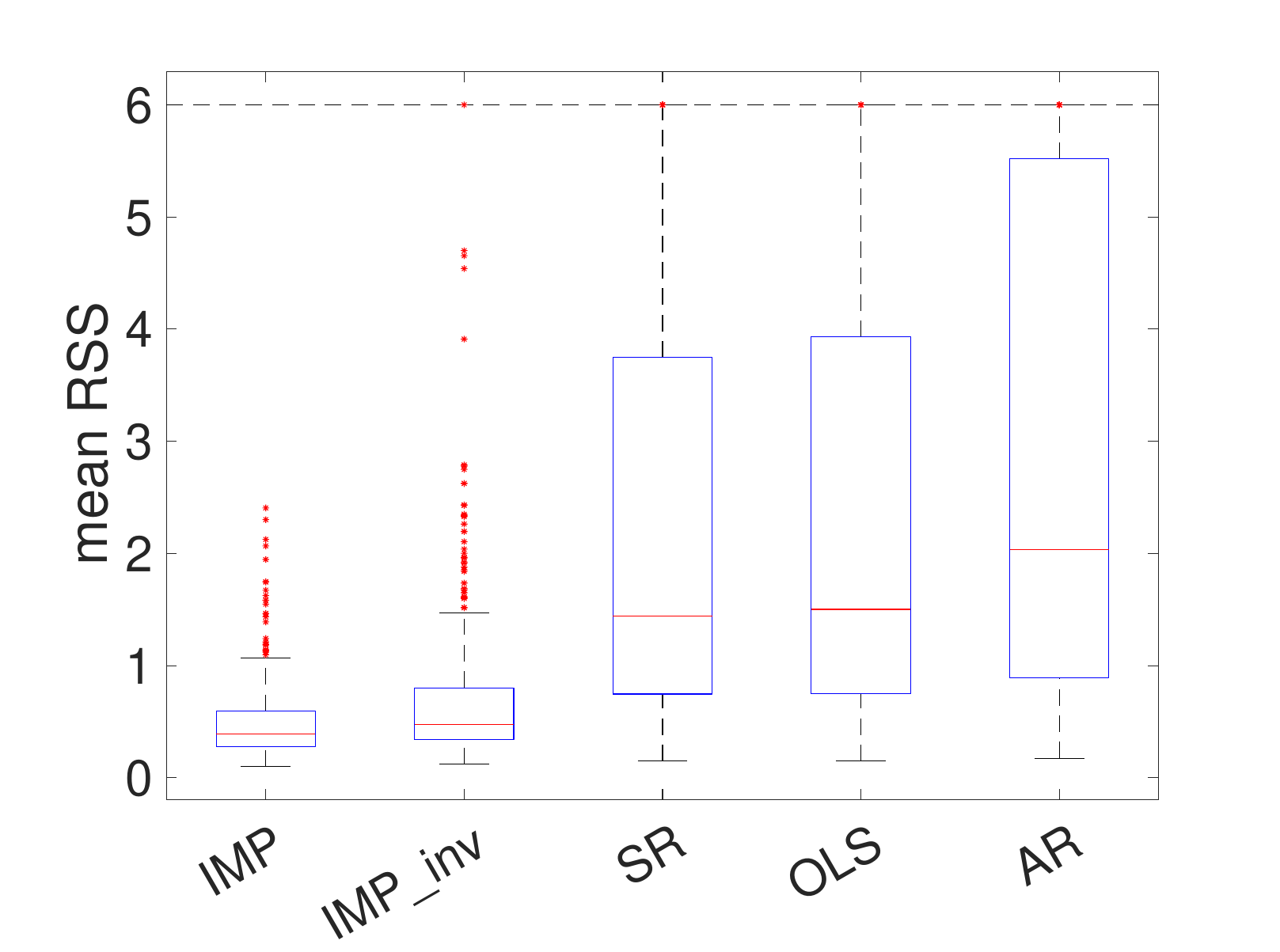}
\caption{{Experiment 7.A.2 with discrete environments: Only $Y$ is intervened.} \label{exp7_2}
 } 
\end{figure}

We consider the response $Y$ to be intervened through both the coefficients and shifts. We randomly select $n_{p} \sim \U\{1,\ldots,|PA(Y)|\}$ of parents of $Y$ to have varying coefficients. For each training environment, we add perturbation terms sampled from $\U[-2,2]$ to the original coefficients. For each test environment, the perturbations are sampled from $\U[-10,10]$. The shift intervention on $Y$ is the same as the shift interventions on $X$ in Section~\ref{sec:int_X}. In this setting, since none of the baseline methods allow interventions on $Y$ through the coefficients, they cannot even improve upon OLS. In Fig.~\ref{exp7_2}, IMP performs slightly better than $\text{IMP}_{\text{inv}}$, which may be due to the fact that the IMP method aims to find all possible IMPs, but the $\text{IMP}_{\text{inv}}$ only looks for IMPs that imply invariance.  To further examine our test procedures for identifying IMPs,  we check if the sufficient conditions in Theorem~\ref{thm1} are satisfied for the estimated IMPs $(\hat{k},\hat{R},\hat{S})$'s, summarized in Table~\ref{tab1_prob}. We observe that the conditions on $(k,S)$ are satisfied for the majority of cases, while the condition on $R$ holds with a noticeably lower empirical probability. Note that our sufficient conditions might be conservative. Moreover, due to the randomly selected coefficients, the intervention strength can be quite weak in some cases, making it challenging to distinguish true IMPs and non-IMPs. Fortunately, the averaging step helps to mitigate inaccurate predictions made by non-IMPs.

\begin{table}[!ht]
    \centering
    \begin{tabular}{c|cccc}
   \thickhline
       & $\hat{P}(\hat{k} \not \in PE)$ & $\hat{P}(PE \subseteq  \hat{R})$  & $\hat{P}(PE \subseteq  \hat{S})$ \\ \thickhline
       \thickhline
       $\text{IMP}$ & 0.9610 (0.0796)& 0.7503 (0.2511)& 0.9376 (0.1079) \\
        \thickhline
       $\text{IMP}_{\text{inv}}$ & 0.9603 (0.0850) & 0.7060 (0.2617)  & 0.9373 (0.1090) \\
        \thickhline
    \end{tabular}
    \caption{ Empirical results on the estimated IMP vs. conditions in Theorem~\ref{thm1} in terms of empirical probability and standard deviation (in parenthesis):  Only $Y$ is intervened.  
    \label{tab1_prob}}
\end{table}

\subsubsection{Interventions on both $X$ and $Y$} 
\label{sec:int_XY}
\begin{figure*}[h]\centering
\begin{subfigure}{}\includegraphics[scale=0.3]{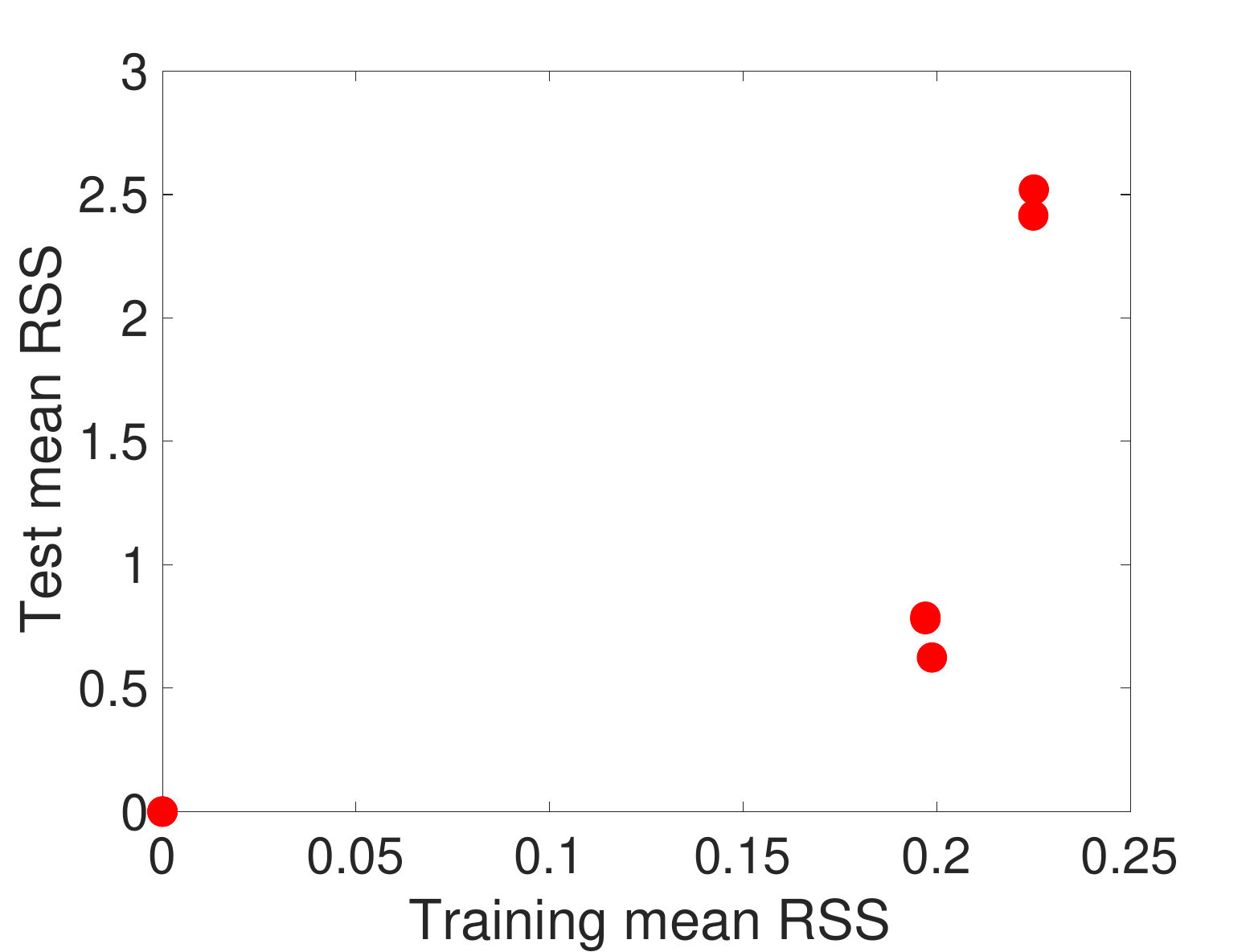}\end{subfigure}
\begin{subfigure}{}\includegraphics[scale=0.3]{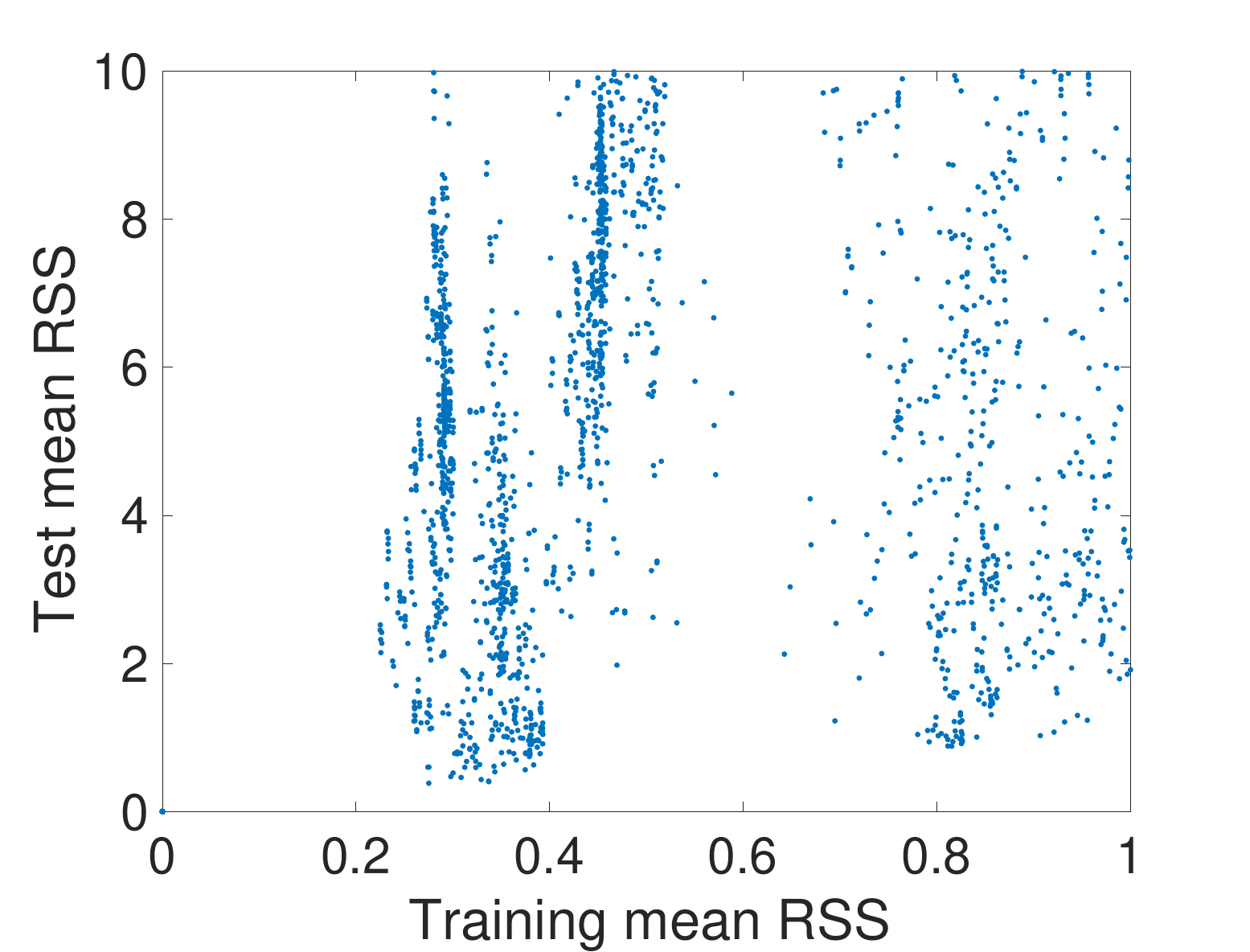}\end{subfigure}
\caption{\label{compare_imp}A generic example from Experiment 7.A.3. We compare the estimated IMPs (left plot in red) with estimated non-IMPs (right plot in blue) in terms of prediction performance in training vs. test environments. For visibility purposes, we only present a portion of estimated non-IMPs with low prediction errors. }\end{figure*}

The setting of interventions on both $X$ and $Y$ is simply a combination of the two settings above. In this challenging setting, our method outperforms the baselines by a large margin.  Since the sufficient conditions in Theorem~\ref{thm:int_xy} are not necessary for some settings as discussed in Remark~\ref{rmk:thm2}, we verify slightly relaxed conditions according to Remark~\ref{rmk:thm2}. The estimation task is more challenging in this setting compared with the setting with interventions only on $X$ or $Y$. The results in Table~\ref{tab_prob2} indicate that a large percentage of estimated IMPs conform to the conditions. In Fig.~\ref{compare_imp}, we present 
an illustration of the difference between the predictions made by the estimated IMPs and other $(k,R,S)$'s that are rejected by our test procedures (which we call the estimated non-IMPs). For the estimated IMP, the gap between the prediction error of the training environments and that of the test environments is relatively small. However, the gap can be huge for the majority of the estimated non-IMPs.

\begin{table}[!ht]
    \centering
 \begin{tabular}{c|cccc}
   \thickhline
       & $\hat{P}(\hat{k} \not \in PE\cup X^{\text{int}})$ & $\hat{P}(A)$  & $\hat{P}(B )$ \\ \thickhline
       \thickhline
       $\text{IMP}$ & 0.8767 (0.2749)& 0.7170 (0.3515)& 0.7003 (0.4412) \\
        \thickhline
       $\text{IMP}_{\text{inv}}$ & 0.8558 (0.3144) & 0.6571 (0.4085) & 0.6853 (0.4460) \\
        \thickhline
    \end{tabular}
    \caption{Empirical results on the estimated IMP vs. conditions in Theorem~\ref{thm:int_xy} in terms of empirical probability and standard deviation (in parenthesis): Both $X$ and $Y$ are intervened. Events $A$ and $B$ are defined by $A = \{PE \subseteq  \hat{R}\}\cap \{(X^{\text{int}}(Y)\cup X^{\text{int}}(X_{\hat{k}}) )\cap \hat{R}=\varnothing\}$ and $B= \{PE \subseteq  \hat{S}\} \cap \{X^{\text{int}}(Y)\cap \hat{S}=\varnothing \}$, respectively. 
    \label{tab_prob2}}
\end{table}

\begin{figure}[h]
\centering
\includegraphics[width=0.55\linewidth]{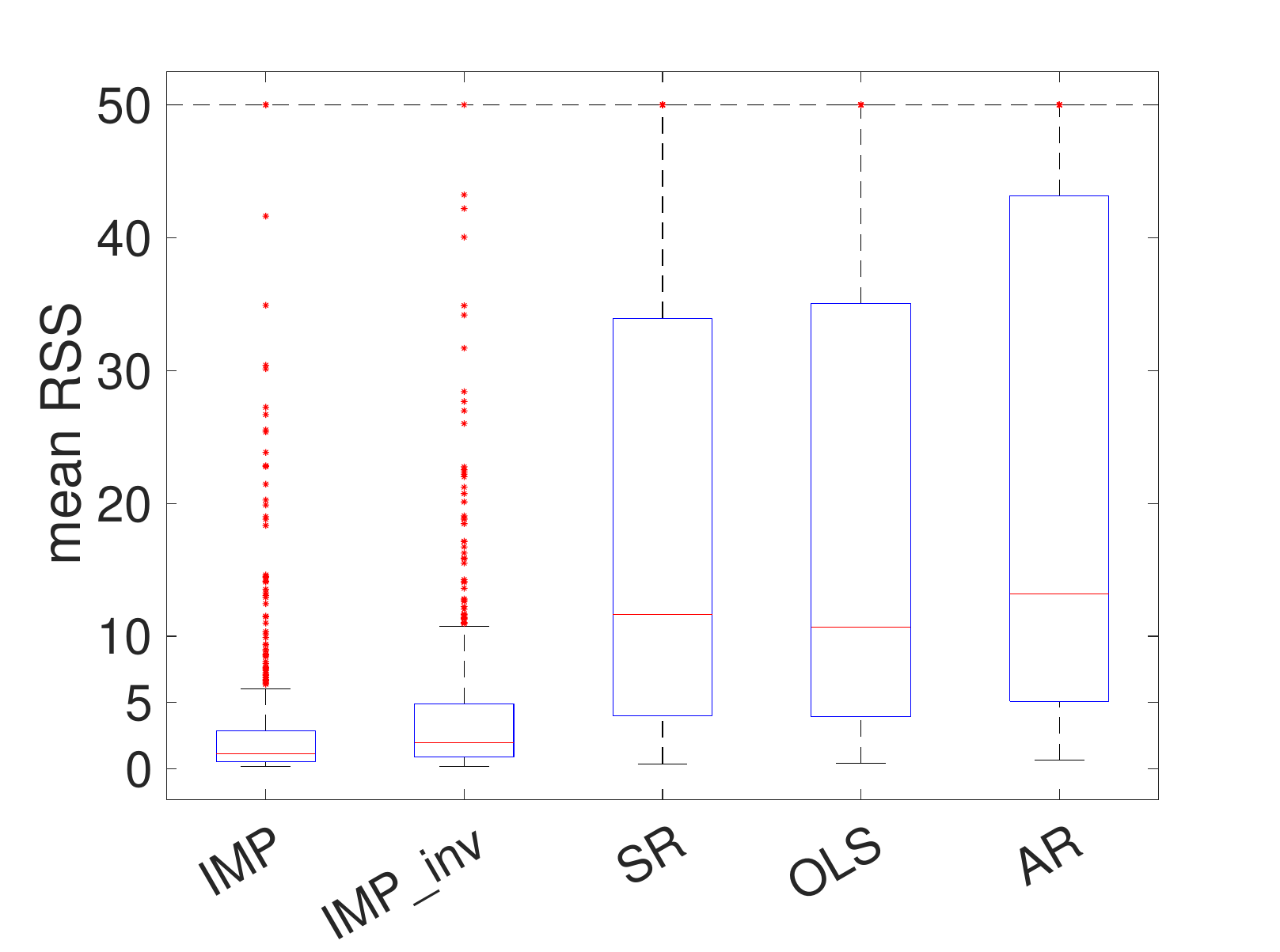}
\caption{{Experiment 7.A.3 with discrete environments: Both $X$ and $Y$ are intervened. } \label{fig1}
 } 
\end{figure}

\subsection{Interventions on both $X$ and $Y$ (continuous)} 
 In this more challenging setting, we only compare with OLS, since AR only considers shift interventions while the other baselines are not developed for continuous environments, and also recall that $\text{IMP}_{\text{inv}}$ is proposed for discrete environments. First, we define $\{U_{1},\ldots,U_{800}\}$ sampled from $\U[0,1]$ and $\{U_{1}^{\tau},\ldots,U_{800}^{\tau}\}$ sampled from $\U[1,2]$ as the training and test environments, respectively. Similar to the setting of discrete environments, we randomly generate the linear SCMs without interventions first and then add interventions to the model. Due to the high computational complexity, we focus on graphs with $5$ nodes where $2$ predictors are intervened. The interventions on the coefficients and shift interventions are defined by adding a perturbation term $a\sin(2\pi w U_i)$, where $w$ is sampled from $\U[0.5,2]$. The parameter $a$ is fixed to be $2$ for the training environments (i.e., the same range as for the discrete case) and $5$ for the test environments. Since the parameter space is much smaller than in the previous experiments, we only generated $100$ data sets. Overall, the performance of our IMP algorithm is similar to that in the discrete setting.


\begin{figure}[h]
\centering
\includegraphics[width=0.55\linewidth]{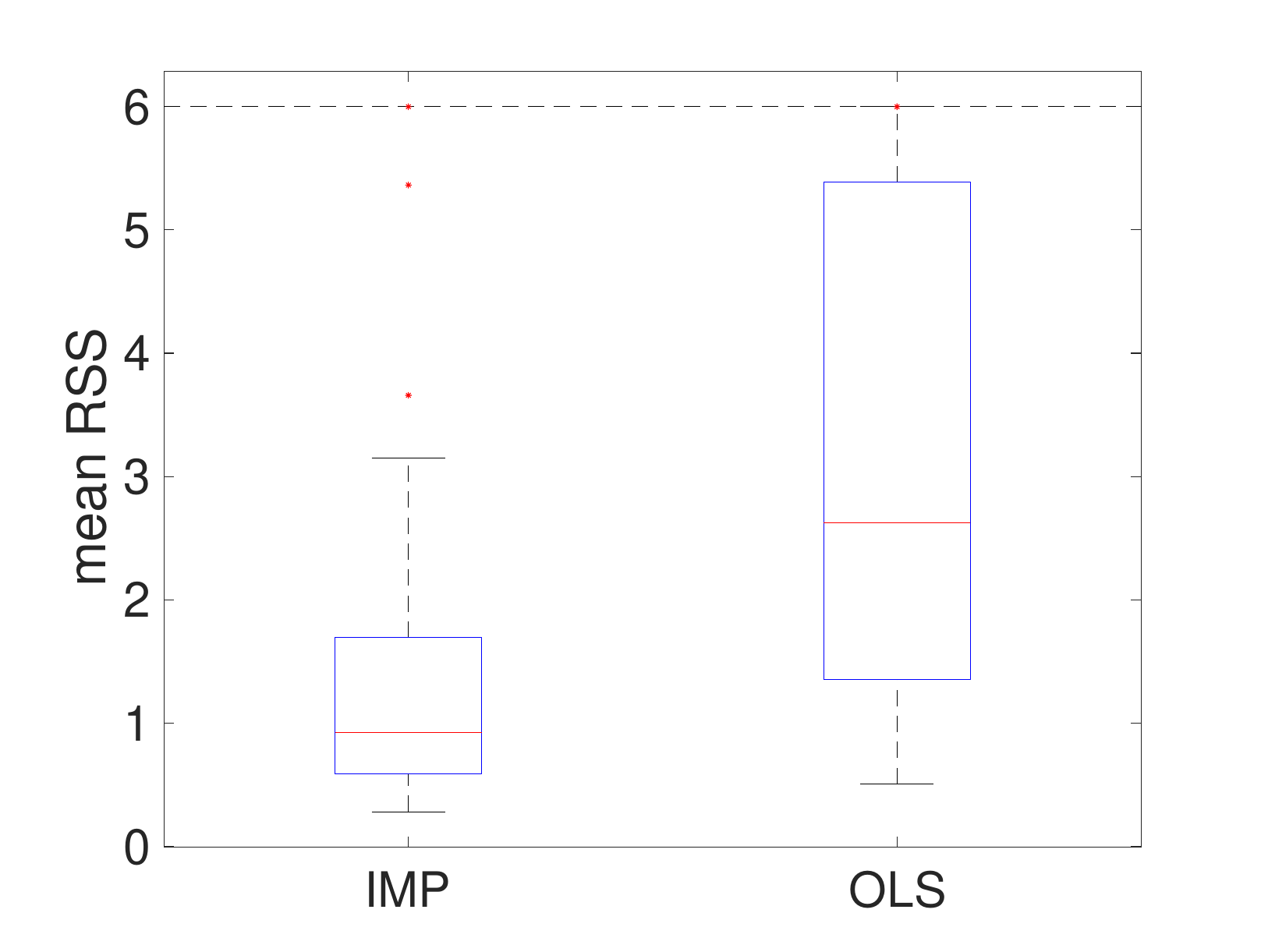}
\caption{{Experiment 7.B with continuous environments: Both $X$ and $Y$ are intervened. }\label{fig1}
 } 
\end{figure}

\subsection{Robustness} \label{exp:robust}

\begin{figure}[h]
\centering
\includegraphics[width=0.55\linewidth]{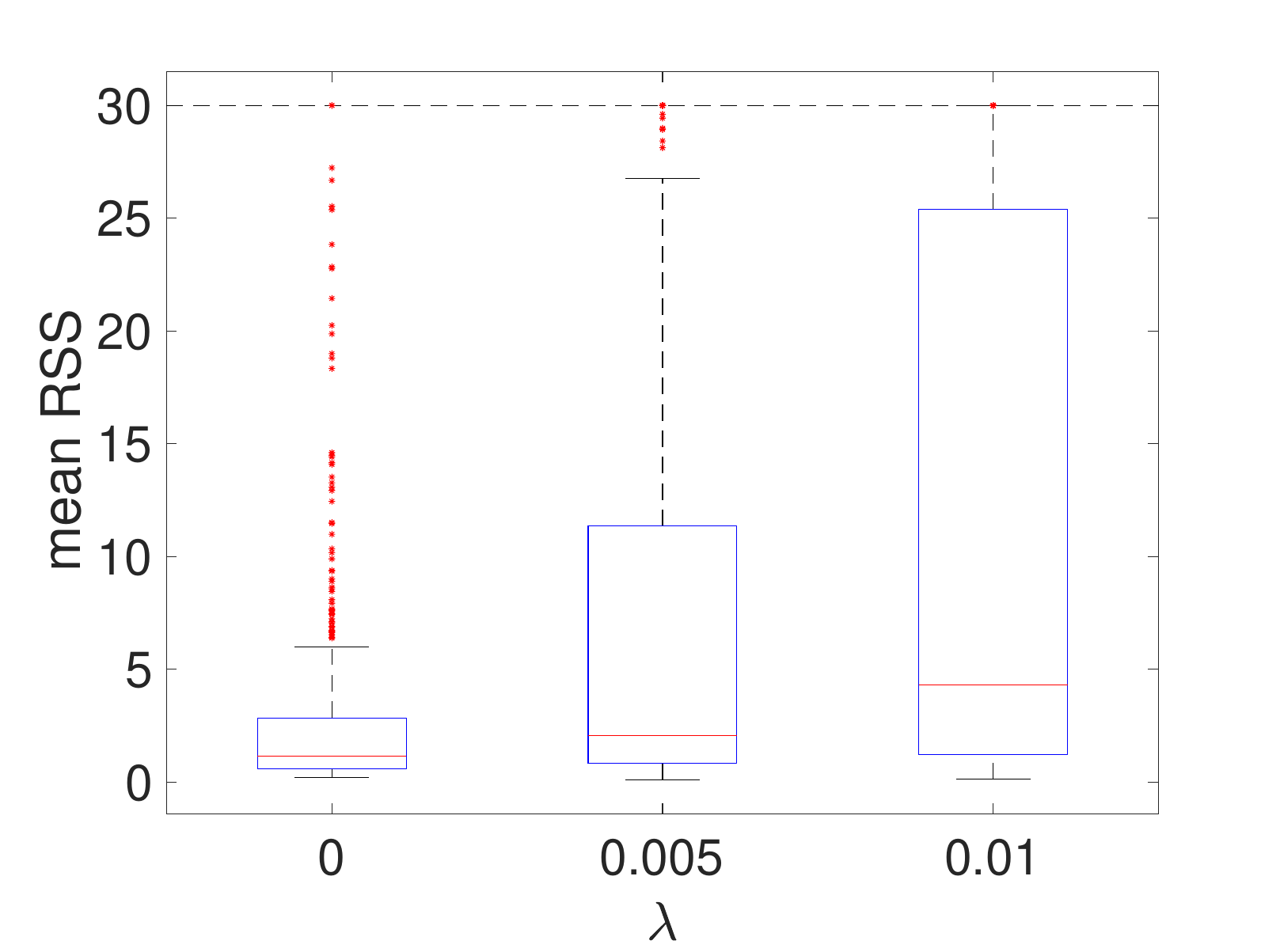}
\caption{Experiment 7.C with discrete environments: $Y$ and every $X_{j}$ are intervened.  \label{fig7_3} 
 } 
\end{figure}

In the previous experiments, we only consider shift interventions on $X$ and interventions on $Y$ other than the noise variance. In this experiment, we consider an extreme case when only a child of $Y$ with the highest causal ordering is not intervened (i.e., Assumption~\ref{assump2} is satisfied), \emph{all other variables} are intervened through \emph{every parameter}. Specifically, a shift intervention or an intervention on the coefficient is defined by adding a perturbation term to the original parameter. The perturbation term is sampled from $\U[-2,2]$ for the training data, and from $\U[-5,5]$ for the test data. The intervened noise variances are sampled from $\U[0.75,1.25]$ for each training environment, and from $\U[0.5,1.5]$ for each test environment. To test how sensitive our method is with respect to Assumption~\ref{assump2} in this challenging setting, we gradually add interventions to the child of $Y$ that is not intervened, where the shifts and coefficient interventions are sampled from $\U[-2\lambda,2\lambda]$ and $\U[-5\lambda,5\lambda]$ for the training and test environments, respectively. The intervened noise variances are sample from $\U[1-0.25\lambda,1+
0.25\lambda]$ for training, and from $\U[1-0.5\lambda,1+
0.5\lambda]$ for testing. The parameter $\lambda \in [0,1]$ controls the intervention strength. Due to the results in Section~\ref{sec:int_XY}, it would not be informative to compare with the baseline methods, so we focus on our IMP method. Note that our $\text{IMP}_{\text{inv}}$ is also not included since IMPs will not imply invariance in this case (see Remark~\ref{rmk:imp_inva}). Overall, as shown in Fig.~\ref{fig7_3}, the median of the mean RSS is not too sensitive with respect to mild interventions on the child that is not intervened, but the variance increases rapidly.  
\subsection{COVID Data Set}

For observational data, there is often no ground truth regarding the set of variables that are intervened. If the data is not collected under a carefully designed experimental setting (e.g., the gene perturbation experiments from~\cite{peters2016causal}), it is reasonable to assume that every variable is more or less intervened, especially the response. To examine our methods under real-world distribution shifts, we consider a COVID data set~\cite{haratian2021dataset} collected at $3142$ US counties from January 22, 2020, to June 10, 2021. The data set consists of 46 predictive features that are relevant to the number of COVID cases. Some of the features are treated as fixed or time-invariant, e.g., population density, age distribution, and education level, and others are temporal features, e.g., temperature, social distancing grade, and the number of COVID tests performed. 

There are plenty of prediction tasks that can be potentially designed for this data set. In our experiment, we focus on the prediction of the number of COVID cases~\footnote{The number of COVID cases is measured in thousands.} using the $12$ temporal features for the time interval from March 1, 2020, to September 30, 2020 ($214$ days). For different regions of the US, features such as population density can differ greatly, making it challenging for a prediction model to perform well on a diverse set of regions. Specifically, we take $5$ major cities from the Western U.S. (Los Angeles, San Francisco, San Diego, Seattle, and Phoenix) as $5$ training environments,  and we examine our methods and the baselines regarding $8$ cities/counties that are mostly from the Eastern U.S. (Baltimore, Boston, Philadelphia, New York County, Queens County~\footnote{Only the data from Manhattan (New York County) and Queens are available for New York City in this data set.}, Houston, Miami, and Chicago). According to the total number of COVID cases in the considered time interval, we partition (using $50$k as the threshold) the cities/counties for testing into two groups in Table~\ref{tbl_covid}, which makes the results more informative as most of the methods perform worse on cities/counties with higher total cases. Since no invariance in terms of mean and variance can be identified by $\text{IMP}_{\text{inv}}$, we focus on the IMP method. Overall, IMP and CIP show similar performance for the lower total cases, while the performance of CIP degrades more over the higher total cases compared with IMP. Since real-world interventions are often beyond shift interventions, AR does not improve upon OLS for this data set. It is noteworthy that no stable sets are found by SR at the significance level of $0.05$, implying that the invariance principle is likely to be violated.  IRM shows less competitive performance compared with other methods, thus its results are not presented in Table~\ref{tbl_covid}.


\begin{table}[!ht]
    \centering
    \begin{tabular}{c|cccccccc}
   \thickhline
        Cities/Counties (total cases) & IMP  & OLS & AR & SR & CIP & DIP & CIRM \\ \thickhline
        Baltimore (18k) & 0.1292  & 0.0937 & 0.0379 & 0.0978 & 0.0184 & 0.9199 & 0.3503 \\
        Boston (26k) & 0.0238  & 0.0503 & 0.0351 & 0.0543 & 0.0192 & 0.5820 & 0.4605 \\ 
        Philadelphia (32k) & 0.1532  & 0.1191 & 0.0701 & 0.0970 & 0.0353 & 1.0407 & 0.4217 \\ 
        NY County (34k)  & 0.1611  & 0.4088 & 0.5606 & 0.3449 & 0.3537 & 2.4055 & 2.6372 \\ \thickhline
        average mean RSS & 0.1168  & 0.1680 & 0.1759 & 0.1485 & \textbf{0.1066} & 1.2370 & 0.9674 \\ \thickhline \thickhline
        Queens County (73k) & 0.1126  & 0.1020 & 0.0882 & 0.1109 & 0.0673 & 0.9389 & 0.9647 \\ 
        Houston (143k) & 1.0884  & 1.2042 & 1.2200 & 1.1673 & 1.3705 & 1.7802 & 1.4982 \\ 
        Chicago (146k) & 0.3687  & 0.4489 & 0.4026 & 0.4099 & 0.4562 & 0.6328 & 0.3499 \\ 
        Miami (171k) & 0.4534  & 0.4621 & 0.5170 & 0.4025 & 0.6011 & 1.4465 & 0.8004 \\ 
        \thickhline
         average mean RSS   & \textbf{0.5058}  & 0.5543 & 0.5548 & 0.5226 & 0.6238 & 1.1996 & 0.9033 \\ \thickhline
    \end{tabular}
    \caption{Experiment results of the COVID data set.\label{tbl_covid}}
\end{table}

\section{Discussion} \label{sec:discuss}
To deal with general interventions on the response, we introduce the IMP that allows for an alternative form of invariance when the traditional invariance principle fails. We provide explicit characterizations of the IMP under different intervention settings and provide asymptotic generalization error analysis for both the discrete environment and the more challenging continuous environment settings, supported by our empirical studies.   

Our work is motivated by allowing general interventions on the response, and it can be extended into several directions. First, we observe a connection between the IMP and self-supervised learning schemes (e.g.,~\cite{devlin2018bert} from natural language processing). Specifically, the prediction of $X_{k}$ using $X_{R}$ corresponds to the pretext (or pretrained) task, and the linear relation in the IMP solves the downstream prediction task. We believe that methods developed through causal modeling including our IMP method will bring new opportunities for self-supervised learning.  We have discussed several heuristic approaches to alleviate the computation complexity issue; it would be worthwhile to analyze such methods and provide theoretical guarantees (e.g., exploiting sparsity beyond the setting in~\cite{du2023generalized}). Our asymptotic analysis of the generalization error is developed given that IMPs are correctly identified. It would be interesting to further analyze the consistency of the set of IMPs $\hat{\mathcal{I}}$ identified by the proposed algorithms or new efficient algorithms.

\bibliographystyle{IEEEtran}
\bibliography{ref.bib}

\newpage

\begin{appendices}

\section{Proof of Proposition~\ref{prop:unique_theta}}
Let $\mathcal{E}^{\text{train}} = \{e_{1},\ldots,e_{n}\}$. For a tuple $(k,S,R)$ that satisfies the IMP, let 
\begin{align*}
    \hat{\boldsymbol{Y}} = (\E[Y^{e_{1}}|X_{S}^{e_{1}}],\ldots,\E[Y^{e_{n}}|X_{S}^{e_{n}}])^{\top},
\end{align*}
\begin{equation}
   \boldsymbol{\tilde{X}} = \begin{bmatrix}
(X^{e_{1}})^{\top}& \E_{l}[X_{k}^{e_{1}}|X_{R}^{e_{1}}]\\
\vdots &  \vdots \\
(X^{e_{n}})^{\top}& \E_{l}[X_{k}^{e_{n}}|X_{R}^{e_{n}}] 
   \end{bmatrix} := \begin{bmatrix}
   \boldsymbol{X} & v
   \end{bmatrix}
 \nonumber,
\end{equation}
where the rows of $\boldsymbol{\tilde{X}}$ are independent. 
According to~\eqref{imp_theta}, we have $\hat{\boldsymbol{Y}} =  \boldsymbol{\tilde{X}}\theta$. Then, if $\E[\boldsymbol{\tilde{X}}^{\top}\boldsymbol{\tilde{X}}]$ is invertible, we have
\begin{align}
  &(\E[\boldsymbol{\tilde{X}}^{\top}\boldsymbol{\tilde{X}}])^{-1}\E[\boldsymbol{\tilde{X}}^{\top} \hat{\boldsymbol{Y}}]
=(\E[\boldsymbol{\tilde{X}}^{\top}\boldsymbol{\tilde{X}}])^{-1}\E[\boldsymbol{\tilde{X}}^{\top}\boldsymbol{\tilde{X}} ]\theta=\theta. \label{determin_theta}
\end{align}
Now, we prove the invertibility. Observe that
\begin{equation}
   \E[\boldsymbol{\tilde{X}}^{\top}\boldsymbol{\tilde{X}}] =  \begin{bmatrix}
   \E[\bs{X}^{\top}\bs{X}] & \E[\bs{X}^{\top}v]\\
   \E[v^{\top}\bs{X}] & \E[v^{\top}v]
   \end{bmatrix},\nonumber
\end{equation}
where $  \E[\bs{X}^{\top}\bs{X}] = \sum_{i=1}^{n}\E[X^{e_{i}}(X^{e_{i}})^{\top}]$ is inveritble since it is a sum of positive-definite matrices. Then, $\E[\boldsymbol{\tilde{X}}^{\top}\boldsymbol{\tilde{X}}]$ is invertible if and only if
\begin{equation}
   \E[v^{\top}v]-  \E[v^{\top}\bs{X}] \E^{-1}[\bs{X}^{\top}\bs{X}] \E[\bs{X}^{\top}v] \neq 0. \nonumber
\end{equation}
This is equivalent to
\begin{equation}
    \E[(v-\boldsymbol{X}\beta)^{\top}(v-\boldsymbol{X}\beta)] \neq 0, \nonumber
\end{equation}
where $\beta:=\E^{-1}[\bs{X}^{\top}\bs{X}] \E[\bs{X}^{\top}v]$.  This is true since there is no $b \in \mathbb{R}^{d}$ such that $v= \boldsymbol{X}b$ almost surely by our assumption in~\eqref{eq_assup_unique}. Therefore, $\theta$ is uniquely determined by~\eqref{determin_theta}.

\section{Problem Formulation Using an environmental random variable}\label{app:sec_note}
By introducing an environmental random variable $E \in \mathcal{E}^{\text{all}}$, we define a mixture of $\mathcal{M}^{e}$'s, $e \in \mathcal{E}^{\text{all}}$, as follows,
\begin{numcases}{\mathcal{M}:}
    X = \gamma(E) Y + B(E)X+\varepsilon_{X}(E) \nonumber\\
    Y = \beta^{\top}(E)X +  \varepsilon_{Y}(E), \nonumber
\end{numcases}
where $E$ in a root node in $\mathcal{G}(\mathcal{M})$ and the noise variables are jointly independent given $E$. We do not specify the distribution of $E$ but assume that $E$ does not have a degenerate distribution. Under this formulation, the invariance property~\eqref{inva_trans} is equivalent to 
\begin{equation}
    Y \independent E  \,|\, \phi(E,X), \label{inva_const_ind}
\end{equation}
and the invariant, first, and second matching properties can be equivalently written as 
\begin{align}
   \E_{l}[Y|X_{S},E=e] &= \lambda\E_{l}[X_{k}|X_{R},E=e]+\eta^{\top}X, \\ 
    \E_{l}[Y|X_{S},E=e] &= \lambda_{Y}\E[Y|X_{PA(Y)},E=e]+\eta_{Y}^{\top}X, \label{fmp_E} \\ 
    \E_{l}[X_{k}|X_{R},E=e] &= \lambda_{X}\E[Y|X_{PA(Y)},E=e]+\eta_{X}^{\top}X, 
\end{align}
for $e \in \mathcal{E}^{\text{all}}$. 
As a special case of $\mathcal{M}$, we
define a mixture of $\mathcal{M}^{e,1}$'s as
\begin{numcases}{\mathcal{M}^{1}:}
X = \gamma Y + B X+ \varepsilon_{X} \label{e_pfq:train_m_X} \\
Y = (\alpha(E)+\beta)^{\top}X +  \varepsilon_{Y},  \label{eq_pf:train_m_Y}
\end{numcases}
where only $Y$ is intervened through the coefficients. 

The proofs of Theorems~\ref{thm1},~\ref{thm:int_xy} and Corollaries~\ref{coro1},~\ref{coro:int_noise_var} will be presented under this formulation. 

\section{Proof of Theorem~\ref{thm1}}
\label{app:thm1_proof}
The problem formulation and necessary notation for the proof are introduced in Appendix~\ref{app:sec_note}. 
For the first part, let $Z(E) = \alpha^{\top}(E)X$ denote an additional node in the acyclic graph $\mathcal{G}(\mathcal{M}^{1})$, then the assignment of $Y$ in~\eqref{eq_pf:train_m_Y} becomes 
\begin{equation}
    Y = Z(E)+ \beta^{\top}X+\varepsilon_{Y} \label{thm1_define_z},
\end{equation}
where $E$ is no longer a parent of $Y$. Note that $Z(E) = \E[Y|X_{PA(Y)},E] - \beta^{\top}X$ as we assume both $X$ and $Y$ have zero means. Since $E$ is a root node, observe that $E$ and $Y$ can be d-connected through only two types of paths as follows,
\begin{enumerate}
    \item  $E \to Z(E) \to Y$, \label{path1_thm}
    \item  $E \to Z(E) \leftarrow X_{i} \to  \cdots \to X_{l} \leftarrow \cdots  \leftarrow Y$,\label{path2_thm}
\end{enumerate}
where $i \in PE$, and $ X_{i} \to  \cdots \to X_{l} \leftarrow \cdots  \leftarrow  Y$ is a V-structure for some $l \in DE(i) \cap DE(Y)$. Note that the second type of path does not exist if Assumption~\ref{assump1} is not satisfied or $CH(i)\setminus Y$ is empty.

We start by showing that the d-separation $Y \independent_{\mathcal{G}} E \,|\, \{Z(E) ,X_{S}\}$ holds given $PE \subseteq S$. First, the first path is immediately blocked by $Z(E)$. Second, for any $s \in CH(i)\setminus Y$, the path $E \to Z(E) \leftarrow X_{i} \to X_{s}$ is blocked by $\{Z(E),X_{i}\}$. Thus, the second path is blocked given $Z(E)$ and $X_{S}$.
According to the Markov property of SCMs~\cite{pearl2009causality}, the d-separation $Y\independent_{\mathcal{G}} E \,|\, \{Z(E) ,X_{S}\}$ implies 
\begin{equation}
    Y \independent E \,|\, \{Z(E) ,X_{S}\}. \label{cond_z_xs}
\end{equation}

Now, we prove that the above conditional independence implies the first matching property~\eqref{fmp_E}. 

By definition, the LMMSE estimators only rely on the (finite) first two moments of the variables. Thus we start with the case when $(X,Y)|_{E=e}$ is jointly Gaussian, for each $e\in \mathcal{E}^{\text{all}}$. First we have 
\begin{align*}
    \E_l[Y|X_{S},E=e] &\overset{(a)}{=}\E[Y|X_{S},E=e] \\
    &\overset{(b)}{=} \E[Y|Z(e),X_{S},E=e], 
    \end{align*} 
where $(a)$ follows from the Gaussian assumption on $(X,Y)|_{E=e}$ and $(b)$ from the fact that $Z(e)$ is a function of $\{X_{S},E=e\}$ given our assumption that $PE \subseteq S$. This implies that 
\begin{align*}
   \E_l[Y|X_{S},E]=\E[Y|Z(E),X_{S},E]\overset{(a)}{=}\E[Y|Z(E),X_{S}],
\end{align*}
where $(a)$ follows from the conditional independence relation~\eqref{cond_z_xs}. Using the Gaussian assumption again, we have $\E[Y|Z(e),X_{S}]=\E_l[Y|Z(e),X_{S}]$. Thus putting all the pieces together, we obtain
\begin{align}
    \E_l[Y|X_{S},E=e]&=\E_l[Y|Z(e),X_{S}]\\
    &= aZ(e) + b^{\top}X_{S},\label{eq:lmmse_thm1}
\end{align}
where $a \in \mathbb{R}$ and $b \in \mathbb{R}^{|S|}$ that are not functions of $E$. When $(X,Y)|_{E=e}$ is non-Gaussian, one can replace it with Gaussian random variables with the matching first and second moments. Then the same argument leading to~\eqref{eq:lmmse_thm1} still holds. Then, the first matching property follows from the fact that $Z(E) = \E[Y|X_{PA(Y)},E] - \beta^{\top}X$. 



Similarly, for the second part, we first show the following d-separation $X_{k}\independent_{\mathcal{G}}\, E \,|\, \{Z(E), X_{R}\}$. Observe that $X_{k}$ and $E$ can be d-connected through two types of paths as follows,
\begin{enumerate}
        \item  $E \to Z(E) \to Y \cdots X_{k}$, \label{path1_thm}
    \item  $E \to Z(E) \leftarrow X_{i}  \cdots X_{k}$,\label{path2_thm}
\end{enumerate}
with $i \in \{j\in\{1,\ldots,d\}:\alpha_{j}(E)\neq 0\}$, where $Y \cdots X_{k}$ denotes any directed path between $Y$ and $X_{k}$, and similarly for $X_{i} \cdots X_{k}$. The two types of paths are immediately blocked by $\{Z(E),X_{R}\}$ under our assumption that $PE\subseteq R$. We thus have the following when $X|_{E=e}$ is jointly Gaussian,  
\begin{align}
    \E_l[X_{k}|X_{R},E=e] &= \E_l[X_{k}|Z(e),X_{R},E=e] \nonumber\\
    &= \E_l[X_{k}|Z(e),X_{R}] \nonumber\\
    &= cZ(e)+d^{\top}X_{R} \label{eq_z_smp}
\end{align}
for some $c \in \mathbb{R}$ and $d \in \mathbb{R}^{ |R|}$ do not depend on $E$. The non-Gaussian cases can be handled in the same way as before, and thus the second property follows again from $Z(E) = \E[Y|X_{PA(Y)},E] - \beta^{\top}X$. Observe that we have $\lambda_{X}=c$ in the second matching property and Assumption~\ref{assump_one_child} is a necessary condition for $\lambda_{X}\ne 0$.    

Finally, given $R \subseteq S$, we have that~\eqref{eq_z_smp} with $c\neq 0$ (i.e., $\lambda_{X}\neq 0$) provides a one-to-one mapping between $\{Z(E),X_{S}\}$ and $\{\E_l[X_{k}|X_{R},E],X_{S}\}$. Therefore, the conditional independence~\eqref{cond_z_xs} is equivalent to
\begin{equation}
       Y \independent E \,\big|\,\{\E_l[X_{k}|X_{R},E],\,X_{S}\}. \end{equation}
 This implies that, using our previous notation, $\phi_{e}(X^{e})= (X_{S}^e,\, \E_{l,\Pcal_{e}}[X_{k}|X_{R}])^{\top}$ satisfies the invariance property~\eqref{inva_trans}, which implies\footnote{We use the shorthand $\phi_{e}(X)$ for $\phi_{e}(X^e)$ since the expectation is with respect to $\Pcal_{e}$.} $ \E_{\Pcal_{e}}[Y|X_{S}] = \E_{\Pcal_{e}}[Y|\phi_{e}(X)]$. Effectively, $\E_{\Pcal_{e}}[Y|\phi_{e}(X)]$ serves as a representation of $\E_{\Pcal_{e}}[Y|X_{S}]$ that is invariant. Note that $\Phi$ is not empty when $\lambda_{X}\neq 0$ holds with $R \subseteq S$. This implies that any $\phi\in \Phi$ with $S = \{1,\ldots,d\}$ minimizes $\mathcal{L}_{\text{test}}(f_{\phi})$, since the optimality of $\phi \in \Phi$ only relies on the corresponding $S$.

\begin{remark}\label{rem_int_noise_var}
When $\varepsilon_{Y}$ is  replace by $\varepsilon_{Y}(E)$, we consider the following two cases.
\begin{enumerate}
    \item The mean of $\varepsilon_{Y}^{e}$ is a function of $e$, and its variance is a constant. 
    \item The variance of $\varepsilon_{Y}^{e}$ is a function of $E$, and its mean can be either a function of $e$ or a constant.  
\end{enumerate}

For the first case, we can introduce $X_{d+1}:=1$ that is a parent of $Y$ with $\alpha_{d+1}:= \E[\varepsilon_{Y}(E)|E]$. Additionally, $X_{d+1}$ is a parent of every $X_{j}$ such that $\varepsilon_{X,j}$ has a non-zero mean. Specifically, the coefficient of $X_{d+1}$ in the assignment of $X_{j}$ will be $\E[\varepsilon_{X,j}]$. Thus, the problem reduces to the setting when $\varepsilon_{X}^{e}$ and $\varepsilon_{Y}^{e}$ have zero means, which has been proved in Theorem~\ref{thm1}. For the second case, however, the varying variance of $\varepsilon_{Y}(E)$ cannot be separated as the mean, thus $E$ is always a parent of $Y$ and the path $E \to Y$ cannot be blocked by $Z$ or any $X_{S} \subseteq \{X_{1},\ldots,X_{d}\}$, i.e., the proof of Theorem~\ref{thm1} breaks down.
\end{remark}

\section{Proof of Proposition~\ref{prop_l} }
The following lemma is a slight extension of Lemma 3.6 from~\cite{pfister2021stabilizing}, where we consider linear models with dependent noise variables rather than linear SCMs considered in~\cite{pfister2021stabilizing}.
\smallskip

\begin{lemma}\label{lemma_tech}
Consider $V \in \mathbb{R}$ and $X= (X_{1},\ldots,X_{p})^{\top}  \in \mathbb{R}^{p}$ satisfying a linear model,
\begin{equation}
  \begin{bmatrix}
  V \\
  X
  \end{bmatrix}
  =   \begin{bmatrix}
0 & h^{\top}\\
g & A
  \end{bmatrix}
  \begin{bmatrix}
  V \\
  X
  \end{bmatrix}+  
  \begin{bmatrix}
  e_{V} \\
  e_{X} 
  \end{bmatrix} := \tilde{A} \begin{bmatrix}
  V \\
  X
  \end{bmatrix}+e,  \label{linear_m}
\end{equation}
 where $e_{V} \in \mathbb{R}$, $g,h, e_{X} \in \mathbb{R}^{p}$, and $A \in \mathbb{R}^{p \times p}$.  Assume that $I-\tilde{A}$ is invertible, the population OLS estimator when regressing $V$ on $X$ is given by
  \begin{align}
   \theta^{\text{ols}}
   &= \left[c - g^{\top}\tilde{\Sigma}^{-1} \left(I-c \sigma^{2}gg^{\top} \tilde{\Sigma}^{-1} \right)v  \right]h 
   + \left(I-A^{\top}\right)\left[c\sigma^2 \tilde{\Sigma}^{-1}g+\tilde{\Sigma}^{-1} \left(I-c \sigma^{2}gg^{\top} \tilde{\Sigma}^{-1} \right)v  \right], \nonumber
\end{align}
where $c :=  \left(1+  \sigma^{2}g^{\top}\tilde{\Sigma}^{-1}g \right)^{-1}$, $\tilde{\Sigma} := \Sigma + vg^{\top}+gv^{\top}$, and
\begin{equation}
    \Cov(e,e) :=     \begin{bmatrix}
     \sigma^{2} &  v^{\top}\\
     v & \Sigma
    \end{bmatrix}, \nonumber
\end{equation}
with $\sigma^{2} := \Var(e_{V})$, $\Sigma := \Cov(e_{X},e_{X})$, and $v := \Cov(e_{X},e_{V})$. 
\end{lemma}
\smallskip

\begin{proof}
Let $P=[0_{1 \times d},I_{d\times d}]$, we continue from Equation~A.7 in~\cite{pfister2021stabilizing} as follows,
\begin{align}
    \theta^{\text{ols}} &= \Cov^{-1}(X,X) \Cov(X,V) \nonumber \\ 
    & = \left\{P(I-\tilde{A})^{-1}
    \begin{bmatrix}
     \sigma^{2} &  v^{\top}\\
    v & \Sigma
    \end{bmatrix} 
    \left[(I-\tilde{A})^{-1}\right]^{\top}P^{\top} \right\}^{-1}
P(I-\tilde{A})^{-1}\begin{bmatrix}
   \sigma^{2} \nonumber\\
   v
      \end{bmatrix}+h,
\end{align}
and it has been shown that 
\begin{equation}
    P(I-\tilde{A})^{-1} = [w,M], \nonumber
\end{equation}
for $M = (I-A-gh^{\top})^{-1}$ and $w = Mg$. Then,
\begin{align}
     \theta^{\text{ols}} &= \left[ \sigma^{2}ww^{\top} + wv^{\top}M^{\top} + Mvw^{\top} + M\Sigma M^{\top}\right]^{-1}   \left( \sigma^{2}w+Mv \right)+h \nonumber\\
   &:= \left[ \sigma^{2}ww^{\top} + M \tilde{\Sigma} M^{\top}\right]^{-1}\left( \sigma^{2}w+Mv \right) +h, \nonumber
\end{align}
where $\tilde{\Sigma} := \Sigma + vg^{\top}+gv^{\top}$ and it was computed in~\cite{pfister2021stabilizing} using the Sherman-Morrison formula~\cite{bartlett1951inverse} that
\begin{align}
    & \left[ \sigma^{2}ww^{\top} + M \tilde{\Sigma} M^{\top}\right]^{-1} \nonumber\\
    =&  (I-A-gh^{\top})^{\top}\tilde{\Sigma}^{-1} \left(I-c \sigma^{2}gg^{\top} \tilde{\Sigma}^{-1} \right)M^{-1}, \nonumber
\end{align}
with $c :=  \left(1+  \sigma^{2}g^{\top}\tilde{\Sigma}^{-1}g \right)^{-1}$.  
Then, some simple algebra leads to
\begin{align}
   \theta^{\text{ols}} &=  h + c \sigma^{2} (I-A-gh^{\top})^{\top}\tilde{\Sigma}^{-1}g 
     +  (I-A-gh^{\top})^{\top}\tilde{\Sigma}^{-1} \left(I-c \sigma^{2}gg^{\top} \tilde{\Sigma}^{-1} \right)v  \nonumber\\
   &= \left[c - g^{\top}\tilde{\Sigma}^{-1} \left(I-c \sigma^{2}gg^{\top} \tilde{\Sigma}^{-1} \right)v  \right]h 
    + \left(I-A^{\top}\right)\left[c\sigma^2 \tilde{\Sigma}^{-1}g+\tilde{\Sigma}^{-1} \left(I-c \sigma^{2}gg^{\top} \tilde{\Sigma}^{-1} \right)v  \right]. \nonumber
\end{align}
\end{proof}
\smallskip

\subsection{Proof of Proposition~\ref{prop_l}}
For each tuple $(k,R,S), k \in \{j \in MB(Y): \alpha_{j}^{e} = 0\}$. $R=-k:=\{1,\ldots,d\}\setminus k$, $S= \{1,\ldots,d\}$, we prove that $\lambda_{X}\ne 0$. Equivalently, we prove that  $\E_{l}[X_{k}|X_{-k},E=e]$  is a non-constant function of $e$.

First, recall that, when the target variables $Y$ is unobserved, the relations between the predictors in $\mathcal{M}^{e,1}$, are described by 
\begin{equation}
    X =\left( \gamma(\beta + \alpha^{e})^{\top} + B \right)X+ \gamma \varepsilon_{Y}+\varepsilon_{X}. \nonumber
\end{equation}
Now, we rewrite the above equation in the same form as the linear model~\eqref{linear_m} as follows,
\begin{align}
  \begin{bmatrix}
  X_{k} \\
  X_{-k}
  \end{bmatrix}
  =   \begin{bmatrix}
0 & \gamma_{k}(\beta+\alpha^{e})_{-k}^{\top}+B_{k,-k} \\
\gamma_{-k}\beta_{k}+B_{-k,k} &\gamma_{-k}(\beta+\alpha^{e})_{-k}^{\top} + B_{-k,-k}
  \end{bmatrix}
  \begin{bmatrix}
  X_{k} \\
  X_{-k}
  \end{bmatrix} +  
  \varepsilon_{X}+\gamma \varepsilon_{Y}, 
  \nonumber
\end{align}
where the top-left element of the coefficient matrix is zero, i.e., $\gamma_{k}(\beta_{k}+\alpha_{k}^{e})+B_{k,k} =0$ since $\alpha_{k}^{e}=0$ (by assumption), $B_{k,k}=0$ (due to acyclicity), and $\gamma_{k}\beta_{k}=0$ (since $X_{k}$ cannot be both a child and a parent of $Y$). Now, by Lemma~\ref{lemma_tech}, the population OLS estimator when regressing $X_{k}$ on $X_{-k}$ given $E=e$ is
 \begin{align}
   \theta^{\text{ols},k}(e)
   &= \left[c - g^{\top}\tilde{\Sigma}^{-1} \left(I-c \sigma^{2}gg^{\top} \tilde{\Sigma}^{-1} \right)v  \right]h 
   + \left(I-A^{\top}\right)\left[c\sigma^2 \tilde{\Sigma}^{-1}g+\tilde{\Sigma}^{-1} \left(I-c \sigma^{2}gg^{\top} \tilde{\Sigma}^{-1} \right)v  \right] \nonumber\\
   & := ah + (I-A^{\top})b, \label{ols_k_u}
\end{align}
where $h = \gamma_{k}(\beta+\alpha^{e})_{-k}+B_{k,-k}^{\top}$,   $v= \gamma_{k}\sigma_{Y}^{2}\gamma_{-k}$, $g = \beta_{k}\gamma_{-k}+B_{-k,k}$, $A = \gamma_{-k}(\beta+\alpha^{e})_{-k}^{\top} + B_{-k,-k}$, $\Sigma = \Cov(N_{X_{-k}},N_{X_{-k}})$, $\tilde{\Sigma} = \Sigma + vg^{\top}+gv^{\top}$, and $c =  \left(1+  \sigma^{2}g^{\top}\tilde{\Sigma}^{-1}g \right)^{-1}$. Note that $a \in \mathbb{R}$ and $b \in \mathbb{R}^{(d-1)}$ are not functions of $e$ and
\begin{equation}
    a = g^{\top}b+c(1+\sigma^{2}g^{\top}\tilde{\Sigma}^{-1}g) = g^{\top}b+1. \label{rela_ab}
\end{equation}

\begin{enumerate}[leftmargin=*]
\item $k \in CH(Y)$: First, observe that $\beta_{k}=0$ implies $g=B_{-k,k}$. By plugging $h$ and $A$ into~\eqref{ols_k_u}, we obtain
\begin{align}
     \theta^{\text{ols},k}(e) &= (a-\gamma^{\top}_{-k}b)\alpha_{-k}^{e}+a\gamma_{k}\beta_{-k} 
     +(I-\beta_{-k}\gamma_{-k}^{\top}-B_{-k,-k}^{\top})b, \nonumber
\end{align}
where $\alpha_{-k}^{e}$ in the first term is non-vanishing only if 
\begin{equation}
    a-\gamma^{\top}_{-k}b = 1+(B_{-k,k}-\gamma_{-k})^{\top}b \neq 0 \nonumber,
\end{equation}
where we use~\eqref{rela_ab}. 
    \item $k \in PA(Y)$: Observe that $v=0_{(d-1)}$ and $\gamma_{k}=0$, then
       \begin{align}
   \theta^{\text{ols},k}(e) &= c\alpha^{e}+c\beta+ c\sigma^{2} \left(I-A^{\top}\right)\Sigma^{-1}g \nonumber\\
   &= c(1+\sigma^{2}\beta_{k}\gamma_{-k}^{\top}\Sigma^{-1}\gamma_{-k} 
+\sigma^2 \gamma_{-k}^{\top}\Sigma^{-1}B_{-k,k})\alpha_{-k}^{e}+ c\beta
  +c\sigma^2(I-\beta_{-k} \gamma_{-k}^{\top}-B_{-k,-k}^{\top})\Sigma^{-1}g,
  \nonumber
\end{align} 
where the first term is not vanishing only if 
\begin{equation}
 1+\sigma^{2}\beta_{k}\gamma_{-k}^{\top}\Sigma^{-1}\gamma_{-k}+\sigma^2 \gamma_{-k}^{\top}\Sigma^{-1}B_{-k,k}\neq 0.   \nonumber 
\end{equation}
    \item $k \in \{j: \exists i\in CH(X_{j}) \text{ such that }i \in CH(Y) \}$:  Again, we have $v=0_{(d-1)\times 1}$ and $\gamma_{k}=0$. Additionally, we have $\beta_{k}=0$, then $\alpha_{-k}^{e}$ in $ \theta^{\text{ols},k}(e)$ is non-vanishing only if 
\begin{equation}
 1+\sigma^2 \gamma_{-k}^{\top}\Sigma^{-1}B_{-k,k}\neq 0.   \nonumber 
\end{equation}    

\end{enumerate}

\section{Proof of Corollary~\ref{coro:int_noise_var}}
\label{app:coro2}
First, we extract the assignments of $(X_{k},X_{R})$'s from~\eqref{eq_only_X} with $E=e$, where $k \in CH(Y)$ and $k \not \in DE(X_{i})$ for any $i \in CH(Y)\setminus k$ and $R = \{1,\ldots,d\} \setminus DE(Y)$ as follows, 
\begin{align}
  \begin{bmatrix}
  X_{k} \\
  X_{R}
  \end{bmatrix}
  &=   \begin{bmatrix}
0 & \gamma_{k}(\beta+\alpha^{e})_{R}^{\top}+B_{k,R} \\
0_{|R| \times 1} & B_{R,R}
  \end{bmatrix}
  \begin{bmatrix}
  X_{k} \\
  X_{R}
  \end{bmatrix} +  
    \begin{bmatrix}
 \varepsilon_{X,k}+\gamma_{k} \varepsilon_{Y}^{e}  \\
 \varepsilon_{X,R}
  \end{bmatrix},
  \nonumber
\end{align}
where the zeros in the coefficient matrix are due to the fact that all the descendants of $X_{k}$ are excluded from $X_{R}$ since $DE(X_{k}) \subseteq DE(Y)$ and $R = \{1,\ldots,d\} \setminus DE(Y)$. Note that $X_{k}$ is a child of $Y$ that is not a descendant of any other child of $Y$, thus any removed node $j \in \{1,\ldots,d\}\setminus \{k,R\}$ can not be the parent of any remaining node $i \in \{k,R\}$.

Now, using Lemma~\ref{lemma_tech}, the population OLS estimator when regressing $X_{k}$ on $X_{R}$ given $E = e$ is
\begin{equation}
    \theta^{\text{ols}}(e) =  \gamma_{k}(\beta+\alpha^{e})_{R}+B_{k,R}^{\top}.  \nonumber
\end{equation}
This implies 
\begin{equation}
    \E_{l}[X_{k}|X_{R}, E=e] =  \gamma_{k}(\beta+\alpha^{e})^{\top}X+B_{k,\cdot}X, \nonumber
\end{equation}
where we use the fact that $\beta_{j}=\alpha^{e}_{j} = B_{k,j} = 0$ for $j \not \in R$ (recall that $\{1,\ldots,d\} \setminus R$ does not contain either the parents of $X_{k}$ or parents of $Y$). Therefore,
\begin{equation}
     \E_{l}[X_{k}|X_{R},E=e]  = \gamma_{k}\E[Y|X_{PA(Y)},E=e]+ B_{k,\cdot}X,  \nonumber
\end{equation}
which is the second matching property.
\smallskip

\begin{remark}\label{rmk:int_var_coro2}
Observe that $\theta^{\text{ols}}(e)$ does not depend on $B_{R,R}$ or the covariance of $\varepsilon_{X,R}$, thus the second matching property holds even if every $ X_{j} \in X_{R}$ is intervened.  
\end{remark}

\section{Proof of Theorem~\ref{thm:int_xy}}
 For the first part, we only need to prove that $Y$ and $E$ are d-connected only through the arrow $E \to Y$ when conditioning on $X_{S}$,  $S \subseteq \{1,\ldots,d\}\setminus X^{\text{int}}(Y)$ such that $PA(Y)\subseteq S $. The rest is simply the proof of the first part of Theorem~\ref{thm1}. Since $E$ is a root node, $Y$ and $E$ can only be d-connected through the two types of paths,
\begin{enumerate}
    \item  $E \to \cdots \to Y$,
    \item $Y \to \cdots \to X_{j} \leftarrow \cdots \leftarrow E$,    
\end{enumerate}
where $j \in DE(Y)\cap DE(E)$ , and $E \to \cdots \to Y$ denotes any directed path from $E$ to $Y$. For the first type of paths, since $PA(Y) \subseteq S$, the only unblocked path when conditioning on $X_{S}$ is $E \to Y$. For the second type of paths that are V-structures, we have $j \not\in S$ since $X^{\text{int}}(Y)$ contains all the intervened children of $Y$ and the descendants of such children, implying that the V-structures are always blocked when conditioning on $X_{S}$. Thus, the only unblocked path between $E$ and $Y$ is $E \to Y$ when conditioning on $X_{S}$. 

For the second part, similarly, for $k \in \{1,\ldots,d\}\setminus \{PE \cup X^{\text{int}}(Y) \}$ and  $R \subseteq \{1,\ldots,d\} \setminus\{k, X^{\text{int}}(X_{k})\cup X^{\text{int}}(Y) \}$ such that $PA(X_{k})\setminus Y \subseteq R$, we prove that $X_{k}$ is d-connected with $E$ only through paths that contain the subpath $E \to Y$ when conditioning on $X_{R}$ (i.e., the second type of path below). Following the same idea as in the first part, $Y$ and $E$ are d-connected only through the arrow $E \to Y$ when conditioning on $X_{R}$ since $X^{\text{int}}(Y) \cap R =\varnothing$ and $PA(Y) \subseteq R$. Then, since $X_{k}$ is not intervened (i.e., $E \not \in PA(X_{k})$), the variables $E$ and $X_{k}$ can only be d-connected through three types of paths as follows,
\begin{enumerate}
    \item $E \to \dots \to X_{k}$ that does not contain $Y$,
    \item  $E \to Y  \to  \cdots  \to X_{k}$,
    \item $E \to \cdots \to X_{i} \leftarrow \cdots \leftarrow X_{k}$,
\end{enumerate}
where $i \in R$ and we use the fact that $k \not \in X^{\text{int}}(Y)$ (i.e., the node $k$ is not an intervened child of $Y$ or a descendant of an intervened child of $Y$). The first type of path is blocked immediately due to $PA(X_{k})\setminus Y \subseteq R$. 

To handle the third type of paths, we will need two technical results. (I) If $i \in R$, then $j \in R$ for any $j \in PA(i)$ such that $j \neq k$. This can be proved by contradiction. If $j \not \in R$ (i.e., $j \in X^{\text{int}}(Y) \cup X^{\text{int}}(X_{k})$), then we have $i \in X^{\text{int}}(Y) \cup X^{\text{int}}(X_{k})$ by the definition of $ X^{\text{int}}(Y)$ and $ X^{\text{int}}(Y)$, i.e., $ i \not \in R$. 
(II) For $i\in R$, observe that $X_{i}$ can only be a child of $X_{k}$, otherwise the path $X_{i} \leftarrow X_{j} \leftarrow \cdots  \leftarrow X_{k}$ is blocked by $X_{j} \in PA(X_{i})$ when conditioning on $X_{R}$, since $i \in R$ implies $j \in R$. 

Now, we proved that the third type of paths are always blocked when conditioning on $X_{R}$, by focusing on $E \to \cdots \to X_{i}$. For any $X_{i} \in CH(X_{k})$, the subpath $E \to \dots \to X_{i}$ cannot be $E \to X_{i}$, since $i \in R$ implies that $X_{i}$  is not an intervened child of $X_k$. Finally, the path $E \to \cdots\to X_{l} \to X_{i}$ is blocked by $X_{l}\in PA(X_{i})$ when conditioning on $X_{R}$, since $i \in R$ implies that $l \in R$.

\begin{remark}\label{rmk:thm2}
   We require $PA(Y)$ to be included $S$ in order to block every path from intervened ancestors to $Y$, but this is not necessary when the ancestors are not intervened. While it is important to include $PE$ in $S$ to handle interventions on $Y$. Also, including $PE$ will block paths from intervened ancestors that pass through variables in $PE$. Similar arguments apply to $R$ as well.  
\end{remark}

\section{Semi-parametric Varying Coefficient Models and Profile Least-Squares Estimation} \label{PLS_estimate} 
First, we introduce the semi-parametric varying coefficient model following most of the notation in~\cite{fan2005profile}. Consider $Y \in \mathbb{R}$, $U \in \mathcal{U}$, and two vectors of predictors $W = (W_{1},\ldots,W_{p})^{\top}$ and $Z=(Z_{1},\ldots,Z_{q})^{\top}$ such that $Z_{j}$'s have invariant coefficients,  a semi-parametric varying coefficient model over $(U,Y,W,Z)$ is defined by
\begin{equation}
        Y = M + \beta^{\top}Z + N,  \quad M =\alpha^{\top}(U) W, \label{svc_y}
\end{equation}
where $N$ is independent of $(U,W,Z)$. 
\smallskip

We briefly introduce the profile least-squares estimator of $\beta$ proposed in~\cite{fan2005profile}. Denote $n$ \iid samples of $(U,Y,Z,N)$ as $\bs{U} = (U_{1},\ldots,U_{n})^{\top}$, $\bs{Y} = (Y_{1},\ldots,Y_{n})^{\top}$, $\bs{W} = (\bs{W_{1}},\ldots,\bs{W_{n}})^{\top}$, $\bs{W}_{i}=(W_{i1}, \ldots, W_{ip})^{\top}$, $\bs{Z} = (\bs{Z}_{1},\ldots,\bs{Z}_{n})^{\top}$, $\bs{Z}_{i}=(Z_{i1}, \ldots, Z_{iq})^{\top}$, and $\bs{N} = (N_{1},\ldots,N_{n})^{\top}$. We thus have $\bs{M}= (M_{1},\ldots,M_{n})^{\top}$ with $M_{i}= \alpha^{\top}(U_{i})\bs{W}_{i}$. Let $\bs{K}_{u} = \text{diag}(K_{h}(U_{1}-u),\ldots,K_{h}(U_{n}-u))$ for some kernel function $K_{h}(\cdot) = K(\cdot/h)/h$ with bandwidth $h$, and
\begin{equation}
    \bs{\tilde{W}}_{u} = 
    \begin{bmatrix}
    \bs{W_{1}}^{\top} & \frac{U_{1}-u}{h}\bs{W_{1}}^{\top}\\
    \vdots & \vdots \\
    \bs{W_{n}}^{\top}& \frac{U_{n}-u}{h}\bs{W_{n}}^{\top}
    \end{bmatrix} \in \mathbb{R}^{n \times 2p}. \nonumber
\end{equation}
 For $u$ in a neighborhood of $u_{0} \in \mathcal{U}$, assume that each function $\alpha_{i}(u)$ in~\eqref{svc_y} can be approximated locally by the first-order Taylor expansion $\alpha_{i}(u) \approx \alpha_{i} + b_{i} (u-u_{0})$. 
Then the varying coefficients $\alpha(U_{1}), \ldots, \alpha(U_{n})$ can be estimated by solving the weighted local least squares problem
\begin{align}
  &\min_{\{a_{i},b_{i}\}}   \sum_{k=1}^{n}\left[\bs{Y}_{k} - \sum_{j=1}^{q}\beta_{j}Z_{kj} -\sum_{i=1}^{p}(a_{i}+b_{i}(U_{k}-u))W_{ki} \right]^{2} 
  \cdot K_{h}(U_{k}-u), \label{wls}
\end{align}
which has the solution
\begin{align}
    &[\hat{a}_{1}(u),\ldots, \hat{a}_{p}(u), h\hat{b}_{1}(u),\ldots, h\hat{b}_{p}(u)] 
    = (\bs{\tilde{X}}_{u}^{\top}\bs{K}_{u}\bs{\tilde{X}}_{u})^{-1}\bs{\tilde{X}}_{u}^{\top}\bs{K}_{u}(\bs{Y}-\bs{Z}\beta),  \nonumber
\end{align}
for $u \in \{U_{1},\ldots,U_{n}\}$. Then, each variable $\bs{M}_{i}$ can be estimated by 
$\bs{\hat{M}}_{i} = \sum_{j=1}^{p}\hat{a}_{j}(u)\bs{X}_{ij}$, and thus the vector $\bs{M}$ can be estimated by 
\begin{align}
    &\bs{\hat{M}}(\beta)= \nonumber\\ 
    & 
    \begin{bmatrix}
        [\bs{W}_{1}^{\top} \quad  0_{1\times p}]\{\bs{\tilde{W}}_{u_{1}}^{\top}\bs{K}_{u_{1}}\bs{\tilde{W}}_{u_{1}}\}^{-1}\bs{\tilde{W}}_{u_{1}}^{\top}\bs{K}_{u_{1}}\\
        \vdots \\
                [\bs{W}_{n}^{\top} \quad 0_{1\times p}]\{\bs{\tilde{W}}_{u_{n}}^{\top}\bs{K}_{u_{n}}\bs{\tilde{W}}_{u_{n}}\}^{-1}\bs{\tilde{W}}_{u_{n}}^{\top}\bs{K}_{u_{n}}\\
    \end{bmatrix}(\bs{Y}-\bs{Z}\beta) \nonumber\\
     & := A(\bs{Y}-\bs{Z}\beta), \label{my_hat_t1}
\end{align}
which depends on the unknown parameter $\beta$ that will be estimated below. Substituting $\bs{\hat{M}}(\beta)$ into the vector form of~\eqref{svc_y}, we obtain $(I-A)\bs{Y} = (I - A)\bs{Z}\beta + \bs{N}$. The profile least-squares estimator of $\beta$ is given by
\begin{align}
\hat{\beta} &=  \{\bs{Z}^{\top}(I-A)^{\top}(I-A)\bs{Z}\}^{-1}  \cdot\bs{Z}^{\top}(I-A)^{\top}(I-A)\bs{Y}. \label{pld_by}
\end{align}
Finally, by replacing $\beta$ in~\eqref{my_hat_t1} with  $\hat{\beta}$, the final form of the estimator for $\bs{M}$ is given by 
\begin{align}
\bs{\hat{M}} &= A(\bs{Y}-\bs{Z}\hat{\beta} ).  \label{my_hat}
\end{align}

\section{Technical Lemmas for the Proof of Theorem~\ref{thm_general}}\label{app:tech_lem}
The semi-parametric varying coefficient model and profile least-squares estimation are introduced in Appendix~\ref{PLS_estimate}. First, we present some technical assumptions and two technical lemmas from~\cite{fan2005profile}. Let $c_{n}'  = \left\{ \frac{\log(1 /h)}{nh}\right\}^{1/2}$ and $c_{n}  = c_{n}'+h^2$.  
\begin{enumerate}[leftmargin=*]
    \item $U$ has a bounded support $\mathcal{U}$ and has density function $f(\cdot)$ that is Lipschitz continuous and bounded away from $0$.
    \item For each $U_{i}$, the matrix $\E[W^{\top}W|U_{i}]$ is non-singular, and the matrices $\E[W^{\top}W|U_{i}]$, $\left(\E[W^{\top}W|U_{i}]\right)^{-1}$, and $\E[WZ^{\top}|U_{i}]$ are Lipschitz continuous. 
    \item $\alpha_{1}(u), \ldots, \alpha_{p}(u)$ have continuous second derivatives.  
    \item  $K(\cdot)$ is a symmetric density function.
   \item There exists an $s>2$ such that $\E[||X||^{2s}]<\infty$, $\E[||Z_{Y}||^{2s}]<\infty$, and $\E[||Z_{V}||^{2s}]<\infty$ and there exists $\varepsilon<2- s^{-1}$ such that $n^{2\varepsilon-1}h \to \infty$. 
    \item The bandwidth $h$ satisfies $nh^{8} \to 0$ and $nh^{2}/ (\log(n))^{2} \to \infty$. 
\end{enumerate}
\smallskip
\begin{lemma}[\!\cite{mack1982weak}]
\label{tech_lemma}
    Let $(U_{1},Y_{1}), \ldots, (U_{n},Y_{n})$ be \iid random vectors in $\mathbb{R}^{2}$. Assume that $\E[|Y|^{s}] <\infty$ and $\sup_{x}\int |y|^{s}f(u,y)dy <\infty$, where $f(u,y)$ is the density of $(U,Y)$. Let $K$ be a bounded positive function with a bounded support that satisfies a Lipschitz condition. Given that $n^{2\varepsilon-1}h \to \infty$ for some $\varepsilon < 1- s^{-1}$, then
    \begin{align}
        &\sup_{u} \left|\frac{1}{n}\sum_{i=1}^{n}K_
        {h}(U_{i}-u)Y_{i}-\E[K_{h}(U_{i}-u)Y_{i}]\right|=O_p(c_n'). \nonumber
    \end{align}
\end{lemma}

\begin{lemma}[\!\cite{fan2005profile}]
\label{asy_norm}
Under assumptions $1)\sim 6)$, $\sqrt{n}  (\hat{\beta}- \beta) \xrightarrow{d} \mathcal{N}(0,\Sigma)$ as $n \to \infty$, where $\Sigma = \Var(N)\cdot C^{-1}$ with 
\begin{align*}
    C=\E[ZZ^{\top}] - \E \big[ \E[ZW^{\top}|U]\E[WW^{\top}|U] \E[WZ^{\top}|U] \big]. 
\end{align*}


\end{lemma}

In the following lemma, we provide the rates of several quantities needed for the proof of Theorem~\ref{thm_general}. 


\begin{lemma}\label{lem_mhat}
Under the same assumptions as in Lemma~\ref{asy_norm},
\begin{align}
R_{1} &= \frac{1}{n}(\hat{\bs{M}}-\bs{M})^{\top}(\hat{\bs{M}} -\bs{M}) = O_{p}(c_{n}^{2} \vee n^{-1}), \nonumber\\
R_{2} &= \frac{1}{n}(\hat{\bs{M}}-\bs{M})^{\top}[\bs{W},\bs{M}] =O_{p}(c_{n} \vee n^{-1/2}), \nonumber\\
R_{3} &= \frac{1}{n}(\hat{\bs{M}}-\bs{M})^{\top}\bs{Z} = O_{p}(c_{n} \vee n^{-1/2}), \nonumber\\
R_{4} &= \frac{1}{n}(\hat{\bs{M}}-\bs{M})^{\top}\bs{N} = O_{p}(c_{n}). \nonumber
\end{align}
\end{lemma}

\noindent\begin{proof} First,~\eqref{my_hat} and the vector form of~\eqref{svc_y} give
\begin{align}
    \bs{\hat{M}} - \bs{M} &= A(\bs{Y}-\bs{Z}\hat{\beta}) -\bs{M} \nonumber \\
    &=  (A-I)\bs{M} + A\bs{Z}(\beta - \hat{\beta}) + A\bs{N}. \nonumber
\end{align}
Observe that $R_{2} \sim R_{4}$ can be defined through
\begin{align}
      I_{1} &=\frac{1}{n} \bs{M}^{\top}(A^{\top}-I)\bs{P},   \nonumber\\
    I_{2} &=\frac{1}{n} (\beta-\hat{\beta})^{\top}\bs{Z}^{\top}A^{\top} \bs{P}, \nonumber\\
     I_{3} &=\frac{1}{n} \bs{N}^{\top}A^{\top} \bs{P}, \nonumber
\end{align}
where $\bs{P}$ can be replaced by $\bs{W}$, $\bs{Z}$, $\bs{M}$, or $\bs{N}$ (note that the rows of $\bs{P}$ are \iid). We also have $R_{1} = \sum_{i=4}^9 I_{i}$, where
\begin{align}
    I_{4} &=\frac{1}{n} \bs{M}^{\top}(A^{\top}-I)(A-I)\bs{M},
    \nonumber\\
     I_{5} &=\frac{1}{n} (\beta-\hat{\beta})^{\top}\bs{Z}^{\top}A^{\top}A\bs{Z}(\beta-\hat{\beta}), \nonumber\\
      I_{6} &=\frac{1}{n} \bs{N}^{\top}A^{\top}A\bs{N}  \nonumber,\\
     I_{7} &=\frac{1}{n}\bs{M}^{\top}(A^{\top}-I) \bigg\{ A\bs{Z}(\beta - \hat{\beta})+ A\bs{N} \bigg\},\nonumber\\
     I_{8} &=\frac{1}{n}(\beta-\hat{\beta})^{\top}\bs{Z}^{\top}A^{\top} \bigg\{(A-I)\bs{M}+ A\bs{N}  \bigg\}\nonumber,\\
    I_{9} &=\frac{1}{n}\bs{N}^{\top}A^{\top} \bigg\{    (A-I)\bs{M}+A\bs{Z}(\beta - \hat{\beta})   \bigg\}\nonumber.
\end{align}
It can be shown that $  I_{1}, I_{3} =O_{p}(c_{n})$ (we use this as a shorthand for $I_{1}=O_{p}(c_{n}), I_{3} =O_{p}(c_{n})$, as $I_{1}$ and $I_{3}$ are not equal), $I_{2} = O_{p}(n^{-1/2})$, $ I_{5} = O_{p}(n^{-1})$, $I_{4}, I_{6} =  O_{p}(c_{n}^{2})$, $I_{8} = O_{p}(c_{n}n^{-1/2})$, $ I_{7}, I_{9} = O_{p}(c_{n}^{2} \vee c_{n}n^{-1/2})$, which implies $R_{1} = O_{p}(c_{n}^{2} \vee n^{-1})$, $R_{2}, R_{3}=O_{p}(c_{n} \vee n^{-1/2})$ and $R_{4} = O_{p}(c_{n})$. The techniques for proving the rates of $I_{1} \sim I_{9}$ are similar; observe that all the components in $I_{4} \sim I_{9}$ are already computed to obtain $I_{1} \sim I_{3}$, thus we only provide the proof for $I_{1} \sim I_{3}$ for simplicity of presentation. 

Using Lemma~\ref{tech_lemma}, it has been shown in~\cite{fan2005profile} that $\bs{\tilde{W}}_{u_{i}}^{\top}\bs{K}_{u_{i}}\bs{\tilde{W}}_{u_{i}}$ can be equivalently expressed as 
\begin{align}
\begin{bmatrix}
B_{1} & B_{2}\\
B_{3} & B_{4}
\end{bmatrix} 
=n f(U_{i}) \E[WW^{\top}|U_{i}] \otimes 
    \begin{pmatrix}
    1 & 0 \\
    0 & \mu_{2}
    \end{pmatrix}\{1+ O_{p}(c_{n})\}, \label{XtKXt}
\end{align}
where $\mu_{2} = \int_{\mathcal{U}} u^{2}K(u)du$, and the four block matrices are $B_1=S_0(U_i)$, $B_2=B_3=S_1(U_i)$, and $B_4=S_2(U_i)$, with respect to
 \begin{align}
    S_{k}(U_i) &= \sum_{j=1}^{n}\left(\frac{U_{j}-U_{i}}{h}\right)^{k} \bs{W}_{j}\bs{W}_{j}^{\top}K_{h}(U_{j}-U_{i}).   \nonumber 
\end{align}
 Since the techniques for proving~\eqref{XtKXt}, omitted  in~\cite{fan2005profile}, will be used repeatedly in the rest of this paper, we provide the proof for completeness. Applying Lemma~\ref{tech_lemma}, it holds uniformly in $u \in \mathcal{U}$ that $S_{k}(u)$ can be expressed as
  \begin{align}
&\E\left[\left(\frac{U-u}{h}\right)^{k}\E[ WW^{\top}|U]K_{h}(U-u)\right] \{1+ O_{p}(c_{n}')\}\nonumber\\
    =&  \int_{\mathcal{V}} v^{k}\E[ WW^{\top}|U = hv+u]K(v) f(hv+u)dv  \cdot\{1+ O_{p}(c_{n}')\} \label{var_rep}\\
    =&  \int_{\mathcal{V}} v^{k}K(v)\left[\E[ WW^{\top}|U=u]+ vO(h)\right] 
  \cdot\left[ f(u)+vO(h) \right] dv  \{1+ O_{p}(c_{n}')\}\label{lipsch} \\
    =&
   \begin{cases}
       \E[ WW^{\top}|U=u] f(u)  \mu_{k} \{1+ O_{p}(c_{n})\}, & k \text{ even} \label{cases_eq}\\
    O(h)+O_{p}(c_{n}'),  & k \text{ odd} 
\end{cases} 
\end{align}
where~\eqref{var_rep} is due to the change of variable $V= (U-u)/h$, \eqref{lipsch} uses the Lipschitz continuity assumptions on $\E[XX^{T}|U]$ and $f(\cdot)$, and~\eqref{cases_eq} is by the symmetry of the kernel function $K(\cdot)$.
Similarly, we obtain
\begin{align}\label{XtKX}
    &\bs{\tilde{W}}_{u_{i}}^{\top}\bs{K}_{u_{i}}\bs{M} \nonumber\\
    =& \begin{bmatrix}
   \sum_{j=1}^{n}\bs{W}_{j}\bs{W}_{j}^{\top}\alpha(U_{j})K_{h}(U_{j}-U_{i})\\ \sum_{j=1}^{n}\frac{U_{j}-U_{i}}{h} \bs{W}_{j}\bs{W}_{j}^{\top} \alpha(U_{j})K_{h}(U_{j}-U_{i})
    \end{bmatrix} \nonumber\\
    =& n f(U_{i}) \E[WW^{\top}|U_{i}]\alpha(U_{i}) \otimes
    \begin{pmatrix}
    1 & 0 
    \end{pmatrix}^{\top}\{1+ O_{p}(c_{n})\}.
\end{align}
Recall the expression of $A$ in~\eqref{my_hat_t1}, we have
\begin{align}
 &(A-I)\bs{M} \nonumber\\
   =&\begin{bmatrix}
        [\bs{W}_{1}^{\top} \quad  \bs{0}]\{\bs{\tilde{W}}_{u_{1}}^{\top}\bs{K}_{u_{1}}\bs{\tilde{W}}_{u_{1}}\}^{-1}\bs{\tilde{W}}_{u_{1}}^{\top}\bs{K}_{u_{1}}\bs{M}\\
        \vdots \\
       [\bs{W}_{n}^{\top} \quad \bs{0}]\{\bs{\tilde{W}}_{u_{n}}^{\top}\bs{K}_{u_{n}}\bs{\tilde{W}}_{u_{n}}\}^{-1}\bs{\tilde{W}}_{u_{n}}^{\top}\bs{K}_{u_{n}}\bs{M}\\
    \end{bmatrix} -\bs{M} \nonumber 
\end{align}
Using~\eqref{XtKXt} and~\eqref{XtKX}, we obtain
\begin{align}
    I_{1} &= \frac{1}{n} \bs{M}^{\top}(A^{\top}-I)\bs{P} \nonumber\\
    &=\frac{1}{n} \sum_{i=1}^{n} (M_{i}\{1+O_{p}(c_{n})\}-M_{i}) \bs{P}_{i}=O_{p}(c_{n}), \nonumber
\end{align}
where $\bs{P}_{i}$ denotes the $i^{\text{th}}$ row of $\bs{P}$ and the last equality is due to the law of large numbers.
For $I_{2}$,  similarly as above, we compute
\begin{equation}
    A\bs{Z} =
    \begin{bmatrix}
    \bs{W_{1}}^{\top}\left\{\E[WW^{T}|U_{1}]\right\}^{-1}\E[WZ^{\top}|U_{1}]\\
    \vdots\\
     \bs{W_{n}}^{\top}\left\{\E[WW^{T}|U_{n}]\right\}^{-1}\E[WZ^{\top}|U_{n}]
    \end{bmatrix}\{1+O_{p}(c_{n})\}. \nonumber
\end{equation}
Since $\beta-\hat{\beta} =  O_{p}(n^{-1/2})$ by Lemma~\ref{asy_norm}, we obtain 
\begin{align}
        I_{2} &=\frac{1}{n} (\beta-\hat{\beta})^{\top}\bs{Z}^{\top}A^{\top} \bs{P} \nonumber \\
        &=(\beta-\hat{\beta})^{\top}\frac{1}{n} \sum_{i=1}^{n}  \bs{W_{i}}^{\top}\left\{\E[WW^{T}|U_{i}]\right\}^{-1} 
        \cdot\E[WZ^{\top}|U_{i}]\{1+O_{p}(c_{n})\}\bs{P}_{i}  \nonumber\\
        &= O_{p}(n^{-1/2}), \nonumber
\end{align}
where, again, the last equality is due to the law of large numbers. Finally, for
\begin{align}
   A\bs{N}  &= \begin{bmatrix}
        [\bs{W}_{1}^{\top} \quad  \bs{0}]\{\bs{\tilde{W}}_{u_{1}}^{\top}\bs{K}_{u_{1}}\bs{\tilde{W}}_{u_{1}}\}^{-1}\bs{\tilde{W}}_{u_{1}}^{\top}\bs{K}_{u_{1}}\bs{N}\\
        \vdots \\
                [\bs{W}_{n}^{\top} \quad \bs{0}]\{\bs{\tilde{W}}_{u_{n}}^{\top}\bs{K}_{u_{n}}\bs{\tilde{W}}_{u_{n}}\}^{-1}\bs{\tilde{W}}_{u_{n}}^{\top}\bs{K}_{u_{n}}\bs{N} \\
    \end{bmatrix}, \nonumber
\end{align}
the same argument for~\eqref{cases_eq} leads to 
\begin{align}
    \bs{\tilde{W}}_{u_{i}}^{\top}\bs{K}_{u_{i}}\bs{N}=  n f(U_{i}) \E[WN^{\top}|U_{i}] \{1+ O_{p}(c_{n})\}, \nonumber
\end{align}
where $\E[WN^{\top}|U_{i}]=0$ since $N$ is independent of $W$ and $U_{i}$, and $N$ has a zero mean. Thus, by the law of large numbers,
\begin{align}
        I_{3} &=\frac{1}{n} \bs{N}^{\top}A^{\top} \bs{P} =\frac{1}{n} \sum_{i=1}^{n} O_{p}(c_{n})\bs{P}_{i}= O_{p}(c_{n}).\nonumber
\end{align}

\end{proof}

\section{ Proofs of Theorem~\ref{thm_general} and Corollaries~\ref{coro_thm_gen} and~\ref{discrete_IMP}}\label{pf_age} 

To reuse model~\eqref{svc_y} for the prediction of $Y^{\tau}$, the main challenge comes from $M$ that changes with $U$ (while $\beta$ remains invariant). First, we introduce some notation for the proofs. We define $(\bs{Y}, \bs{W}, \bs{W}_i)$ and $(\bs{Z}, \bs{Z}_{i}, \bs{M}, \bs{N})$ in the same way as in~Appendix~\ref{PLS_estimate}. Similarly, we define $\bs{Z}_{V}, \bs{M}_{V}$, $\bs{N}_{V}$, and all the corresponding data matrices for the test data (e.g., $\bs{Y}^{\tau}$), and let $\sigma^{2} = \E[(N^{\tau})^{2}]$. With this notation, the IMP~\eqref{invar_coeff} implies that
\begin{equation}
    M = \lambda M_{V} + \zeta^{\top}W := [W,M_{V}]w, \label{imp_for_m}
\end{equation} 
for some $\zeta \in \mathbb{R}^{p}$. Let $\bs{\hat{W}}_{V} := [\bs{W},\bs{\hat{M}}_{V}]$ and $\bs{\hat{W}}_{V}^{\tau} := [\bs{W}^{\tau},\bs{\hat{M}}_{V}^{\tau}]$. Then, the OLS estimator of $w$ according to the above equation is given by 
\begin{equation}
    \hat{w} = (\bs{\hat{W}}_{V}^{\top}\bs{\hat{W}}_{V})^{-1}\bs{\hat{W}}_{V}^{\top}\bs{\hat{M}}. \label{estimate_w}
\end{equation}
We predict $\bs{Y}^{\tau}$ using the continuous IMP estimator
\begin{equation}
   \bs{ \hat{Y}}^{\tau} = \bs{\tilde{W}}_{V}^{\tau} \hat{w} + \bs{Z}^{\tau}\hat{\beta}, \label{y_hat_t}
\end{equation}
where $\hat{\beta}$, $\bs{\hat{M}}$,  and $\bs{\hat{M}}_{V}$ are provided in Appendix~\ref{PLS_estimate}. 

\smallskip
\begin{lemma}\label{rate_theta}
Under assumptions 1)$\sim$ 4)  in Appendix~\ref{app:tech_lem}, it holds that
\begin{equation}
     \hat{w}- w = O_{p}(c_{n} \vee n^{-1/2}).  \nonumber
\end{equation}
\end{lemma}

\smallskip
\begin{proof}
Using the fact that
\begin{equation}
    \bs{\hat{W}}_{V} = [\bs{W},\bs{M}_{V}]  + [\bs{0},\bs{\hat{M}}_{V} - \bs{M}_{V}], \nonumber
\end{equation}
we obtain
\begin{align}
&\frac{1}{n}\bs{\hat{W}}_{V}^{\top}  \bs{\hat{W}}_{V} \nonumber\\
     & = \frac{1}{n} [\bs{W},\bs{M}_{V}]^{\top} [\bs{W},\bs{M}_{V}]+\frac{1}{n} [\bs{W},\bs{M}_{V}] ^{\top} [\bs{0},\bs{\hat{M}}_{V} - \bs{M}_{V}] \nonumber\\ 
     &\hspace{1em}+ \frac{1}{n}[\bs{0},\bs{\hat{M}}_{V} - \bs{M}_{V}]^{\top}[\bs{W},\bs{M}_{V}]  +\frac{1}{n} [\bs{0},\bs{\hat{M}}_{V} - \bs{M}_{V}]^{\top} 
    \cdot  [\bs{0},\bs{\hat{M}}_{V} - \bs{M}_{V}] \nonumber\\
    &= \E\bigg[(W^{\top},M_{V})^{\top}(W^{\top},M_{V})\bigg] \{1+ O_{p}(c_{n}\vee n^{-1/2})\}, \nonumber
\end{align}
where we use $R_{1}$ and $R_{2}$ from Lemma~\ref{lem_mhat} and the law of large numbers. Similarly,
\begin{align}
\frac{1}{n}\bs{\hat{W}}_{V}^{\top}\bs{\hat{M}}  &=\frac{1}{n} [\bs{W},\bs{M}_{V}]^{\top}\bs{M}+ \frac{1}{n} [\bs{W},\bs{M}_{V}]^{\top}(\bs{\hat{M}}-\bs{M})\nonumber\\ 
    &\hspace{5em}+\frac{1}{n} [\bs{0},\bs{\hat{M}}_{V} - \bs{M}_{V}]^{\top}(\bs{\hat{M}}-\bs{M})  
    + \frac{1}{n} [\bs{0},\bs{\hat{M}}_{V} - \bs{M}_{V}]^{\top}\bs{M} \nonumber\\
    &= \E\bigg[(W^{\top},M_{V})^{\top}M\bigg]\{ 1+O_{p}(c_{n} \vee n^{-1/2})\}.  \nonumber
\end{align}
Note that the rate is not directly implied by Lemma~\ref{lem_mhat}, but it can be proved using the same techniques demonstrated in the proof Lemma~\ref{lem_mhat}. Therefore, using~\eqref{imp_for_m}, 
\begin{align}
    \hat{w}  = (\bs{\hat{W}}_{V}^{\top}\bs{\hat{W}}_{V})^{-1}\bs{\hat{W}}_{V}^{\top}\bs{\hat{M}}  =  w \{1+O_{p}(c_{n} \vee n^{-1/2})\}. \nonumber
\end{align}
\end{proof}

\smallskip

\subsection{Proof of Theorem~\ref{thm_general}}
Using~\eqref{y_hat_t} and $\bs{Y}^{\tau} = \bs{M}^{\tau}+\bs{Z}^{\tau}\beta+\bs{N}^{\tau}$, we derive
\begin{align}
    & \bs{\hat{Y}^{\tau}} - \bs{Y}^{\tau} \nonumber
   \\ =&  [\bs{W}^{\tau}, \bs{\hat{M}}^{\tau}_{V}] \hat{w} + \bs{Z}^{\tau}\hat{\beta} - [\bs{W}^{\tau}, \bs{M}^{\tau}_{V}] w - \bs{Z}^{\tau}\beta -\bs{N}^{\tau} \nonumber\\
   =&[\bs{0}, \bs{\hat{M}}^{\tau}_{V} - \bs{M}^{\tau}_{V}] (\hat{w}- w) + 
   [\bs{W}^{\tau}, \bs{M}^{\tau}_{V}] (\hat{w}-w) 
  +  
   [\bs{0}, \bs{\hat{M}}^{\tau}_{V} - \bs{M}^{\tau}_{V}] w +
   \bs{Z}^{\tau}(\hat{\beta}  -\beta)
   -\bs{N}^{\tau}.  \nonumber
\end{align}
Then, the generalization error is given by
\begin{align}
   \frac{1}{m}(\bs{\hat{Y}^{\tau}} - \bs{Y}^{\tau})^{\top}(\bs{\hat{Y}^{\tau}} - \bs{Y}^{\tau}) = \sum_{i=1}^{10} J_{i}
    \nonumber
\end{align}
where
\begin{align}
J_{1} &= \frac{1}{m} (\hat{w}- w)^{\top} [\bs{0}, \bs{\hat{M}}^{\tau}_{V} - \bs{M}^{\tau}_{V}]^{\top}
 [\bs{0}, \bs{\hat{M}}^{\tau}_{V} - \bs{M}^{\tau}_{V}] (\hat{w}- w),  \nonumber\\ 
     J_{2} &= \frac{1}{m} (\hat{w}-w)^{\top} [\bs{X}^{\tau}, \bs{M}^{\tau}_{V}]^{\top}  [\bs{X}^{\tau}, \bs{M}^{\tau}_{V}] (\hat{w}-w), \nonumber \\
      J_{3} &= \frac{1}{m}   w^{\top} [\bs{0}, \bs{\hat{M}}^{\tau}_{V} - \bs{M}^{\tau}_{V}]^{\top} [\bs{0}, \bs{\hat{M}}^{\tau}_{V} - \bs{M}^{\tau}_{V}] w,  \nonumber \\ 
       J_{4} &=\frac{1}{m} (\hat{\beta}  -\beta)^{\top}  (\bs{Z}^{\tau})^{\top}\bs{Z}^{\tau}(\hat{\beta}  -\beta), \nonumber \\
        J_{5} &= \frac{1}{m}(\bs{N}^{\tau})^{\top}\bs{N}^{\tau} \nonumber, \\ 
         J_{6} &= \frac{1}{m}(\hat{w}-w)^{\top}  [\bs{0}, \bs{\hat{M}}^{\tau}_{V} - \bs{M}^{\tau}_{V}]^{\top}\bigg\{
          [\bs{W}^{\tau}, \bs{M}^{\tau}_{V}]  \nonumber\\
          &\hspace{5em} \cdot (\hat{w}-w)+  [\bs{0}, \bs{\hat{M}}^{\tau}_{V} - \bs{M}^{\tau}_{V}] w +
   \bs{Z}^{\tau}(\hat{\beta}  -\beta)
   -\bs{N}^{\tau} \bigg\},  \nonumber\\
     J_{7} &= \frac{1}{m}(\hat{w}-w)^{\top} [\bs{W}^{\tau}, \bs{M}^{\tau}_{V}]^{\top} \bigg\{
     [\bs{0}, \bs{\hat{M}}^{\tau}_{V} - \bs{M}^{\tau}_{V}] \nonumber\\
     &\hspace{5em}\cdot (\hat{w}-w)+  [\bs{0}, \bs{\hat{M}}^{\tau}_{V} - \bs{M}^{\tau}_{V}] w +
   \bs{Z}^{\tau}(\hat{\beta}  -\beta)
   -\bs{N}^{\tau}\bigg\},  \nonumber\\
    J_{8} &= \frac{1}{m} w^{\top}[\bs{0}, \bs{\hat{M}}^{\tau}_{V} - \bs{M}^{\tau}_{V}]^{\top} \bigg\{[\bs{0}, \bs{\hat{M}}^{\tau}_{V} - \bs{M}^{\tau}_{V}] \nonumber\\
    &\hspace{5em}\cdot(\hat{w}- w)+ 
   [\bs{W}^{\tau}, \bs{M}^{\tau}_{V}] (\hat{w}-w)
  +\bs{Z}^{\tau}(\hat{\beta}  -\beta)
   -\bs{N}^{\tau}  \bigg\}, \nonumber\\
   J_{9} &=\frac{1}{m}(\hat{\beta}  -\beta)^{\top}(\bs{Z}^{\tau})^{\top} \bigg\{[
   \bs{0}, \bs{\hat{M}}^{\tau}_{V} - \bs{M}^{\tau}_{V}] (\hat{w}- w)\nonumber\\
   &\hspace{5em}+ 
   [\bs{W}^{\tau}, \bs{M}^{\tau}_{V}] (\hat{w}-w)+
   [\bs{0}, \bs{\hat{M}}^{\tau}_{V} - \bs{M}^{\tau}_{V}] w +\bs{N}^{\tau} \bigg\}, \nonumber\\
   J_{10} &= -\frac{1}{m} (\bs{N}^{\tau})^{\top} \bigg\{[\bs{0}, \bs{\hat{M}}^{\tau}_{V} - \bs{M}^{\tau}_{V}] (\hat{w}- w)  \nonumber\\
   &\hspace{5em}+[\bs{W}^{\tau}, \bs{M}^{\tau}_{V}]
  (\hat{w}-w)+  
   [\bs{0}, \bs{\hat{M}}^{\tau}_{V} - \bs{M}^{\tau}_{V}] w +
   \bs{Z}^{\tau}(\hat{\beta}  -\beta)\bigg\} \nonumber.
\end{align}
We can show the following rates of these terms through simple applications of Lemmas~\ref{asy_norm},~\ref{lem_mhat},~\ref{rate_theta}, along with the law of large number and the central limit theorem. $J_{2},J_{3} = O_{p}(c_{n}^2 \vee n^{-1})$, $J_{4} = O_{p}(n^{-1})$, $J_{5} =\sigma^{2} + O_{p}(m^{-1/2})$,
\begin{align}
    J_{10} &= O_{p}(c_{m} \vee m^{-1/2})\cdot O_{p}(c_{n} \vee n^{-1/2}) 
    +O_{p}(c_{n} \vee n^{-1/2}) +O_{p}(c_{m} \vee m^{-1/2}) + O_{p}(n^{-1/2}), \nonumber
\end{align}
and $J_{1},J_{6} \sim J_{9}$ have higher orders compared with either $O_{p}(c_{n} \vee n^{-1/2})$ or $O_{p}(c_{m} \vee m^{-1/2})$ in $J_{10}$. This completes the proof of Theorem~~\ref{thm_general}. 


\subsection{Proof of Corollary~\ref{coro_thm_gen}}
We use $l$ as the shorthand notation for $l_n$ in the proof. Since the estimation of $\beta, \bs{M},$ and $w$ are based on the labeled data, we have $\hat{\beta}-\beta = O_{p}(l^{-1/2})$ in Lemma~\ref{asy_norm} and $R_{1}, R_{2}, R_{3} = O_{p}(c_{l} \vee l^{-1/2})$ and $R_{4} = O_{p}(c_{l})$ in Lemma~\ref{lem_mhat}. In the proof of Lemma~\ref{rate_theta}, the rate of $\bs{\hat{W}}_{V}^{\top}  \bs{\hat{W}}_{V}$ remains the same since the estimation of $\bs{M}_{V}$ only uses the unlabeled data but the rate of $ \bs{\hat{W}}_{V}^{\top}\bs{\hat{M}}$ now depends on $l$. This observation implies $  \hat{w}- w = O_{p}(c_{l} \vee l^{-1/2})$. Now, observe that $J_{2} = O_{p}(c_{l}^2 \vee l^{-1})$, $J_{3} = O_{p}(c_{m} \vee m^{-1/2})$, $J_{4} = O_{p}(l^{-1})$, $J_{5} =\sigma^{2} + O_{p}(m^{-1/2})$, and $J_{10} = O_{p}(c_{l} \vee l^{-1/2}) +O_{p}(m^{-1/2})$. Since $\max(\frac{l}{n},\frac{l}{m}) \to 0$ as $\min(n,m) \to \infty$, we get $\sum_{i=1}^{10}J_{i} = O_{p}(c_{l} \vee l^{-1/2})$.  


\subsection{Proof of Corollary~\ref{discrete_IMP}}
According to the definition of the discrete IMP estimator, all the coefficients are treated as varying coefficients (i.e., $\beta=\bs{0}$ and $\beta_{V}=\bs{0}$), and
 $\bs{M}$ is estimated by performing the OLS for each environment and then putting the estimates into one vector, thus $R_{1} = O_{p}(a_{n}^{2})$ and $R_{2}, R_{4} = O_{p}(a_{n})$ in Lemma~\ref{lem_mhat}, with $a_{n} = (\min_{e \in \mathcal{E}^{\text{train}}} n_{e})^{-1/2}$ by the asymptotic normality of the OLS estimators ( note that $X$ and $Z$ are assumed to have finite fourth moments in Lemma~\ref{asy_norm}). Accordingly, we have $\hat{w}- \hat{w} = O_{p}(a_{n})$ in Lemma~\ref{rate_theta} using the law of large numbers. Setting $Z^{\tau}= \bs{0}$ in $J_{1} \sim J_{10}$, we obtain $J_{2},J_{3} = O_{p}(a_{n}^2)$, $J_{4}=0$, $J_{5} = \sigma^{2} +  O_{p}(m^{-1/2})$, $J_{10} = O_{p}(a_{n}) +O_{p}(a_{m})$, and the other terms have higher orders compared with $ O_{p}(a_{n})$ or $ O_{p}(a_{m})$. Similar to Theorem~\ref{thm_general}, the asymptotic generalization error is dominated by $J_{10}$. 
\end{appendices}


%

\ifCLASSOPTIONcaptionsoff
  \newpage
\fi

\end{document}